\newcommand{\red}[1]{{\color{black}#1}}

\documentclass[draft]{agujournal2019}
\usepackage{url} 
\usepackage{lineno}
\usepackage[inline]{trackchanges} 
\usepackage{soul}
\usepackage{lscape}
\usepackage{amsmath}
\usepackage{amssymb}
\draftfalse

\journalname{Geochemistry, Geophysics, Geosystems}

\begin{document}

%
%

\title{The origin of Earth's mantle nitrogen: primordial or early biogeochemical cycling?}

%
%




\authors{H. Kurokawa\affil{1}, M. Laneuville\affil{1}, Y. Li\affil{1}, N. Zhang\affil{1}, Y. Fujii\affil{2}, H. Sakuraba\affil{3}, C. Houser\affil{1}, H. J. Cleaves II\affil{1,4,5}}


\affiliation{1}{Earth-Life Science Institute, Tokyo Institute of Technology}
\affiliation{2}{Division of Science, National Astronomical Observatory of Japan}
\affiliation{3}{Department of Earth and Planetary Sciences, Tokyo Institute of Technology}
\affiliation{4}{Blue Marble Space Institute of Science}
\affiliation{5}{Earth and Planets Laboratory, Carnegie Institution of Washington}




\correspondingauthor{H. Kurokawa}{hiro.kurokawa@elsi.jp}




\begin{keypoints}
\item We studied two possible origins for mantle nitrogen: the primordial and recycling scenarios.
\item The primordial scenario requires accretion of excess nitrogen and impact erosion during late accretion.
\item The recycling scenario needs efficient N fixation and biotic processing on early Earth.
\end{keypoints}

%
%

%
%


\begin{abstract}
Earth's mantle nitrogen (N) content is comparable to that found in its N-rich atmosphere. Mantle N has been proposed to be primordial or sourced by later subduction, yet its origin has not been elucidated. Here we model N partitioning during the magma ocean stage following planet formation and the subsequent cycling between the surface and mantle over Earth history using argon (Ar) and N isotopes as tracers. The partitioning model, constrained by Ar, shows that only about 10\% of the total N content can be trapped {in the solidified mantle} due to N's low solubility in magma and low partitioning coefficients in minerals in oxidized conditions supported from geophysical and geochemical studies. A possible solution for the primordial origin is that Earth had about 10 times more N at the time of magma ocean solidification. We show that the excess N could be removed by impact erosion during late accretion. The cycling model, constrained by N isotopes, shows that mantle N can originate from efficient N subduction, if the sedimentary N burial rate on early Earth is comparable to that of modern Earth. Such a high N burial rate requires biotic processing. Finally, our model provide\red{s} a methodology to distinguish the two possible origins with future analysis of the surface and mantle N isotope record.
\end{abstract}

\section*{Plain Language Summary}

Nitrogen (N) is the main component of Earth's atmosphere, and essential for life. The atmospheric N content influences Earth's climate and capability to retain its surface water. Primary biological production is limited by bio-available N as well as phosphorous on modern Earth. It has been recently recognized that Earth's interior contains N comparable to that found in its atmosphere, and thus its origin is important for our understanding of Earth-life co-evolution. We modeled N partitioning in Earth's molten stage and long-term cycling after Earth's solidification. Two scenarios are proposed from our modeling. One is that Earth's mantle acquired its modern N content in the earlier stage due to an excess amount of N Earth accreted, which was later lost to space following asteroid impacts. Another is that Earth's mantle acquired N via subduction of N-rich sediments, which requires the sedimentary N burial rate on early Earth comparable to the modern value sustained by biological activity. The two scenario\red{s} can be tested with future analysis of the geochemical record of surface and mantle N.

%
%

%


%
%
%
%

\section{Introduction}
\label{sec:introduction}

Nitrogen (N) is the main component of Earth's atmosphere and essential for the biosphere as it is an integral part of the building blocks of life. Though molecular nitrogen (N$_2$) is transparent in the infrared, its presence in the atmosphere increases the warming effect of existing greenhouse gases by pressure broadening of their absorption lines \cite{Goldblatt+2009}. The N inventory in the atmosphere is also important \red{for sustaining} water on planets by the cold-trap mechanism \cite{Wordsworth+Pierrehumbert2014}. Primary biological production in modern aquatic environments is mostly limited by either fixed N or phosphorus \cite{Zerkle2018}. Though deficits in biologically-available N can be mitigated by microbial N fixation on modern Earth, it might have been a limiting factor for early life on Earth.

Though N was once thought to predominantly reside in the atmosphere and biosphere, it can indeed become incorporated into the geosphere: minerals and rocks \cite<e.g,>[]{Johnson+Goldblatt2015}. Combined sample analyses with N-$^{40}$Argon (Ar) geochemistry \red{\cite{Marty1995}} suggest that present Earth's mantle contains a comparable mass of N as the atmosphere \cite{Marty+Dauphas2003,Halliday2013,Bergin+2015,Johnson+Goldblatt2015}. A mantle which hosts N at a similar level as found in the atmosphere has great potential for influencing the surface environment through ingassing and outgassing of this mantle reservoir. Determining whether mantle N is a primordial feature set during the magma ocean stage or evolved with the subduction of sediments by plate tectonics is essential to our understanding of Earth's formation and early biogeochemical cycling.


Mantle N has been proposed to be sourced by subduction of N-bearing sediments and the altered oceanic lithosphere \cite{Marty+Dauphas2003,Barry+Hilton2016,Johnson+Goldblatt2018,Stueken+2021}, which would also lead to secular evolution of atmospheric N partial pressure ($p_{\rm N_2}$) \cite{Zerkle+Mikhail2017,Laneuville+2018}. Notably, such subducting N on modern Earth has ${\rm \delta ^{15}N \sim +6}$\textperthousand \ \cite{Cartigny+Marty2013,Halama+2014,Johnson+Goldblatt2018}, where ${\rm \delta ^{15}N = [(^{15}N/^{14}N)_{sample}/(^{15}N/^{14}N)_{standard}} - 1] \times 1000$. Thus, efficient subduction should lead to $^{15}$N enrichment in the mantle. Carbon-bearing deep mantle diamonds have been found to host N with almost exclusively positive ${\rm \delta}^{15}$N values \cite{Regier2020}, which might be consistent with an origin from subducted oceanic sediments.

However, mantle N sampled from mid-ocean ridge basalt (MORB) is rather depleted in $^{15}$N (${\rm \delta ^{15}N}\sim-5$\textperthousand), which is known as N isotopic disequilibrium \red{\cite{Marty1995,Cartigny+Marty2013}}. In this regard, {several} previous studies hypothesized that mantle N is predominantly a remnant of planet formation \cite{Cartigny+Marty2013,Mallik+2018,Labidi+2020}. In this primordial origin scenario, the isotopic disequilibrium is attributed to atmospheric escape, which enriches the {residual} atmosphere in $^{15}$N. Another study focusing on N isotopes proposed a recycling origin scenario in which the N isotope composition of sediments has changed over time \cite{Marty+Dauphas2003}. We note that here the prefix \textquotedblleft re\textquotedblright \ is given because, in this scenario, the modern mantle N is a component once degassed to the atmosphere in the magma ocean stage. 

Ar is another tracer of N partitioning and cycling \red{\cite{Marty1995}}. In oxidized magma, where N exists in the form of N$_2$, N and Ar behave similarly \cite<e.g.,>[]{Marty+Dauphas2003}. However, under reducing conditions N exists as NH$_4^+$, which can substitute for K$^{+}$ in mineral phases, allowing for more efficient recycling of N than Ar. The large difference in $^{40}$Ar/$^{36}$Ar ratios between the modern atmosphere and mantle implies that the two reservoirs have largely stopped exchanging Ar with each other since early catastrophic degassing, and radiogenic $^{40}$Ar derived from $^{40}$K has continued to accumulate in the mantle \cite<>[and references therein]{Ozima+Podosek2002}.

Previous studies which argued either primordial or recycling scenarios have greatly contributed to our understanding of the origin of mantle N, but they have missed some important behavior. N partitioning between the atmosphere and magma ocean has been experimentally investigated by several studies \cite{Libourel+2003,Boulliung+2020}, which concluded that mantle N could be primordial. However, these studies have not considered N degassing upon magma ocean solidification, which might limit N partitioning into the solidified magma significantly \cite{Hier-Majumder+Hirschmann2017}. Models which treated long-term N cycling after the magma ocean solidification did not trace N isotopic exchange between different reservoirs \cite{Stueken+2016ABb,Johnson+Goldblatt2018} or introduced a simplified assumption on the N isotopic ratio of the upper mantle \cite{Tolstikhin+Marty1998} or sediments \cite{Labidi+2020}. These points are covered by this study to further test the two scenarios (see Section \ref{subsec:discussion:comparison} for more detailed comparison with these previous studies).

Here we test the primordial and recycling origin scenarios using numerical models for N partitioning and cycling coupled with Ar and N isotopes. Section \ref{sec:methods} presents our models. Section \ref{sec:results} shows the model results. We discuss the origin of N in the mantle and the implications for the future analysis of N isotope record of early Earth and for other rocky planets in Section \ref{sec:discussion}, and conclude in Section \ref{sec:conclusions}.

\section{Methods}
\label{sec:methods}

\begin{figure}
    \centering
    \includegraphics[width=\linewidth]{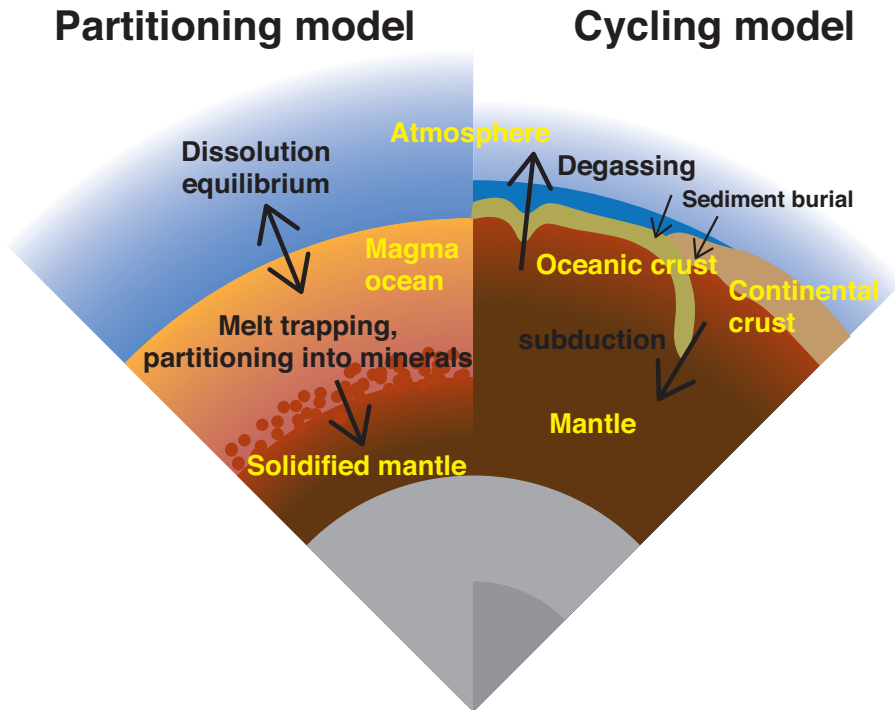}
    \caption{Schematic view of our models. Reservoirs and exchange fluxes are written in yellow and black, respectively.}
    \label{fig:model}
\end{figure}

\begin{table}
    \centering
    \begin{tabular}{llll}
         Reservoir & N content [PAN] & ${\rm \delta^{15}N}$ [\textperthousand] & $^{36}$Ar content [PA36Ar] \\ \hline
        Atmosphere & 1 & 0 & 1 \\
        Mantle & 0.41--2.53$^{\rm a}$ & -10 to -2$^{\rm b}$ & 2.8--5.7$\times 10^{-3}$ $^{\rm c}$ \\
        {Continental crust} & {0.25--0.43$^{\rm d}$} & {+1 to +10$^{\rm e}$} & -- \\
        {Oceanic crust} & {0.038--0.15$^{\rm f}$} & {+4 to +10$^{\rm g}$} & -- \\
    \end{tabular}
    \caption{Constraints on our model results. 1 {present atmospheric N} (PAN) = $4.0\times 10^{18}$ kg. 1 PA36Ar = $2.1\times 10^{14}$ kg. a: \citeA{Marty+Dauphas2003,Halliday2013,Bergin+2015}. b: \citeA{Marty+Dauphas2003}. c: \citeA{Bekaert+2020}. {d: \citeA{Rudnick+Gao2003,Johnson+Goldblatt2015,Johnson+Goldblatt2017}. e: \citeA{Busigny+Bebout2013}. f: \citeA{Johnson+Goldblatt2015}. g: \citeA{Cartigny+Marty2013}.}}
    \label{tab:constraints}
\end{table}

\subsection{Model overview and constraints}
\label{subsec:model_constraints}

We modeled N partitioning in the magma ocean stage and N cycling after magma ocean solidification separately (Figure \ref{fig:model}).  Our first model {tested} whether N in the modern mantle {could} be primordial by computing the N content trapped in the mantle upon solidification. Partitioning {was} also computed for $^{36}$Ar because the discrepancy in N/$^{36}$Ar ratios between the present atmosphere and mantle has been used to discuss the origin of mantle N \cite{Marty+Dauphas2003}. Our second model {tested} whether mantle N can originate from secular subduction. This cycling model explicitly calculates $^{15}$N/$^{14}$N ratios in the surface reservoirs (the atmosphere and continental and oceanic crust) and mantle, which can be used as another constraint.

Table \ref{tab:constraints} summarizes the constraints on model results: N contents, $\delta^{15}$N values, and $^{36}$Ar contents in the four largest reservoirs: the atmosphere, continental and oceanic crust, and mantle. We adapted a N content estimate for the modern mantle derived from N-$^{40}$Ar geochemistry by \citeA{Bergin+2015}: $0.41$--$2.53$ times present atmospheric N $4.0\times 10^{18}$ kg \cite<hereafter PAN,>[]{Johnson+Goldblatt2015}. This range brackets independent estimates by \citeA{Marty+Dauphas2003} ($0.41$--$1.27$ PAN) and \citeA{Halliday2013} ($0.42$--$1.12$ PAN). We {adapted} the N isotopic composition of the modern mantle of ${\rm \delta^{15} N} = -10$ to $-2$\textperthousand \ as informed from MORB \cite{Marty+Dauphas2003}. We note that \citeA{Johnson+Goldblatt2015} derived a higher estimate ($2$--$10$ PAN) by assuming the existence of the \textquotedblleft high-N mantle,\textquotedblright \ which is associated with some oceanic island basalts (OIB) and OIB-related xenoliths. This proposed component is isotopically distinct \cite<$\delta^{15}$N = +5\textperthousand,>[]{Johnson+Goldblatt2015}. The high-N estimate {was} not used in our main results as its presence is not assumed in the other studies, but we discuss it further in Section \ref{subsec:discussion:uncertainties}. The $^{36}$Ar amounts of modern atmosphere and mantle {were} adapted from the estimate of \citeA{Bekaert+2020}.

\subsection{Partitioning model}
\label{subsec:model_partitioning}

\begin{table}
    \centering
    \begin{tabular}{lll}
        \hline
        density & $\rho$ & PREM $^{\rm a}$ \\
        difference between solidus and liquidus & $\Delta T$ & F-peridotitic model $^{\rm b}$ \\
        heat capacity & $C_p$ & 1000 [${\rm J\ kg^{-1}\ K^{-1}}$] $^{\rm c}$ \\
        entropy change on melting & $\Delta S$ & 300 [${\rm J\ kg^{-1}\ K^{-1}}$] $^{\rm c}$ \\
        disaggregation melt fraction & $\phi_{\rm c}$ & 0.3 $^{\rm c}$ \\
        compaction time & $\tau$ & 4.35 Myr $^{\rm c}$ \\
        solidification time & $t_{\rm s}$ & 0.2--7 Myr $^{\rm d}$ \\
        solubility of N & $S_{\rm N}$ & Equation \ref{eq:SN} $^{\rm e}$ \\
        partitioning coefficient of N & $D_{\rm N}$ & 0.0073, 0.020, 0.0030 $^{\rm f}$ \\
        solubility of Ar & $S_{\rm Ar}$ & 0.2 [ppm/MPa] $^{\rm g}$ \\
        partitioning coefficient of Ar & $D_{\rm Ar}$ & 0.0011 $^{\rm h}$ \\ \hline
    \end{tabular}
    \caption{Parameter values used in our partitioning model. a: \citeA{Dziewonski+Anderson1981}. b: \citeA{Monteux+2016}. c: \citeA{Hier-Majumder+Hirschmann2017}. d: \citeA{Hamano+2013,Salvador+2017,Nikolaou+2019}. e: Used \citeA{Boulliung+2020} data (see \ref{sec:appendix_SN}). f: For upper mantle, transition zone, and lower mantle from \citeA{Yoshioka+2018}. g: Model of \citeA{Iacono+2010}, assuming a bulk silicate Earth (pyrolite) magma composition \cite{McDonough+Sun1995} and 1,700 K. h: \citeA{Heber+2007}.} \label{tab:partitioning}
\end{table}

Our partitioning model calculated the N and $^{36}$Ar amounts trapped in the solidified mantle upon magma ocean cooling after core-mantle separation (Figure \ref{fig:model}). \red{Our model assumes bottom-up solidification, but the case for a basal magma ocean is discussed in Section \ref{subsec:discussion:uncertainties}.} In contrast to previous studies which modeled N partitioning \cite{Libourel+2003,Boulliung+2020,Gaillard+2022}, degassing caused by magma solidification, which limits N trapping into the mantle, was considered. We did not calculate N isotopes in this first model. Equilibrium degassing enriches the residual melt in $^{15}$N by $\sim$+1\textperthousand\ \cite<e.g.,>[]{Dalou+2019}. As we discuss quantitatively in Section \ref{subsec:discussion:Norigin1}, however, atmospheric escape processes enrich atmospheric $^{15}$N easily and should be responsible for the observed isotopic disequilibrium between the current atmosphere and mantle in the primordial origin scenario.

\subsubsection{N partitioning between reservoirs}

Our method is based on the melt-trapping model of \citeA{Hier-Majumder+Hirschmann2017}. The relevant parameters are summarized in Table \ref{tab:partitioning}. Dissolution equilibrium {was} assumed between the atmosphere and magma ocean for each time step, given as \red{(see \ref{sec:appendix_Miatm} for derivation)},
\begin{equation}
    M^{i}_{\rm atm} = \frac{M^{i}_{\rm atm+mo}}{1+{(\bar{m}/m_i)\cdot}S_i M_{\rm mo} g / A_{\rm E}}, \label{eq:Miatm}
\end{equation}
\begin{equation}
    M^{i}_{\rm mo} = M^{i}_{\rm atm+mo} - M^{i}_{\rm atm}, \label{eq:Miatmmo}
\end{equation}
where $M^{i}_{\rm j}$ is the mass of element $i$ (N or $^{36}$Ar) in the reservoir $j$ (atm = atmosphere, mo = magma ocean, and sm = solidified mantle), $m_i$ is the molecular mass in the atmosphere (28 and 36 amu for N$_2$ and $^{36}$Ar), \red{$\bar{m}$ is the mean molecular mass of atmospheric gases,} $S_i$ is the solubility, $g$ is the gravitational acceleration at the surface, and $A_{\rm E}$ is the total surface area of Earth. The value without a superscript $i$ to identify an element is the mass of the reservoir itself (e.g., the mass of magma ocean $M_{\rm mo}$). \red{Here we assumed that the volatile abundances in the magma ocean are constant in depth to derive Equation \ref{eq:Miatm}. This assumption is justified because the viscosity of the magma ocean is so low that the mixing time scale is much shorter than the cooling timescale \cite<the orders of days to weeks vs. Myrs,>[]{Solomatov2015,Nikolaou+2019,Sossi+2020}.} The mean molecular \red{mass $\bar{m}$} was assumed to be 28 amu, assuming CO as the dominant gas. This assumption is valid for Earth's magma ocean \cite{Sossi+2020}. While $S_{\rm ^{36}Ar}$ was assumed to be constant, $S_{\rm N}$ is dependent on N partial pressure, $p_{\rm N_2}$ ($= (\bar{m}/m_{\rm N_2})\cdot M_{\rm atm}^{\rm N}g/A_{\rm E}$, Equation \ref{eq:SN}). Thus, Equations \ref{eq:Miatm} and \ref{eq:SN} were iteratively solved to obtain $S_{\rm N}$ and $M_{\rm atm}^{\rm N}$. Our N solubility model, based on \citeA{Boulliung+2020}, assumed atmospheric N is in the form of N$_2$ (\ref{sec:appendix_SN}), which is expected for the oxygen fugacity range considered here \cite<e.g., that of>[]{Schaefer+Fegley2017}. As magma solidifies at the bottom of magma ocean, the volatile content in the solidified mantle increases as,
\begin{equation}
    \frac{d M^{i}_{\rm sm}}{dt} = [F_{\rm tl} + D_i(1-F_{\rm tl})] C_i \frac{d M_{\rm sm}}{dt}, \label{eq:dMidt}
\end{equation}
where $t$ is the time, $F_{\rm tl}$ is the trapped melt fraction, $D_i$ is the partitioning coefficient between minerals and silicate melt, and $C_i$ is the concentration of the element $i$ in the magma ocean ($C_i \equiv M^i_{\rm mo}/M_{\rm mo}$). The first and second terms in the bracket of Equation \ref{eq:dMidt} correspond to melt trapping and partitioning into minerals, respectively. The trapped mass is subtracted from the rest of the system, as,
\begin{equation}
    \frac{d M^{i}_{\rm atm+mo}}{dt} = - \frac{d M^{i}_{\rm sm}}{dt}.
\end{equation}
The trapped melt fraction as a function of cooling rate is defined as the fraction of the solidified mass trapped as interstitial melt relative to the total solidified mass (solidified interstitial melt plus crystallized minerals), and given by \cite{Hier-Majumder+Hirschmann2017},
\begin{equation}
    F_{\rm tl} = - \frac{\phi_{\rm c} \tau}{\Delta T} \frac{dT_{\rm p}}{dt},\label{eq:Ftl}
\end{equation}
where $\phi_{\rm c}$ is the disaggregation melt fraction, $T_{\rm p}$ is the mantle potential temperature, $\Delta T$ is the difference between the solidus and the liquidus temperatures at the bottom of magma ocean, and $\tau$ is the compaction time of the mushy layer at the bottom of magma ocean. \red{The upper limit of $F_{\rm tl} = 0.3$ is given by the melt fraction for rheological transition \cite{Hier-Majumder+Hirschmann2017}}.

\subsubsection{Thermal evolution}

The thermal evolution of Earth was solved to compute $dT_{\rm p}/dt$ in Equation \ref{eq:Ftl}, as given by \cite{Hier-Majumder+Hirschmann2017},
\begin{equation}
    \frac{dT_{\rm p}}{dt} = - L_{\rm int} \biggl[ C_p (M_{\rm E}-M_r(a)) - 4\pi \rho T_{\rm p} \Delta S a^2 \frac{d a}{d t} \biggr]^{-1},
\end{equation}
where $L_{\rm int}$ is the intrinsic luminosity, $M_{\rm E}$ is Earth's mass, $M_r$ is the enclosed mass, $\rho$ and $C_p$ are the density and the heat capacity of the magma, and $a$ is the radius of the solidification front. We derived a fitting formula for $a$ to reproduce Figure 2 of \citeA{Hier-Majumder+Hirschmann2017} as,
\[
  a(T_{\rm p}) = \begin{cases}
  R_{\rm E}\ (T_{\rm p} < 1810\ {\rm K}), \\
  -1.587\times 10^{-1} T_{\rm p}^3 + 8.83571\times 10^2 T_{\rm p}^2 -1.641\times 10^6 T_{\rm p} + 1.0234\times 10^9\ [{\rm m}]\ \\ (1810\ {\rm K} < T_{\rm p} < 2100\ {\rm K}), \\
  R_{\rm core} (2100\ {\rm K} < T_{\rm p}),
  \end{cases}
\]
where $R_{\rm E}$ and $R_{\rm core}$ are the radii of Earth and the core, respectively. Computing the intrinsic luminosity $L_{\rm int}$ requires a radiative-convective model for the atmospheric structure with complete description of molecular composition, cloud properties, etc. \cite<e.g.,>[]{Salvador+2017,Nikolaou+2019}. Instead, we assumed a constant value for $L_{\rm int}$ estimated from the given solidification time $t_{\rm s}$. 

\subsubsection{Initial conditions and parameters}

The redox state of the mantle influences N solubility and, consequently, the trapped N content. Earth's magma ocean in its final stage is thought to be more oxidized than $\log_{10}f_{\rm O_2,\Delta{\rm IW}} = -1$ \cite{Badro+2015,Armstrong+2019,Sossi+2020}, where $f_{\rm O_2,\Delta{\rm IW}}$ is the oxygen fugacity relative to that of the iron-w\"{u}stite buffer. Following \citeA{Dalou+2017}, we {considered} three cases -- oxidized ($\log_{10}f_{\rm O_2,\Delta{\rm IW}} = -1$), {intermediate} ($\log_{10}f_{\rm O_2,\Delta{\rm IW}} = -2$), and {reduced} ($\log_{10}f_{\rm O_2,\Delta{\rm IW}} = -3.5$) cases (Table \ref{tab:partitioning}), where the latter two are shown for comparison with the realistic, oxidized model. 

We considered $t_{\rm s} = 0.2$--$7$ Myrs, which covers the range of estimates in previous studies which simulated the magma ocean solidification \cite{Hamano+2013,Salvador+2017,Nikolaou+2019}.

A fully molten magma ocean ($T_{\rm p} = 2100$ K) was assumed for the initial condition. The assumption maximizes the amount of trapped N. Initially the total N content (Table \ref{tab:constraints}) was assumed to be in the atmosphere and magma ocean with the dissolution equilibrium. As we focus on the evolution after the major accretion and core-mantle separation, we did not model partitioning into alloy which would determine the total N content in concert with atmospheric loss during the major accretion stage \cite{hirschmann2016,Grewal+2019,Grewal+2021,Sakuraba+2021} before the magma ocean solidification stage we considered here.

\subsection{Cycling model}
\label{subsec:model_cycling}

\begin{figure}
    \centering
    \includegraphics[width=0.8\linewidth]{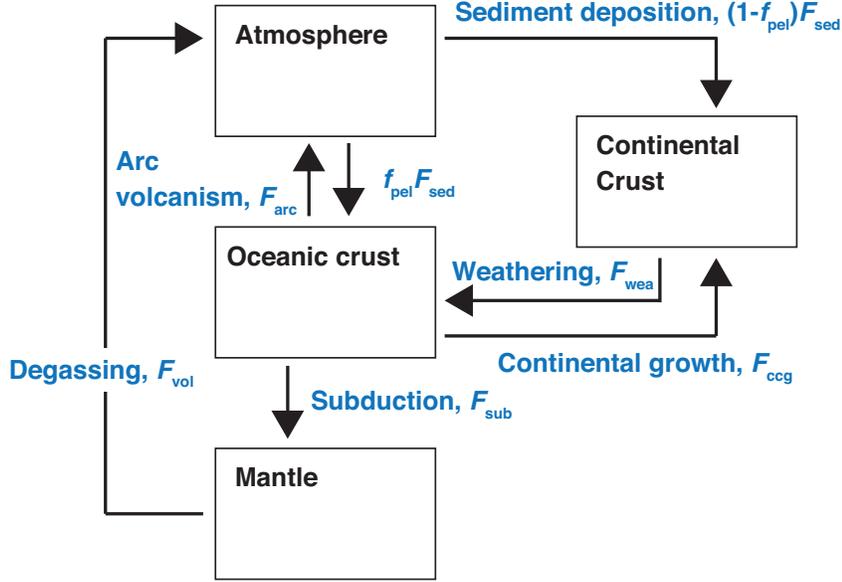}
    \caption{A schematic image of our N cycling model.}
    \label{fig:cycling_model}
\end{figure}

To test the recycling origin scenario, our N cycling model simulated the secular evolution of N masses and isotope ratios in the four major reservoirs: the atmosphere, continental and oceanic crust, and mantle (Figure \ref{fig:cycling_model}). Our model explicitly treats more reservoirs than a recent study considering isotopic evolution \cite{Labidi+2020} but is simplified compared to the N cycling models that were developed based on geophysics and (bio-)geochemistry \cite{Stueken+2016,Laneuville+2018,Johnson+Goldblatt2018}. Our model was designed to understand i) the requirement for realizing the recycling origin, and ii) its isotopic signatures as explained below.

Evolution of N masses and isotopic ratios ($^{15}$N/$^{14}$N) in the reservoirs was calculated by using the following equations \cite{Kurokawa+2018EPSL},
\begin{equation}
    \frac{d M_i}{dt} = \sum_{\rm sources} F_k - \sum_{\rm sinks} F_k,
\end{equation}
\begin{equation}
    \frac{d}{dt} (M_iI_i) = \sum_{\rm sources} (F_k \alpha_k I_{i'}) - \sum_{\rm sinks} (F_k \alpha_k I_i),
\end{equation}
where $M_i$ and $I_i$ are the mass and isotopic ratio of N in the reservoir $i$ (atm: atmosphere, cc: continental crust, oc: oceanic crust, and ma: mantle), and $F_k$ and $\alpha_k$ are the flux of the process $k$ and its isotopic fractionation factor. We considered net isotopic fractionation in sediment deposition (the atmosphere to sediments on the crust, through fixed N in the oceans). The other processes are known to induce relatively small isotopic fractionation only \cite<typically $\simeq$1\textperthousand,>[]{Busigny+Bebout2013,Cartigny+Marty2013}. A potential caveat here is that N isotopic fractionation in the subduction zones and in the mantle has not been fully understood, which is discussed in Section \ref{subsec:discussion:uncertainties}.

\subsubsection{Sediment burial}

Surface N cycling evolved through time \cite<e.g.,>[]{Shen+2006,Stueken+2016,Ader+2016}. On modern Earth, biology mediates primary N fixation from N$_2$, and dissolved nitrate in the ocean is the main form of bio-available N. In contrast, abiotic N fixation might have been a dominant source of fixed N on early Earth \cite{Navarro+2001}, especially before the development of biological N fixation \cite{Stueken+2015}, and nitrate would be limited because of limited biological oxidative nitrification \red{in} anaerobic environments. A fraction of fixed N is continuously removed from the N pool as sedimentary deposits, which is ultimately transported to the mantle via subduction through time (Figure \ref{fig:cycling_model}).

Because how the sedimentary N deposition rate changed through time is poorly constrained, we simply parameterized the flux as,
\begin{equation}
    F_{\rm sed}(t) = F_{{\rm sed}, 0} \times 10^{C_{\rm sed}(t_0-t)/t_0},
\end{equation}
where $t_0$ = 4.5 Gyr, $F_0$ is the sediment deposition (N burial) flux on modern Earth \cite<0.455 PAN/Gyr,>[]{Stueken+2016ABb} and $C_{\rm sed}$ is the power-law parameter and can be either positive or negative. \red{This parameterization does not require explicit modeling of the size of biosphere, which influences surficial N cycling, but we discuss the implications for the early biosphere by utilizing the obtained values of $C_{\rm sed}$ (Section \ref{subsec:discussion:Norigin2}).} The majority of sediments \red{is} deposited onto continental shelves (modeled as a part of the continental crust), and the rest onto the oceanic crust as pelagic sediments. We assumed the fraction of pelagic sediments, $f_{\rm pel}$, to be 0.038 \cite{Berner1982,Stueken+2016ABb}. 

Sedimentary N can be incorporated into rocks after deposition. This process was not explicitly included in our model as the crust and overlying sediments are treated as a single reservoir, but it was implicitly considered as an internal process.

Because there exists an isotopic disequilibrium within the N cycle \cite{Cartigny+Marty2013}, a successful recycling model should take into account the evolution of the surficial N cycle which controls the net fractionation factor between atmospheric and sedimentary N. The net isotopic fractionation factor from the atmosphere to the sediments was modeled as occurring over four stages. 

In Period 1 ($>$3.2 Ga), $\Delta {\rm ^{15}N} = -9$\textperthousand \ was assumed (by definition, $10^3 \ln \alpha_{\rm sed} = \Delta {\rm ^{15}N}$). This was motivated by lower mean ${\rm \delta^{15}N}$ values ($-3.6$\textperthousand) of organic matter in {the} Eo-paleoarchean \cite<e.g.,>[]{Marty+Dauphas2003,Shen+2006}. The adapted factor ($\sim-9$\textperthousand) is the lowest value of the sedimentary record (assuming ${\rm \delta^{15}N}=0$\textperthousand \ for the atmosphere) as well as the value typically expected for abiotic N-fixation via lightning or photochemistry \cite{Moore+1977,Navarro+2001,Kuga+2014,Stueken+2021}. Anaerobic processing after N fixation (and thus absence of nitrate) was implicitly assumed in this stage, as we did not consider $^{15}$N enrichment due to denitrification. Abiotically-fixed, $^{15}$N-depleted ammonia \cite{Moore+1977,Navarro+2001} would have been consumed by chemosynthetic bacteria in hydrothermal systems under anaerobic conditions in this period \cite{Shen+2006}. 

In Period 2 (3.2--2.75 Ga), $\Delta {\rm ^{15}N} = 0$\textperthousand. The environment is still anaerobic, but biotic N fixation (diazotrophy) began and became the major source of bio available N in this period \cite{Stueken+2015}. 

During Period 3 (2.75--2.4 Ga), $\Delta {\rm ^{15}N} = +10.8$\textperthousand. Aerobic N cycling (coupled nitrification, denitrification, and assimilation) started in this period \cite{Garvin+2009}. Dissimilatory denitrification leads to residual nitrate being kinetically enriched in $^{15}$N, and the isotopically enriched nitrate is then reduced to ammonium and converted to organic matter and finally trapped in sediments \cite<e.g.,>[]{Stueken+2016}. Larger N fractionation as recorded in organic matter compared to the modern value can be explained by limited amounts of nitrate \cite{Shen+2006}. 

In Period 4 ($<$2.4 Ga), $\Delta {\rm ^{15}N} = +6$\textperthousand. The isotopic shift during nitrification-denitrification-assimilation would have decreased to the modern average value due to the formation of a sizable global nitrate pool \cite{Shen+2006}. 

\subsubsection{Continental weathering}

Continental N is gradually transported to the oceanic crust via weathering. The weathering flux is assumed to be proportional to $M_{\rm cc}$ as,
\begin{equation}
    F_{\rm wea} = \frac{M_{\rm cc}}{\tau_{\rm wea}},
\end{equation}
where \red{$\tau_{\rm wea}$} is the timescale of weathering. The timescale depends on atmospheric O$_2$ partial pressure, and here modeled as \cite{Johnson+Goldblatt2018},
\begin{equation}
    \tau_{\rm wea} = \tau_{{\rm wea}, 0} \cdot \biggl[ a_{\rm wea} + (1-a_{\rm wea}) \frac{PAL_{\rm O_2}}{PAL_{\rm O_2}+k_{\rm wea}} \biggr]^{-1}, \label{eq:PAL}
\end{equation}
where $\tau_{{\rm wea}, 0}$ is the weathering timescale on modern Earth ($\tau_{{\rm wea}, 0} = 0.3\ {\rm Gyr}$), $PAL_{\rm O_2}$ is the atmospheric O$_2$ partial pressure scaled to present atmospheric levels for which we used the model of \citeA{Johnson+Goldblatt2018} developed based on geochemical reconstruction \cite<e.g.,>[]{Lyons+2014}, $a_{\rm wea}$ is the fraction of available material weathered under anoxic conditions ($a_{\rm wea} = 0.1$), and $k_{\rm wea}$ is the weathering constant ($k_{\rm wea} = 10^{-3}$), respectively \cite{Johnson+Goldblatt2018}. \red{The dependence on $PAL_{\rm O_2}$ is not critical for the surface-mantle N exchange, but it is important to reproduce the Archean low $p_{\rm N_2}$ record \cite<>[see Section \ref{subsec:results:cycling}]{Marty+2013,Som+2012,Som+2016}.}

\subsubsection{Subduction, arc volcanism, and continental growth}

Sedimentary N on the oceanic crust is eventually conveyed to subduction zones. N is then either returned to the atmosphere via arc volcanism ($F_{\rm arc}$), incorporated into the continental crust ($F_{\rm ccg}$), or transported to the mantle ($F_{\rm sub}$). We modeled these fluxes as,
\begin{equation}
    F_{\rm sub, tot} \equiv F_{\rm sub} + F_{\rm arc} + F_{\rm ccg} = \frac{M_{\rm oc}}{\tau_{\rm sub}},
\end{equation}
where $\tau_{\rm sub}$ is the subduction timescale. We assumed $\tau_{\rm sub} = 0.1\ {\rm Gyr}$, following \citeA{Stueken+2016ABb}. Fast plate tectonics on early Earth, which leads to the short $\tau_{\rm sub}$, was assumed in the past, but recent studies suggest that the plate speed is nearly constant through time \cite{Korenaga2003,Bradley+2008,Condie+2015,Pehrsson+2016,Korenaga2018}.

We note that hydrothermal addition of N to the oceanic crust is not explicitly included in our cycling model. In reality, N in the altered crust can also transport N to the mantle \cite{Mitchell+2010,Halama+2014,Johnson+Goldblatt2018}. However, this N in the altered crust is likely sourced from sedimentary organic material deposited to the crust \cite{Halama+2014}, and thus the conversion does not change the global N balance. 

The ratio $F_{\rm sub}$:$F_{\rm arc}$:$F_{\rm ccg}$ ($\equiv 1-f_{\rm arc}-f_{\rm ccg}$:$f_{\rm arc}$:$f_{\rm ccg}$) is highly uncertain \cite{Johnson+Goldblatt2018,Catling+Zahnle2020}. We evaluated $f_{\rm arc}$ and $f_{\rm ccg}$ as follows. First, we assumed $f_{\rm arc} = f_{\rm ccg}$, following \citeA{Johnson+Goldblatt2018}. Then, We evaluated $f_{\rm ccg}$ by considering the mass balance. Assuming a steady state for $M_{\rm cc}$ and $M_{\rm oc}$ leads to,
\begin{equation}
    M_{\rm cc} = F_{\rm sed} \tau_{\rm wea} \biggl( \frac{1}{1-f_{\rm ccg}} -f_{\rm pel} \biggr),\ M_{\rm oc} = \frac{F_{\rm sed} \tau_{\rm sub}}{1-f_{\rm ccg}}. \label{eq:MccMoc_steady}
\end{equation}
Provided the short $\tau_{\rm wea}$ after the GOE and $\tau_{\rm sub}$, the steady state assumed here to evaluate $f_{\rm ccg}$ is shown to be valid with our cycling model where the time evolution is explicitly solved (Section \ref{subsec:results:cycling}). In order to reproduce the modern $M_{\rm cc}$ and $M_{\rm oc}$ (Table \ref{tab:constraints}) with Equation \ref{eq:MccMoc_steady}, $f_{\rm ccg}$ needs to be $\simeq 0.5$. Thus, we assumed $f_{\rm ccg} = f_{\rm arc} = 0.45$, and the rest, $1-f_{\rm ccg}-f_{\rm arc} = 0.1$, is transported to the mantle.

\subsubsection{Mantle degassing}

The flux of volcanic N degassing from the mantle with time depends on the mantle N content, $M_{\rm ma}$, the thermal evolution \cite{Keller+Schoene2018,Korenaga2018}, and the redox state \cite{Mikhail+Sverjensky2014,Zerkle+Mikhail2017} which might have become oxidized over time \cite{Aulbach+Stagno2016,Nicklas+2018,Nicklas+2019,Aulbach+2019}.

We parameterized the volcanic degassing flux from the mantle as,
\begin{equation}
    F_{\rm vol} = \frac{M_{\rm ma}}{\tau_{\rm vol}},
\end{equation}
where $\tau_{\rm vol}$ is the degassing timescale, which was parameterized as,
\begin{equation}
    \tau_{\rm vol} = \tau_{{\rm vol}, 0} \times 10^{C_{\rm vol}(t_0-t)/t_0}.
\end{equation}
The power-law parameter, $C_{\rm vol}$, can be either positive or negative. The mantle N degassing timescale, \red{$\tau_{\rm vol,0}$}, on modern Earth was estimated from the mean value of the estimates of modern mantle N content 1.47 PAN (Table \ref{tab:constraints}) divided by the modern N degassing flux \cite<0.018 PAN/Gyr,>[]{Busigny+2011}.

\subsubsection{Initial conditions, parameters, and classification of results}
\label{subsubsec:cycling_parameters}

We performed Monte Carlo simulations to find parameter sets which satisfy the constraints on N masses and isotopic compositions of present reservoirs (Table \ref{tab:constraints}). The model runs which are consistent with these observations are hereafter called \textquotedblleft accepted runs.\textquotedblright \ {The Monte Carlo simulations were repeated until 100 accepted runs were found; this typically required 10$^6$ trials.} In accepted runs, we classified those with $M_{\rm ma}(t = 0\ {\rm Ga})\times0.5>M_{\rm ma}(t = 4.5\ {\rm Ga})$ as \textquotedblleft accepted recycling-origin runs,\textquotedblright \ and the others as \textquotedblleft accepted primordial-origin runs.\textquotedblright

\red{There are five Monte Carlo parameters. The total N content is varied from 1.844 to 3.964 PAN (Table \ref{tab:constraints}). The initial mantle N content is from 0 PAN to the total N content. The N isotopic ratio of initial mantle is from $-$40 to 0\textperthousand\ as found in enstatite chondrites \cite<>[and references therein]{Piani+2020}. We note that such low $\delta^{15}$N values have been reported for some rare diamonds from Earth's mantle, which possibly record the signature of Earth's N source \cite{Palot+2012}. The subduction and degassing parameters $C_{\rm sub}$ and $C_{\rm vol}$ are both from $-$5 to 5}. 

The other parameters were fixed during the Monte Carlo simulations, but their influences were investigated separately. i) The onset time of biological N fixation. Our nominal case assumed that biological N fixation (Phase 2) started at 3.2 Ga. This is based on the N isotope record which suggests that biological N fixation had already started by that time \cite{Stueken+2015}. Phylogenetic data also suggests that the last universal common ancestor (LUCA) had this function \cite{Weiss+2016}. Thus, we tested the case where Period 2 started 4.5 Ga as another endmember. ii) The onset time of the plate tectonics. The onset time of the plate tectonics on Earth is controversial \cite{Korenaga+2013}. In the nominal case, we assumed the initiation time for plate tectonics, $t_{\rm pt, init}$, to be 3.5 Ga, following \citeA{Johnson+Goldblatt2018}. We also tested $t_{\rm pt, init}$ = 4.5 Ga. iii) The onset time of N subduction. Our nominal case assumed $f_{\rm ccg} = f_{\rm arc} = 0.45$ and, thus, 10\% of N conveyed to subduction zones is transported to the mantle. Because a hot mantle might have inhibited N subduction prior to 2.5 Ga \cite{Johnson+Goldblatt2018}, we also tested a model where $f_{\rm ccg} = f_{\rm arc} = 0.5$ and $f_{\rm ccg} = f_{\rm arc} = 0.45$ before and after 2.5 Ga, respectively. However, we find that these model variants do not differ significantly (Section \ref{subsec:results:cycling}).

Paleo-atmospheric pressure and N$_2$ partial pressure, $p_{\rm N_2}$, have been constrained from geochemical and geological record \cite{Marty+2013,Som+2012,Som+2016}. These studies suggest low $p_{\rm N_2}$ values in the Archean, which corresponds to 0.64--1.4 PAN at 3.5-3.0 Ga \cite{Marty+2013} and $<$0.64 PAN (assuming that N$_2$ is the dominant atmospheric gas) at 2.7 Ga \cite{Som+2012,Som+2016}. Our model did not adapt these constraints explicitly, but we compared the accepted runs with them. Overall, we find that these low $p_{\rm N_2}$ values are satisfied in many, but not all, accepted runs by N sequestration in the crust \cite<as proposed by>[]{Stueken+2016ABb}.

\section{Results}
\label{sec:results}

\subsection{Partitioning model}
\label{subsec:results:partitioning}

\begin{figure}
    \centering
    \includegraphics[width=\linewidth]{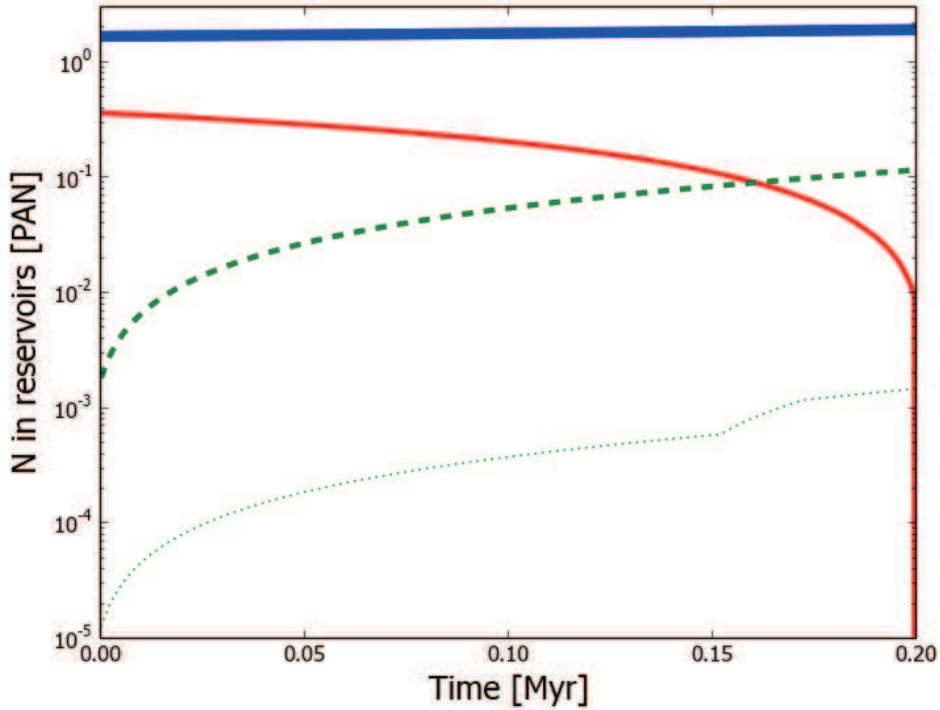}
    \caption{Evolution of the N content in each reservoir in our partitioning model. Blue line: atmosphere. Red line: magma ocean. Green lines: solidified mantle (dashed: trapped melt and minerals, dotted: minerals only). The result for the oxidized case with $t_{\rm s} = 0.2$ Myrs and the total content = 2 PAN is shown.}
    \label{fig:time-massN}
\end{figure}

\begin{figure}
    \centering
    \includegraphics[width=\linewidth]{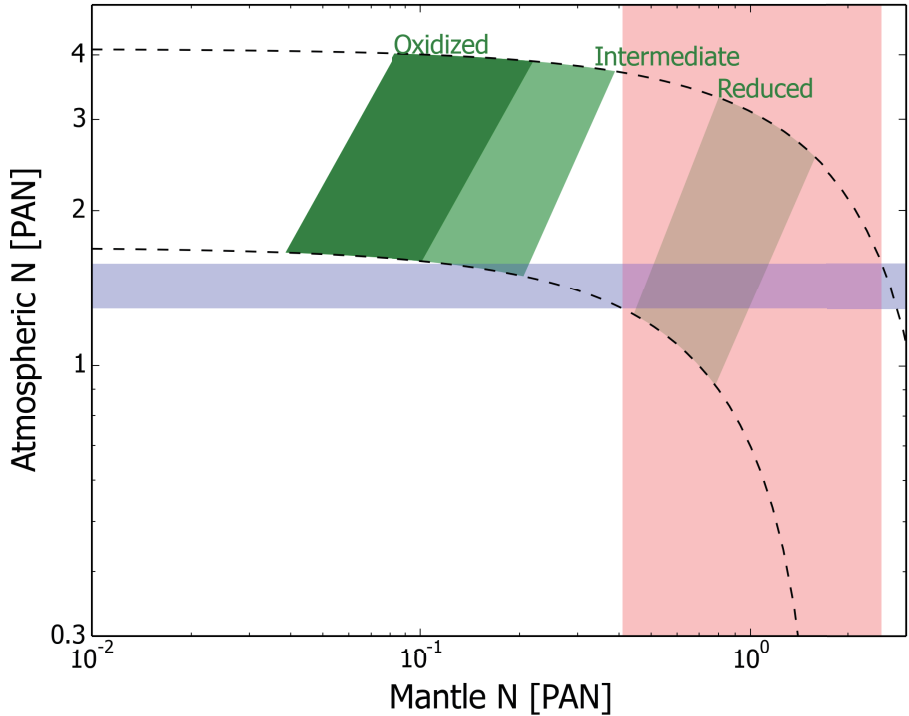}
    \caption{N partitioning between the atmosphere and mantle at the time of magma ocean solidification. The result of one simulation run corresponds to a single point in the figure, and changing two parameters (the solidification time $t_{\rm s}$ and the total N content in the system $M^{\rm N}_{\rm atm+mo+sm}$) draws a {green} patch of solutions. Results are shown for the oxidized, intermediate, and reduced cases (see text) with increasing transparency. Dashed curves show constant total-N (1.698 and 4.110 PAN). The shaded areas are the N amounts in the modern surface (the atmosphere plus continental and oceanic crust, blue) and mantle (red).}
    \label{fig:Npartitioning}
\end{figure}

\begin{figure}
    \centering
    \includegraphics[width=\linewidth]{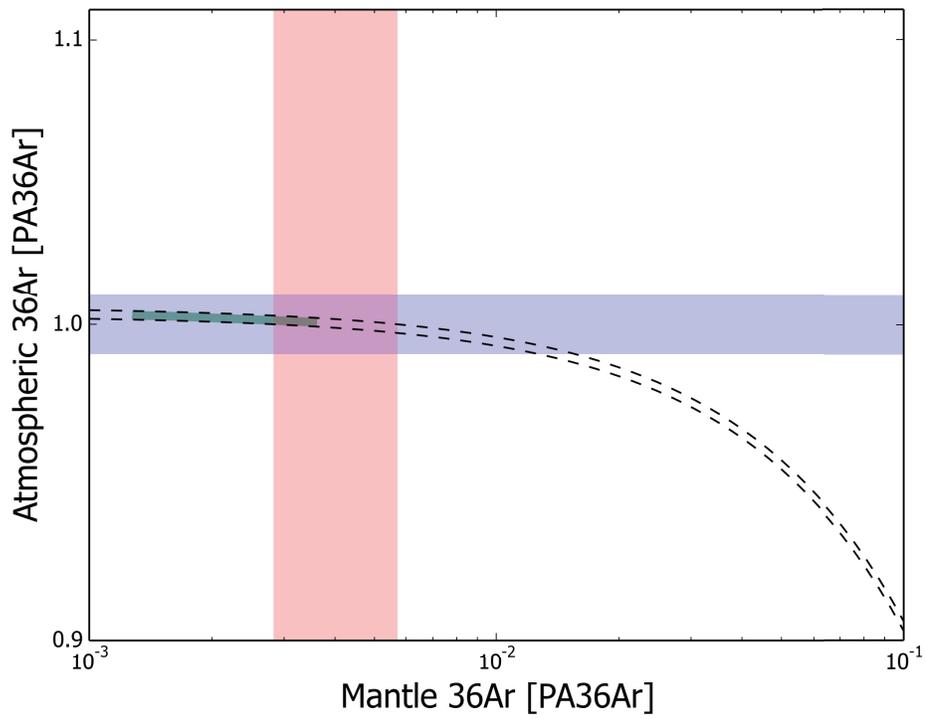}
    \caption{$^{36}$Ar partitioning between the atmosphere and mantle at the time of magma ocean solidification. Dashed curves show constant total-$^{36}$Ar (1.0028 and 1.0057 PA36Ar). The shaded areas are the $^{36}$Ar amounts in the modern atmosphere (blue) and mantle (red).}
    \label{fig:Arpartitioning}
\end{figure}

\begin{figure}
    \centering
    \includegraphics[width=\linewidth]{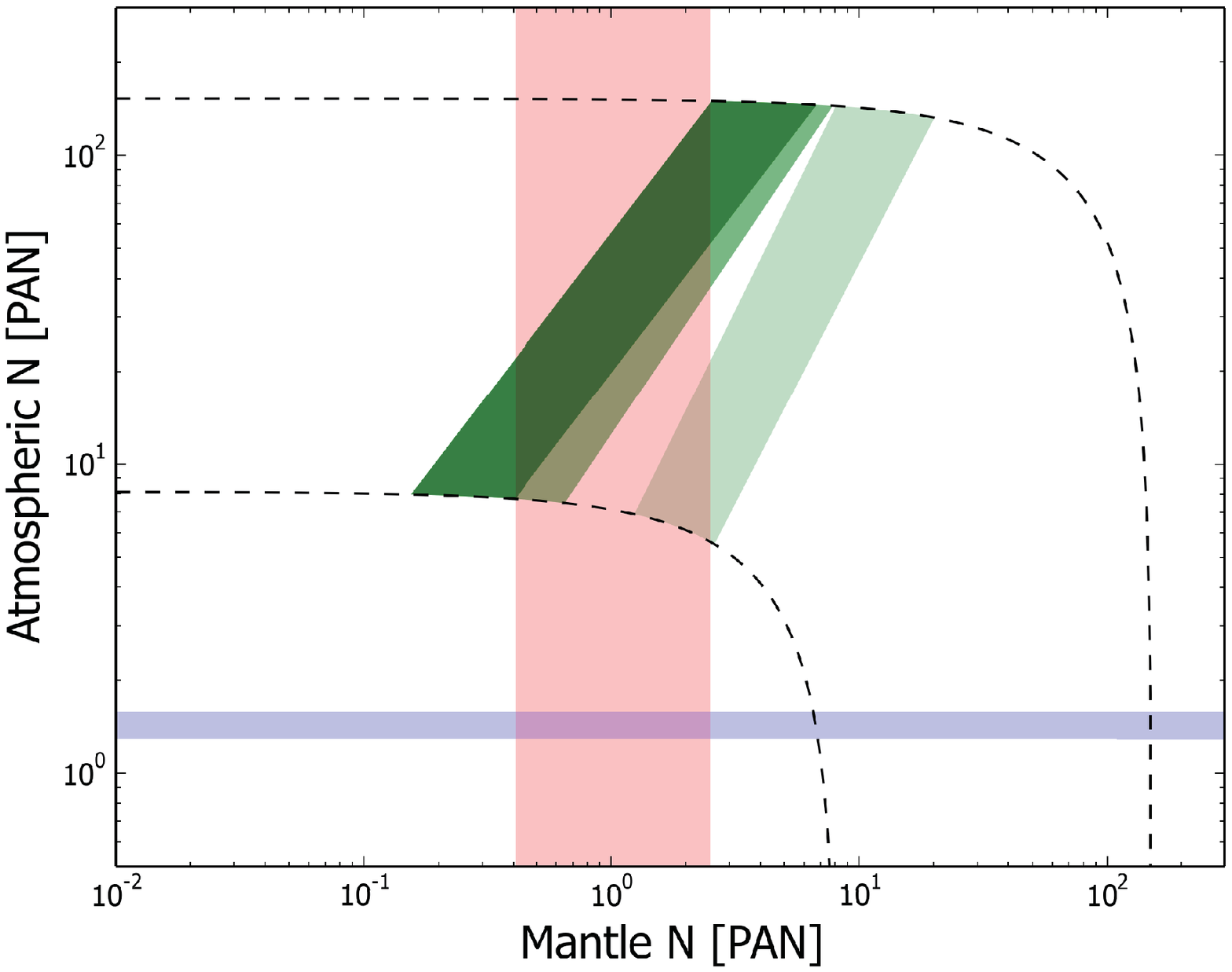}
    \caption{N partitioning between the atmosphere and mantle at the time of magma ocean solidification in the cases where we assumed larger amounts of total N to reproduce the modern \red{mantle} N content with the oxidized magma ocean model. Results are shown for the oxidized, intermediate, and reduced cases (see text) with increasing transparency. Dashed curves show constant total-N (8.11 and 152.53 PAN). The shaded areas are the N amounts in the modern surface (the atmosphere plus continental and oceanic crust, blue) and mantle (red).}
    \label{fig:Npartitioning_excess}
\end{figure}

First, we studied N partitioning in the magma ocean stage to test the primordial-origin scenario. Time evolution with $t_{\rm s}$ = 0.2 Myrs and the total N content = 2 PAN in the oxidized case is shown in Figure \ref{fig:time-massN}. N was chiefly partitioned into the atmosphere owing to its low solubility (Equation \ref{eq:SN} in \ref{sec:appendix_SN}). As the magma ocean solidified, a fraction of N in the magma ocean partitioned into the solidified mantle chiefly in the form of N in the trapped melt. Because of its incompatibility (low mineral-melt partitioning coefficients, $D_{\rm N}$, Table \ref{tab:partitioning}), the contribution of N partitioned into minerals was small. As a consequence, the mantle right after magma ocean solidification is depleted in N (the mantle N content is only $\sim$0.1 PAN). We note that the final mantle N content in this low solubility limit is well approximated by $M^{\rm N}_{\rm mantle}/M_{\rm mantle} \simeq p_{\rm N_2} S_{\rm N} F_{\rm tl}$ \cite<Equation 11 of >[]{Kurokawa+2021} with $F_{\rm tl} = 0.3$ in this rapid solidification case.

Figure \ref{fig:Npartitioning} shows the atmospheric and mantle N contents after magma ocean solidification for a range of total N content and solidification times $t_{\rm s}$ in the oxidized, intermediate, and reduced cases. For a given redox state, increasing the total N content leads the results to move to the upper right perpendicular to the constant total-N curves, while increasing $t_{\rm s}$ to the upper left parallel to the curves. If the atmospheric and mantle N contents obtained in the model match those of the modern surface (N in the atmosphere plus continental and oceanic crust which is thought to be in the atmosphere when the magma ocean solidified) and mantle, the results suggest a primordial origin for mantle N. However, in both oxidized and intermediate cases, the model mantle N content was lower than the modern value. Only the reduced case {reproduced} the lower estimate ($\lesssim$1 PAN) of the modern mantle N content. This is caused by higher N solubilities in reduced conditions due to chemically-bound N in the magma \cite{Libourel+2003}. The mantle N amounts in all cases {were} mainly attributed to trapped interstitial melt (Figure \ref{fig:time-massN}). The contribution of N in minerals {was} smaller by an order of magnitude. A shorter solidification time $t_{\rm s}$ leads to a higher trapped melt fraction $F_{\rm tl}$ and thus more N in the solidified mantle. However, our result with the shortest solidification time ($t_{\rm s} = 0.2$ Myrs) already reached the maximum value ($F_{\rm tl} = 0.3$).

In contrast, the partitioning model reproduced the $^{36}$Ar content of the modern mantle as well as the atmosphere (Figure \ref{fig:Arpartitioning}). Ar behaves similarly to N in the oxidized magma \cite<e.g.,>[]{Marty+Dauphas2003}, and ends up with a similar degree of depletion in the mantle. The agreement between the model and observations is due to the fact that the modern mantle is depleted in $^{36}$Ar by two orders of magnitude with respect to N/$^{36}$Ar ratios compared to the current atmosphere.

To summarize, our partitioning model suggests that the mantle received only $\sim$10\% of the total N when the magma ocean fully solidified, unless the magma was reducing. The reduced magma ocean at the time of solidification is not supported from geophysical and geochemical studies \cite{Badro+2015,Armstrong+2019,Sossi+2020}. 

We also performed model calculations with increased total N content and found that putting the modern N content into the solidified mantle in the oxidized case ended up with 7.7--150 PAN in the atmosphere (Figure \ref{fig:Npartitioning_excess}). The range was derived from uncertainties in the modern mantle N content $M^{\rm N}_{\rm mantle}$ and the magma ocean solidification time $t_{\rm s}$. In Section \ref{subsec:discussion:Norigin1}, we discuss a primordial-origin scenario for the mantle N by considering the excess amount of total N, which could have been removed later by atmospheric escape processes.

\subsection{Cycling model}
\label{subsec:results:cycling}

\begin{landscape}
\begin{figure}
    \centering
    \includegraphics[width=\linewidth]{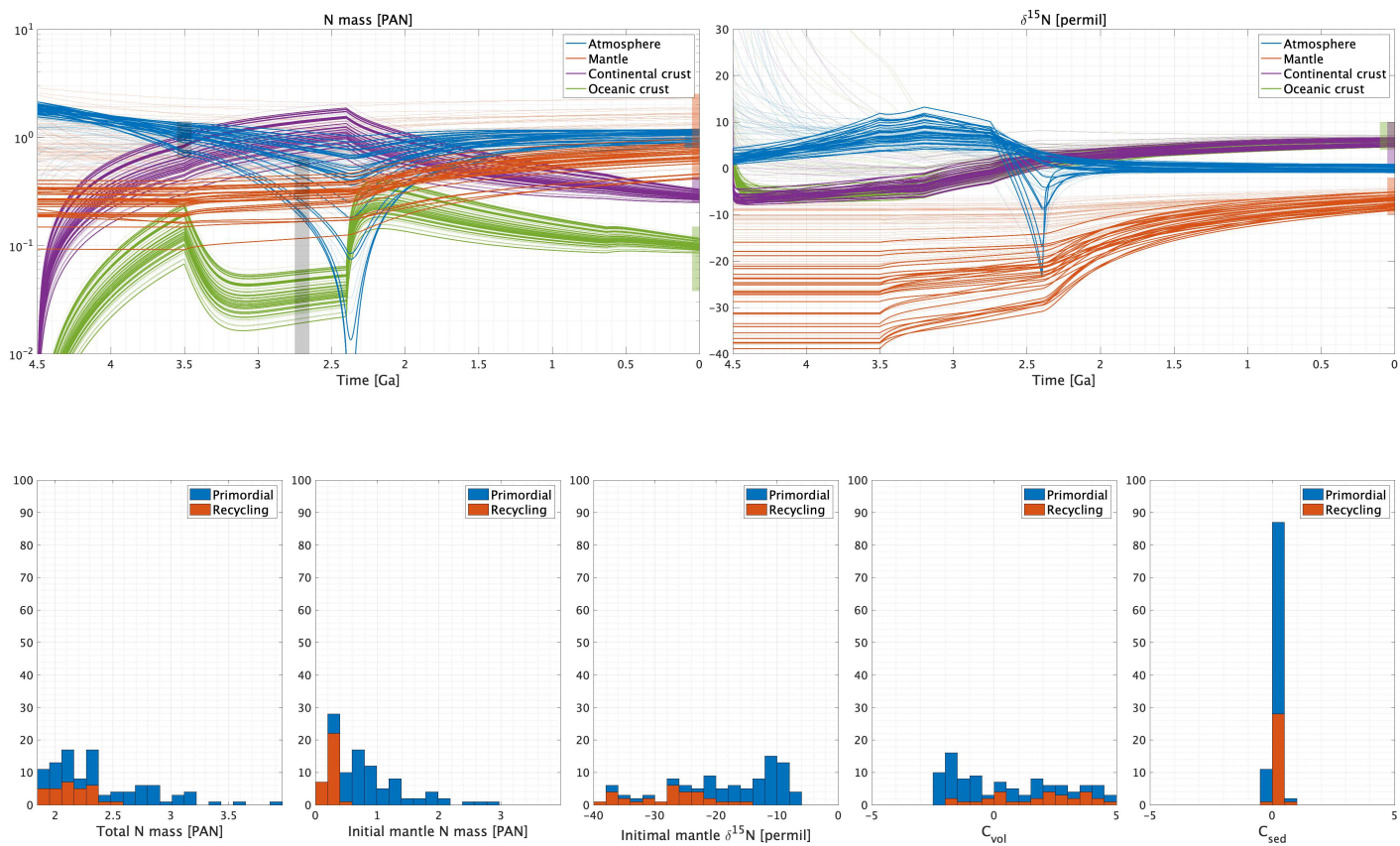}
    \caption{Cycling model results for the nominal case where we assumed the onset times of biological N fixation, plate tectonics, and N subduction to be 3.2, 3.5. and 3.5 Ga, respectively (see Section \ref{subsubsec:cycling_parameters} for details). Top: Evolution of masses (left) and $^{15}$N/$^{14}$N ratios (right) in the atmosphere (blue), continental and oceanic crust (purple and green), and mantle (red). Curves show accepted runs whose outcomes are consistent with modern values (shown along the right y-axis). Thick and thin curves denote accepted recycling- and primordial-origin runs, respectively. Gray boxes indicate {$p_{\rm N_2}$} constraints from \citeA{Som+2016} and \citeA{Marty+2013}. Bottom: Histograms of Monte Carlo parameters of all accepted runs. Recycling- and primordial-origin runs are shown in red and blue, respectively.}
    \label{fig:Ncycle1}
\end{figure}

\begin{figure}
    \centering
    \includegraphics[width=\linewidth]{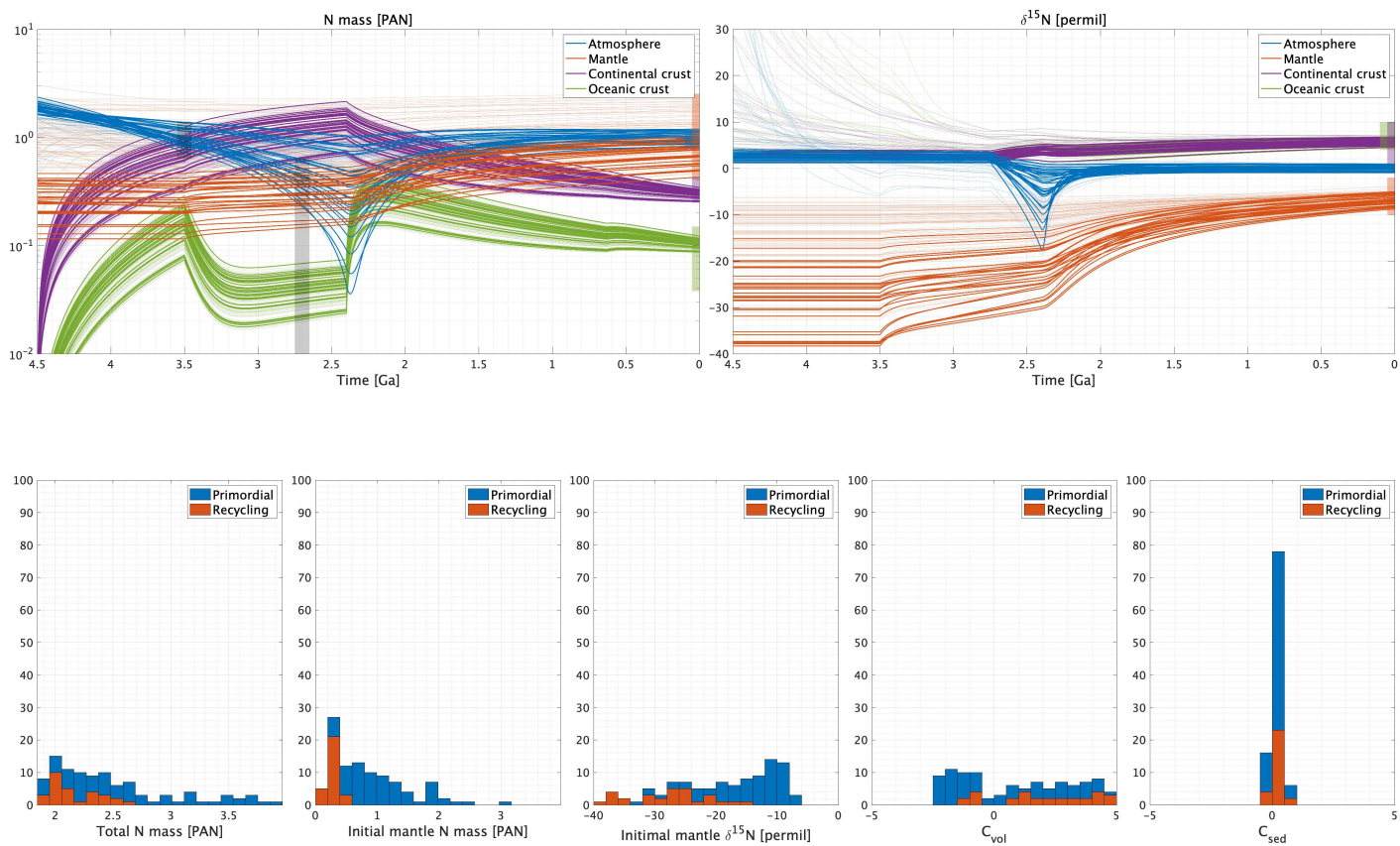}
    \caption{Cycling model results for the case where we assumed Phase 2 (biological N fixation) started 4.5 Ga, while the other settings are the same as in the nominal case (Figure \ref{fig:Ncycle1}). See Figure \ref{fig:Ncycle1} caption for details.}
    \label{fig:Ncycle1_WO_PHASE1}
\end{figure}
\end{landscape}

\subsubsection{Evolution of N contents in reservoirs}

Next, we tested the recycling origin for mantle N with our cycling model. We note that, unless otherwise stated, we refer to the results of the nominal case where we assumed the onset times of biological N fixation, plate tectonics, and N subduction to be 3.2, 3.5. and 3.5 Ga, respectively (see Section \ref{subsubsec:cycling_parameters} for details). The evolution of N masses and isotopic ratios in the atmosphere, continental and oceanic crust, and mantle after the solidification of magma ocean was simulated with the nominal parameter set (Figure \ref{fig:Ncycle1}). Rather than directly adopting the results of the magma-ocean partitioning model, initial conditions of the cycling model were treated as Monte Carlo parameters, and we looked for parameter sets which reproduce the modern N contents and isotopic ratios in the four reservoirs. We compare the results of the two models to discuss possible scenarios for N partitioning and cycling in Section \ref{sec:discussion}. 

The atmospheric N contents in the runs which started from a few PAN show decline with time to reach a minimum at 2.4 Ga (the top-left panel of Figure \ref{fig:Ncycle1}). The decline is caused by sediment deposition and sequestration in the continental and oceanic crust. Notably, the constraints on $p_{\rm N_2}$ \cite{Marty+2013,Som+2012,Som+2016} can be satisfied by many, but not all, accepted runs. The modeled evolution curves also \red{agree} with evidence for a stable atmospheric $p_{\rm N_2}$ reservoir over the Phanerozoic \cite{Berner2006}. The initiation of plate tectonics can be seen as the decline of N contents in the oceanic crust after 3.5 Ga. In contrast, the influence of subduction on the mantle N contents in the recycling-origin runs is limited before 2.4 Ga. In some primordial-origin runs, the mantle N contents even decreased in this period due to degassing.

At 2.4 Ga, which corresponds to the GOE, the rate of continental weathering increased drastically, leading to the decline of N content in the continental crust. The continental N was first transported to the oceanic crust as sediments, and then to the mantle via subduction. Owing to the return N flux via arc volcanism, the atmospheric N contents start to increase with time after the GOE.

\subsubsection{Evolution of N isotopic ratios in reservoirs}

N cycling between the reservoirs is reflected to the evolution of isotopic ratio (the top-right panel of Figure \ref{fig:Ncycle1}). In the recycling-origin runs, the atmospheric $\delta^{15}$N values increase by $\sim$+5\textperthousand \ with time prior to 3.2 Ga. This is caused by deposition of $^{15}$N-depleted sediments in Period 1 ($>$3.2 Ga, Section \ref{subsec:model_cycling}). As the net fractionation factor between the atmosphere and sediments became zero (Period 2, 3.2--2.75 Ga) to positive (Periods 2 and 3, $<$2.75 Ga), the atmospheric $\delta^{15}$N values change their trend to decline with time. The minimum values at 2.4 Ga correlate with the atmospheric N contents at that time. Following the recovery of $p_{\rm N_2}$ after 2.4 Ga, the atmospheric $\delta^{15}$N values approach asymptote of 0\textperthousand. 

The $\delta^{15}$N values of the continental and oceanic crust showed secular increase as a result of the change in $\Delta^{15}$N between the atmosphere and sediments deposited. The rate-determining process of the continental $\delta^{15}$N evolution is continental weathering whose timescale is 3 Gyrs and 0.3 Gyrs before the GOE and at present, respectively \red{(Equation \ref{eq:PAL})}. The timescale of the evolution of the oceanic crust $\delta^{15}$N values is determined by the subduction (0.1 Gyrs) after the onset of the plate tectonics. Because this is much shorter than the timescale of continental weathering, the oceanic crust follows the continents in terms of $\delta^{15}$N evolution.

Mantle $\delta^{15}$N stays constant before 3.5 Ga because no subduction is assumed in the nominal case. After the onset of the plate tectonics, these values start to increase due to subduction of relatively-$^{15}$N-rich surface N. The rate of change increases after 2.4 Ga. This is caused by enhanced continental weathering after the GOE, which leads to more N transported onto the oceanic crust, and consequently, to the mantle.

We note that several primordial origin runs showed high $\delta^{15}$N values for the atmosphere and crust initially. These are runs where the Monte Carlo code chose high initial mantle N contents, so that the mass balance led to high atmospheric $\delta^{15}$N values to keep the bulk $\delta^{15}$N consistent with the modern value. Because the continental and oceanic crust receives atmospheric N as sediments, their $\delta^{15}$N values were also high earlier on. The record of the initial state is eventually lost after $\sim$1 Gyr.

\subsubsection{Monte Carlo parameters in accepted runs}

Histograms of Monte Carlo parameters in accepted runs are shown in the bottom panels in Figure \ref{fig:Ncycle1}. Because our model is not fully physics-based but rather conceptual, all results should not necessarily be treated as {\it realistic} without discussion, but these should be considered as {\it requirements} to reproduce the N contents and isotopic ratios in the modern reservoirs (see Section \ref{subsec:discussion:Norigin2} for discussion on possible scenarios).

In comparing the recycling- and primordial-origin runs, the former show lower total N masses and lower initial mantle $\delta^{15}$N values than the latter. This is caused by the finite ability of the system to transport surface N without increasing the mantle $^{15}$N value too much. The low initial mantle $\delta^{15}$N values in the recycling runs (typically  $-40$ to $-20$\textperthousand) mean that, even they were classified as the recycling origin, the initial mantle condition matters - in terms of the isotopic composition, mantle N cannot be fully of recycling origin. 

The sediment deposition parameter, $C_{\rm sed}$ (positive and negative values mean higher and lower fluxes in earlier time, respectively), was constrained around zero, which means that sediment N burial has been nearly constant, for both recycling- and primordial-origin runs. This is reasonable for the former. Given the modern N subduction flux, $\sim$0.1 PAN/Gyr assumed in our model, it takes $\sim$4 Gyrs to transport $\sim$0.4 PAN (the lowest estimate for the modern mantle N mass) to the mantle. Lower sediment N deposition flux on early Earth (negative $C_{\rm sed}$) would lead to less N subduction, which would not satisfy the constraints of modern mantle N content. 

Why $C_{\rm sed}$ is highly constrained even for the primordial-origin runs is not so intuitive. In order to understand this behavior, we tested an additional case where we omitted the constraints on the N masses and $\delta^{15}$N values for the modern continental and oceanic crust (Figure \ref{fig:Ncycle1_WO_MccMocCONSTRAINT}). In this case, accepted primordial-origin runs include lower $C_{\rm sed}$ values down to $-$5, but these runs show lower N contents in the continental and oceanic crust. This result indicates that modern crustal N masses rule out low $C_{\rm sed}$ values from the primordial-origin runs. Importantly, even in this case where the constraints on crustal N masses are relaxed, the recycling-origin runs showed their $C_{\rm sed}$ values constrained to be around zero, indicating that sediment N deposition flux on early Earth comparable to the modern value is a robust requirement for the recycling-origin scenario.

In contrast to $C_{\rm sed}$, the accepted degassing parameter values, $C_{\rm vol}$ (positive and negative values mean longer and shorter earlier degassing timescale, respectively), were distributed over a wider range. This is likely because mantle degassing has a relatively minor effect on overall N cycling. The modern degassing flux of mantle N was assumed to be 0.018 PAN/Gyr \cite{Busigny+2011}, which is an order of magnitude lower than the N subduction flux.

\subsubsection{Effects of the onset time of biological N fixation, plate tectonics, and N subduction}

The case where we assumed the onset of biological N fixation (Phase 2) at 4.5 Ga is shown in Figure \ref{fig:Ncycle1_WO_PHASE1}. There was no significant difference with the nominal case (Figure \ref{fig:Ncycle1}) in the accepted Monte Carlo parameters. However, the difference of the evolution of $\delta^{15}$N values is notable. In contrast to the nominal case showing atmospheric $\delta^{15}$N values to be $\sim$+5\textperthousand \ prior to 2.4 Ga in the recycling-origin runs, the atmospheric $\delta^{15}$N values stayed nearly constant at 1--4\textperthousand \ in the early biological-N-fixation case. Because there is no net-fractionation between the atmosphere and sediments, the crustal $\delta^{15}$N values stayed equal to the atmospheric values before 2.75 Ga.

Finally, we tested two cases where we assumed the onset of plate tectonics at 4.5 Ga (Figure \ref{fig:Ncycle2}) and all subducted N was returned to the atmosphere and continental crust before 2.5 Ga \red{due to the hotter mantle of early Earth} (Figure \ref{fig:Ncycle3}). Although there were differences in the evolution of N contents in the oceanic crust and mantle, no significant difference was found in the accepted Monte Carlo parameters. This is because the most important mechanisms to determine N cycling are N storage in the crust and efficient continetal weathering to cause N subduction before and after the GOE, respectively, both of which were satisfied in these variants of the nominal case.

\section{Discussion}
\label{sec:discussion}

\subsection{{The origin of mantle N: the primordial excess-N scenario}}
\label{subsec:discussion:Norigin1}

\begin{figure}
    \centering
    \includegraphics[width=\linewidth]{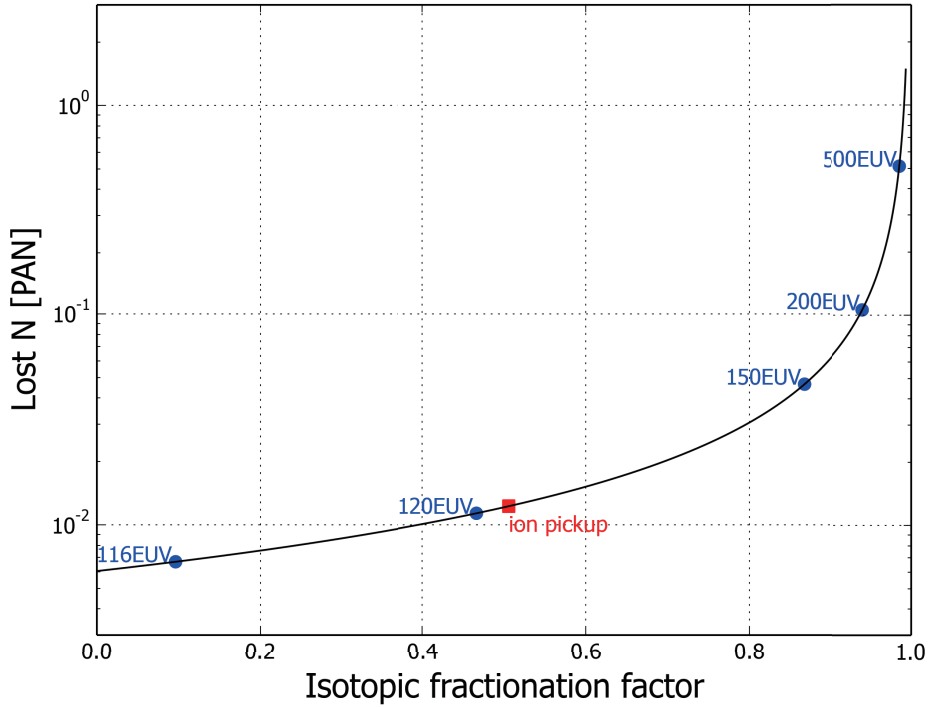}
    \caption{The amount of escaped N required to enrich the remnant atmosphere (1 PAN) in $^{15}$N by $+6$\textperthousand \ as a function of isotopic fractionation factor. Blue circles and a red square are estimates for EUV-driven hydrodynamic escape and solar wind-induced ion pickup, respectively.}
    \label{fig:escape}
\end{figure}

\begin{figure}
    \centering
    \includegraphics[width=\linewidth]{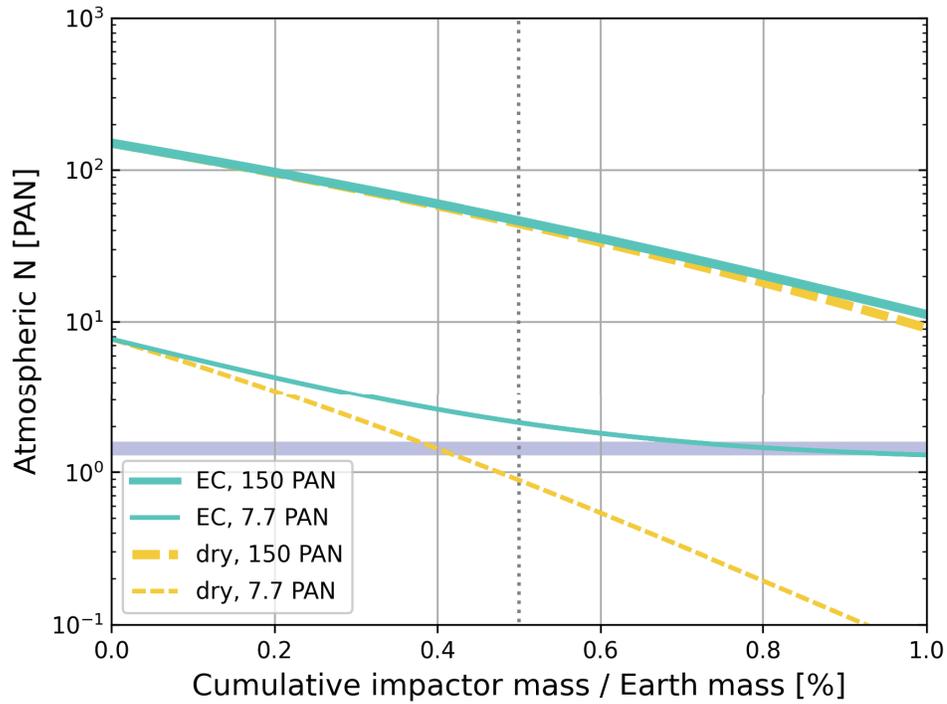}
    \caption{Evolution of atmospheric N content during the late accretion in the primordial N-excess scenario. Green: Enstatite chondrite-like impactors. Yellow: N-depleted \textquotedblleft dry\textquotedblright \ impactors.  Initial N contents are 7.7 PAN (thin curves) and 150 PAN (thick curves), which corresponds to the atmospheric N content required to put N into the mantle comparable to the current mantle in the oxidized magma-ocean scenario (Section \ref{subsec:results:partitioning} and Figure \ref{fig:Npartitioning_excess}). The horizontal blue line is the modern surface N content (the atmosphere and crust). The vertical dashed line is the minimum estimate for the cumulative mass of the late accretion. Note that the thick green and yellow curves are largely overlapping.}
    \label{fig:impacts}
\end{figure}

In Section \ref{subsec:results:partitioning}, we suggests that magma ocean solidification {led} to depletion of N in the mantle, unless the magma ocean was reducing (Figure \ref{fig:Npartitioning}). Both high pressure experiments \cite{Armstrong+2019,Sossi+2020} and coupled geochemical and geophysical constraints \cite{Badro+2015} point to Earth's magma ocean {at the time of solidification} being close to or more oxidizing than IW-1. Therefore, we rule out a scenario that reducing conditions led to the primordial origin of N in the mantle, and focus on the oxidized magma ocean condition.

A possibility we should consider is that primitive Earth right after core-mantle separation might have a higher bulk N content than it currently does, and thus even preferential partitioning into the atmosphere in the oxidized magma ocean condition could end up with an amount of N remaining in the mantle sufficient to match modern observed values. If this were the case, the early atmosphere would have an excess amount of N of 7.7--150 PAN in the oxidized case (Section \ref{subsec:results:partitioning} and Figure \ref{fig:Npartitioning_excess}). We name this case the primordial \textquotedblleft excess-N\textquotedblright \ scenario as a variant of the primordial origin and test whether atmospheric escape processes can remove the excess N in the atmosphere.

We estimated the $^{15}$N-enrichment effect due to extreme ultraviolet (EUV)-driven escape and solar wind-induced ion pickup to see how much atmospheric N should be removed to end up with $\sim+$6\textperthousand\ $^{15}$N-enrichment in the remnant 1 PAN atmosphere (Figure \ref{fig:escape}, see \ref{sec:appendix_escape} for model details). We found that these processes can remove only 0.5 PAN at most even assuming an EUV level 500 times higher than at present, which is significantly smaller than the excess atmospheric N estimated above. This level of EUV corresponds to so-called fast-rotating young Sun model, but previous studies focusing on volatile {contents on Venus \cite{Lammer+2020,Lammer+2020SSR} and the Moon \cite{Saxena+2019}} concluded that the EUV level of the young Sun is likely to be less active than a moderate rotator, whose EUV level was $\simeq$100 times that of the current Sun. For this EUV level, neither $^{14}$N nor $^{15}$N can be removed by EUV-driven escape. Moreover, the contribution of solar wind-induced ion pickup should be limited to 0.01 PAN, otherwise the atmosphere would have excess $^{15}$N.

Next we estimated how much N can be removed by atmospheric erosion due to late accretion impacts, using the model of \citeA{Sakuraba+2019} (Figure \ref{fig:impacts}, see \ref{sec:appendix_impacts} for {model} details). Impact erosion does not fractionate isotopes, as it removes the bulk atmosphere well below the homopause. However, we found that impact erosion cannot remove the excess N (7.7--150 PAN) with the minimum estimate of the late accretion mass (0.5 wt.\% of Earth's total mass) and N-containing impactors (such as enstatite chondrites) that have been suggested from the isotopic analysis of siderophile elements in Earth and the Moon \cite{Dalou+2017,Fischer+Kleine2017,Fischer+2020,Worsham+Kleine2021}. The excess N can be removed {by} completely N-depleted impactors or by a higher cumulative mass of late accreting material \cite{Marchi+2018}. We note that here we assumed asteroid-belt-like impactor-size distribution which leads to efficient impact erosion. Assuming shallower size distribution \cite<more large impactors, as proposed by>[]{Bottke+2010} does not yield sufficient atmospheric loss \cite{Sakuraba+2019,Sakuraba+2021}.

From these results shown above, we conclude that the primordial origin scenario for the mantle N is possible only if Earth at the time of magma ocean solidification had an excess amount of N ($>$7.7 PAN in the atmosphere), and if the N content, size distribution, and total mass of late accretion impactors allowed efficient loss of the excess N in the atmosphere to space by impact erosion. The feasibility of this model also depends on primitive Earth's total N content after major accretion and core-mantle separation. \citeA{Grewal+2021} modeled Earth's accretion and argued that the N content of the bulk silicate Earth (Earth minus the core, hereafter BSE) was determined by partitioning into the core and atmospheric loss during major accretion, which is incompatible with the excess-N scenario proposed here. In their model, the atmospheres on planetesimals in each stage of accretion were assumed to be lost completely. In contrast, \citeA{Sakuraba+2021} simulated Earth's accretion with an impact erosion model \cite{Sakuraba+2019}. They showed that atmospheric loss in the main accretion stage before the late accretion is significant but not complete, and thus Earth formed with excess N, which was removed by late accretion impacts.

\red{The excess-N and impact erosion scenario need to reproduce $^{36}$Ar/N ratios in Earth's atmosphere and mantle.} The agreement of the $^{36}$Ar partitioning model results with $^{36}$Ar in the modern atmosphere and mantle (Figure \ref{fig:Arpartitioning}) suggests that, \red{in contrast to N}, there was no significant excess atmospheric$^{36}$Ar at the time of solidification, \red{requiring preferential N removal. However, impact erosion itself is not a fractionation process. A possibility to reconcile the excess-N scenario with the observed $^{36}$Ar/N ratios is late accretion of bodies having a high $^{36}$Ar/N ratio on average, as proposed for the cometary contribution to Earth’s atmospheric noble gases \cite<but not to major volatile elements: H, C, and N,>[]{Marty+2016,Marty+2017}, while the majority of impactors would be chondritic or more dry bodies. If this was the case, even though impact atmospheric erosion itself is not a selective process, the erosion combined with the replenishment of Ar would lead to a net loss of N preferentially to Ar.} This gives another constraint on the impactor composition in the primordial scenario.

\subsection{The origin of mantle N: the early biogeochemical cycling scenario}
\label{subsec:discussion:Norigin2}

In Section \ref{subsec:results:cycling}, we showed that the recycling origin scenario requires sediment N deposition rates on early Earth being comparable to that of modern Earth. The high {sediment N deposition} flux is, as discussed below, likely to be sustained by biotic activity. 

A simple solution is that high sediment deposition flux at earlier times has been sustained by biological N fixation similar to as on modern Earth (as assumed in Figure \ref{fig:Ncycle1_WO_PHASE1}). This scenario might be consistent with the phylogenetic study which suggests that LUCA was able to fix N \cite{Weiss+2016}.

Alternatively, even if N fixation on early Earth was being sustained by abiotic processes (as assumed in the nominal case, Figure \ref{fig:Ncycle1}), biotic activity is needed to keep high N concentration in sediments. Considering the reasonable range of abiotic N fixation fluxes, adsorbed N content in sediments {was} estimated to be smaller by more than an order of magnitude in abiotic conditions than in modern Earth condition \cite<$\sim$10 ppm vs. $\sim$400 ppm,>[]{Stueken2016AB,Busigny+2011}. This is the maximum estimate as it assumes that all NOx is reduced to NH$_4^+$ during hydrothermal circulation or by reaction with ferrous iron. Another study found that abiotic chemistry significantly limits the NOx abundance in the oceans \cite{Hu+Diaz2019} and thus NH$_4^+$ as well. On modern Earth, the efficient NH$_4^+$ adsorption is caused by a high NH$_4^+$ concentration in pore water of anoxic sediments due to the degradation of organic matter \cite<e.g.,>[]{Stueken2016AB}. Therefore, here we suggest a scenario that the same mechanism also contributed to the high NH$_4^+$ abundance in sediments and thus efficient cycling, and that the modern mantle N was sourced by biogeochemical cycling (either biotic or abiotic N fixation and subsequent biotic processing) developed on early Earth.

\subsection{Model uncertainties and limitation}
\label{subsec:discussion:uncertainties}

As discussed in Section \ref{sec:methods}, \citeA{Johnson+Goldblatt2015} estimated a higher modern mantle N content (2--10 PAN) than the range assumed in our study (0.41--2.53 PAN). The higher N content is more difficult to reconcile with partitioning in the magma ocean stage. In contrast, because the proposed high-N mantle has high $\delta^{15}$N values ($\sim$+5\textperthousand) comparable to modern sediments \cite{Johnson+Goldblatt2015}, it can be sourced by subduction of sediments after the development of the modern aerobic N cycle, if high sedimentary deposition flux is considered \cite<as modeled by>[]{Johnson+Goldblatt2018}. Therefore, the high-N mantle component, if confirmed, supports the recycling origin for mantle N.

\red{Our} cycling model assumes monotonic increase or decrease {for the sediment deposition flux and the mantle degassing timescale for simplicity}, but Earth might have experienced continuous inversions of N subduction and degassing fluxes \cite{Zerkle+Mikhail2017}. For instance, the N subduction flux has likely been lower and higher before and after the emergence of life as it enhances NH$_4^+$ adsorption efficiency (Section \ref{subsec:discussion:Norigin2}). Nevertheless, our conclusion for the requirement for the high sediment N burial rate would be valid as an averaged condition before the onset of aerobic N cycling.

Following \citeA{Stueken+2016ABb}, we assumed that the effects of continental growth \cite<e.g.,>[]{Korenaga2018} on the offshore to onshore ratio of sediment burial ($f_{\rm pel}/(1-f_{\rm pel})$) and on the continental weathering timescale ($\tau_{\rm wea}$) were minor. The reader is referred to \citeA{Stueken+2016ABb} for detailed discussion. Key assumptions of our model were that sediments are mainly deposited to the continental crust (shelves) through Earth's history ($f_{\rm pel} \ll 1$ and is assumed to be constant through time) and the long residence time of sediments on the continental crust before the GOE ($\tau_{\rm wea} \sim$1 Gyr). Thus, unless the continental fraction was extremely small in the Archean to lead $f_{\rm pel} \sim 1$, the continental growth curve would not influence our model results and conclusions significantly. \red{If these assumptions did not hold, the low $p_{{\rm N}_2}$ constraints might not be satisfied in the recycling scenario.}

Our modeling followed the generally-accepted inference that $^{15}$N-enriched sedimentary rocks are dominant in subduction flux of N \cite<e.g.,>[]{Bekaert+2020}, and thus there exists N isotopic disequilibrium between the atmosphere and mantle \cite{Marty+Dauphas2003}. However, \citeA{Mitchell+2010} argued that hydrothermally-altered oceanic crust is the dominant cause of this, and it has a negative ${\rm \delta ^{15}}$N value (-1.8\textperthousand) on average. The negative ${\rm \delta ^{15}}$N value assumed is the case for Mariana trench \cite{Li+2007}, and may not necessarily represent the mean value of the whole oceanic crust. Estimates in other studies have shown positive ${\rm \delta ^{15}}$N values for the average oceanic crust \cite{Busigny+2011,Bekaert+2020}. 

N isotopic fractionation in the subduction zones and in the mantle has not been fully understood, and thus not considered in our cycling model. Devolatilization during subduction leads to an increase in residual $^{15}$N of metamorphic rocks \cite{Haendel+1986}. The effect is limited on present-day Earth \cite{Cartigny+Marty2013}, yet its influence on N cycling through Earth's history remains to be resolved. Additionally, it is not yet known whether N isotopes are fractionated during partial melting reactions in the mantle \cite{Busigny+Bebout2013}. Better understanding of N isotopic fractionation in these processes is needed to further test the recycling scenario.

Atmospheric loss processes (EUV-driven escape, ion pickup, and impact erosion) and replenishment by impactors were not coupled in our model. This would require complete modeling of atmospheric composition and of how escape rates are dependent on it, the latter of which is not available for ion pickup. A possible scenario to support the primordial origin scenario by combining those processes is that, EUV-driven escape or ion pickup first removed the large amount of excess N in the atmosphere, and then impact erosion and replenishment later erased the excess $^{15}$N signature. Though the comparable timescales of those processes ($\simeq$100 Myrs) implies that this is not the case, a future modeling study is needed to fully test this scenario.

\red{Mineral-melt partitioning of N in the magma ocean solidification stage has not been fully understood. We used the partitioning coefficients estimated by \citeA{Yoshioka+2018}. This estimation was based on solubility experiments involving minerals and melts coexisting with a discrete fluid phase, and it is not guaranteed whether the partitioning coefficients are applicable for lower levels of N concentration. However, our partitioning model showed that the contribution of minerals as N hosts is smaller than that of melt trapping in the cumulate by two orders of magnitude (Figure \ref{fig:time-massN}). Thus, unless the actual mineral-melt partitioning coefficients were more than two orders of magnitude higher, the influence of their uncertainties would be minor.}

\red{Our cycling model does not consider possible N supply from the core, but its contribution for the mantle N content seems to be limited from the following reasons. First, the N abundance in the core is too low to cause N exsolution from the outer core due to N saturation induced by cooling and inner core solidification. The N solubility in the core condition is the order of $\sim$10 wt.\% \cite{Speelmanns+2018}. In contrast, geophysical constraints limit the N abundance in the core to be $\ll$ 2.0 wt.\% \cite{Bajgain+2019} and modeling of Earth’s accretion and volatile partitioning has estimated the core N abundance to be $\sim$10--100 ppm \cite{Sakuraba+2021}, both of which are much smaller than the saturation level. Second, the N exchange at the core-mantle boundary (CMB) would be insufficient to supply N from the core to explain the modern mantle N abundance. Occasional observations of core signatures in surface lavas suggest the fraction of silicate Earth which interacted with the core at the CMB may be the order of $\sim$1 wt.\% \cite{Hernlund+2015,Lim+2021}. In order to estimate the maximum contribution of the core, here we assume that this fraction of silicate Earth received core N through the interaction at the CMB and released all N to the entire mantle. To explain the modern mantle N abundance ($\sim$1 ppm), the mantle portion which interacted with the core needs to contain $\sim$100 ppm of N. Although mantle mineral-metal melt partition coefficients have not been directly measured, adapting an estimated value \cite<$\sim$100,>[]{Speelmanns+2018} leads to the required N abundance in the core side to be $\sim$1 wt.\%, which is much higher than the above-mentioned estimates \cite{Bajgain+2019,Sakuraba+2021}.}

\red{Our partitioning model assumed bottom-up solidification of an initially-fully-molten magma ocean, but Earth might have had a long-lived (the order of $\sim$1 Gyr) basal magma ocean \cite{Labrosse+2007}, which could affect N partitioning and trapping in the mantle. Although quantitative modeling of the influence of the basal magma ocean is beyond the scope of our study, we expect that, because of the two-step processes discussed below, its primary effect is to reduce the amount of N trapped in the bulk mantle and, consequently, limit the possibility of the primordial scenario. First, the lower cooling rate of the basal magma ocean compared to that of the surficial magma ocean (the orders of Gyrs vs. Myrs) would reduce the trapped melt fraction $F_{\rm tl}$ (Equation \ref{eq:Ftl}) significantly. This effect would limit the trapped N content in the solidified mantle above the basal magma (We note that for the basal magma ocean its solidification proceeds in a top-down fashion) and the majority of N which was initially in the magma just solidified would be released into the residual basal magma ocean. In the slow-cooling limit, the trapped N content, which is determined by N partitioning into crystallized minerals, would be smaller than that in the fast-cooling limit by two orders of magnitude (see Figure \ref{fig:time-massN} for comparison of N contents in minerals and in trapped melt). Second, N which concentrated in the residual basal magma ocean would ultimately be lost to the underlying core. Recent high-pressure experiments performed in graphite-undersaturated conditions \cite{Grewal+2021} have shown that the metal-silicate melt partitioning coefficient of N is $\sim$20--30 for $\log_{10}f_{\rm O_2,\Delta{\rm IW}} = -2$ \cite<the oxygen fugacity expected for the core-magma ocean boundary,>[]{Armstrong+2019}. As a result, the trapped N content in the bulk mantle would become smaller than that in the case of bottom-up solidification without the basal magma ocean.}

\subsection{Comparison with previous studies}
\label{subsec:discussion:comparison}

\red{Several} studies \red{on N partitioning in the magma ocean stage have} concluded that Earth's mantle N \red{is primordial} \cite{Libourel+2003,Boulliung+2020}, \red{but} these studies did not consider degassing upon solidification. The degassing process significantly limits N in the solidified mantle (Section \ref{subsec:results:partitioning}), which led us to conclude that the primordial origin scenario is only possible if Earth had an excess amount of N (Section \ref{subsec:discussion:Norigin1}). 

\citeA{Labidi+2020} simulated the evolution of N masses and isotopic ratios by using a two-reservoir model (the surface and mantle) where the atmosphere-sediment N isotopic fractionation factor was assumed to be constant through time. Their preferred model supported the primordial origin for mantle N. In contrast, our cycling model uses four reservoirs (the atmosphere, continental and oceanic crust, and mantle) and the isotopic fractionation factor varied with time following the evolution of surface N biogeochemical cycling, and found that the recycling origin was possible with high sedimentary N deposition rates on early Earth. 

Although our N cycling model is simple compared to the two recent \red{models} \cite{Stueken+2016ABb,Johnson+Goldblatt2018}, those most important results were reproduced: sequestration of N in the continental crust in the Archean \cite{Stueken+2016ABb} and efficient N subduction after the GOE \cite{Johnson+Goldblatt2018}. We note that the mechanisms are slightly different. In contrast to \citeA{Stueken+2016ABb} which modeled enhanced crustal weathering after the GOE as the release of N to the atmosphere, our model assumed the weathering as N-containing sediment transport to the oceanic crust, following \citeA{Johnson+Goldblatt2018}. The latter treatment seems to be more consistent with the picture of global N balance which suggests high N flux from crustal rocks to sediments \cite{Houlton+2018}. As a consequence, whereas weathering mainly led to increase of $p_{\rm N_2}$ in \citeA{Stueken+2016ABb}, our model showed efficient N subduction after the GOE. Moreover, efficient N subduction after the GOE in \citeA{Johnson+Goldblatt2018} was caused by an increase in primary biological production at the time, but in our case N transport to the mantle was the result of the release of the continental crust N, which was a product of N fixation before the GOE. This difference in the model assumption allowed our model to meet the low $p_{\rm N_2}$ constraints \cite{Som+2012,Som+2016,Marty+2013} in the Archean, while it was difficult with the nominal model of \citeA{Johnson+Goldblatt2018}.

\subsection{Implications for sample analysis}

Our modeling of N isotopic exchange provides a methodology to distinguish primordial and recycling origins for mantle N. The recycling origin is characterized by large increase ($\gtrsim$+10\textperthousand) in the mantle $\delta^{15}$N values with time, while the primordial origin model showed more limited change (Figures \ref{fig:Ncycle1}, \ref{fig:Ncycle1_WO_PHASE1}, \ref{fig:Ncycle1_WO_MccMocCONSTRAINT}, \ref{fig:Ncycle2}, and \ref{fig:Ncycle3}). The surface (the atmosphere and crust) $\delta^{15}$N record is more complicated because they can vary owing also to N exchange between the surficial reservoirs, but combining them with the mantle trend would help constrain deep N cycling. Moreover, if the surface N cycle had already become slightly aerobic as assumed in our model (Period 3, Section \ref{subsec:model_cycling}), the possible minimum in $p_{\rm N_2}$ around the time of the GOE can be informed from a kink in the surface $\delta^{15}$N values.

Given these results, constraining mantle isotopic evolution informed for example from diamonds \cite<e.g.,>[]{Cartigny+Marty2013} would offer a more global picture of Earth's N cycling and the origin of mantle N. In addition, better determination of the surface N isotopic composition and particularly more data around the time of the GOE will be useful. 

\subsection{Constraints from other volatile elements \red{and isotopes}}

Comparison to cycling of other volatile elements is potentially useful, but care should be taken to the importance of redox control on N. Xenon (Xe) \red{isotope} studies suggest limited subduction before 2.5 Ga \cite{Parai+Mukhopadhyay2018} and a higher degassing rate at the time \cite{Marty+2019}. If N behaved similarly to Xe, these conditions might rule out the recycling-origin scenario. However, the crust is an additional reservoir for N to cause different evolution. Furthermore, because both the surficial and mantle redox states chiefly influence N (but not noble gases), the redox control, in addition to the different stability in their host phases, can lead decoupling of N from noble gases. Such decoupling between volatile elements has already been suggested for sulfur (S) and Xe \cite{Bekaert+2020}. Mass-independent S isotope signatures in the source of various hotspots indicate significant recycling occurred prior to 2.5 Ga \cite{Cabral+2013,Dottin+2020,Bekaert+2020}.

\red{A classical interpretation for elementally-depleted and isotopically-fractionated Xe in Earth's atmosphere as a result of early hydrodynamic escape \cite<e.g.,>[]{Pepin1991} might suggest significant N loss due to the same process and support the presence of excess N. However, this interpretation is not supported by recent measurements of paleo-atmospheric Xe which revealed that Xe isotopic fractionation was progressive through time \cite{Pujol+2011,Avice+2017,Avice+2018,Bekaert+2018,Bekaert2020xenon}. Rather, Xe fractionation was likely due to selective loss of Xe$^+$ ions caused by strong coupling with escaping H$^+$ ions with time \cite{Zahnle2019,Avice+Marty2020}.}

\red{The $^{40}$K-$^{40}$Ar decay system gives another constraints on degassing history as well as continental growth \cite{Ozima+Podosek2002,Pujol+2013,Johnson+Goldblatt2018,Guo+Korenaga2020}. Our cycling model showed that N masses and isotopic ratios in different major reservoirs on modern Earth can be reproduced with a wide range of the degassing parameter, $C_{\rm vol}$ (Section \ref{subsec:results:cycling}), but it can be constrained by $^{40}$Ar data. Combining these volatile elements and their isotopes in a single model would be important for future studies.}

\subsection{Implications for Venus, Mars, and exoplanets}

Venus' atmosphere has about 3 PAN \cite{Johnson+Goldblatt2015}, which is comparable to BSE N content (Table \ref{tab:constraints}). Whether Venus once possessed an habitable condition and an Earth-like volatile cycle is controversial \cite<e.g.,>[]{Way+2020,Krissansen+2021}, but recent three-dimensional atmospheric simulations \cite{Turbet+2021} and accretion modeling focusing on the difference in hydrogen to carbon ratios between Earth and Venus \cite{Sakuraba+2021} suggest Venus never possessed oceans. This lack of oceans would lead to no N cycling between the surface and mantle. Because Venus' magma ocean was likely more oxidized than Earth's due to hydrogen loss to space \cite{Wordsworth2016}, we predict that its mantle right after magma ocean solidification was depleted in N similarly to Earth (Section \ref{subsec:results:partitioning}). Provided that Venus and Earth have similar bulk N contents (namely that the atmosphere is the major reservoir of N for Venus), the comparison with Venus may suggest that N recycling operating only on Earth differed the distribution of N on the two planets. Alternatively, the different distribution might be caused by the cumulative effect of possible active volcanism on Venus. Constraining Venus' volcanic activity with future missions \cite<e.g., ESA's EnVision and NASA's VERITAS and DAVINCI+,>[]{DIncecco2021} may provide better understanding of the partitioning and cycling of N on Venus.

Prolonged N degassing on Mars has been suggested from the atmospheric $^{15}$N/$^{14}$N ratio being close to the steady state between atmospheric escape and degassing \cite{Kurokawa+2018,Lammer+2020SSR}. Mars' small size might have led to a more reducing condition than Earth during its magma ocean stage \cite{Armstrong+2019}. Our partitioning model predicted {$\gtrsim$}10\% of bulk N can be partitioned into the mantle in the reduced case (Section \ref{subsec:results:partitioning}). While N recycling driven by plate tectonics has been absent at least since $\sim$4 Ga on Mars \cite<e.g.,>[]{Grott+2013}, the N-rich mantle might have contributed to prolonged degassing. The atmospheric neon abundance which is likely in balance between atmospheric escape and mantle degassing also suggests abundant volatile elements in Mars' mantle \cite{Kurokawa+2021}.

Our study suggests that the $p_{\rm N_2}$ of extrasolar rocky planets in habitable zones may provide useful constraints not only on present climate \cite{Vladilo+2013,Wordsworth+Pierrehumbert2014} but also on the biogeochemical processes that have acted on such planets. The $p_{\rm N_2}$ of extrasolar rocky planets can in principle be constrained from future observations \cite<e.g.,>[]{Schwieterman+2015,Stueken+2016ABb} through e.g., N$_2$-N$_2$ dimer absorption, collision-induced absorption involving N$_2$, and/or Rayleigh scattering slope, although the last measurement is prone to the degeneracy with other parameters. Combined with the estimates for the amount of N that a planet initially accreted and the amount that has escaped to space, a low $p_{\rm N_2}$ value may indicate efficient subduction of N, which may be attributed to active tectonics (which may not necessarily Earth-like plate tectonics but can be mobile lid and episodic resurfacing for example) with liquid water and perhaps biological N cycling. It would be interesting to look at the $p_{\rm N_2}$ of multiple rocky planets in the same system whose building blocks can be reasonably considered to be similar; they may show interesting diversity in the $p_{\rm N_2}$ depending on their evolution, similar to the case of Venus, Earth and Mars. Our model for the recycling origin scenario for mantle N implies that the $p_{\rm N_2}$ of planets with active tectonics and an anoxic atmosphere anti-correlates with their age, which can be tested with future observations. In this manner, understanding the origin of Earth's mantle N and the surface environments of extrasolar rocky planets is linked with each other.

\section{Conclusions}
\label{sec:conclusions}

Two hypotheses for the origin of N in Earth's mantle were tested with numerical models constrained by Ar and N isotopes. Our model results for N partitioning in the magma ocean stage constrained by Ar showed that low solubility and low partition coefficients between minerals and silicate melts leads to N depletion in the solidified mantle unless the magma ocean was reducing. The highly reducing magma ocean model may be relevant to Mars which shows indirect evidence of prolonged degassing of N, but it is not favored for Earth from high pressure experiments \red{or} geochemical and geophysical constraints. Modeling of impact erosion and isotopic fractionation via EUV-driven escape and ion pick up suggests that the excess atmospheric N, which is required to put sufficient N into the mantle as comparable as the minimum estimate for the modern mantle, can be removed only when efficient net removal is assumed. Our model results for N cycling between surface reservoirs and mantle constrained by N isotopes showed that sequestration of N into the crust before the GOE and subsequent efficient weathering and subduction after the period could have sourced N to the modern mantle. This recycling origin scenario requires high sedimentary N burial flux on early Earth, which was likely sustained by biotic N utilization. The two \red{hypotheses} can be tested by analyzing the surface and mantle N isotope record. Unveiling the origin of Earth's mantle N is linked with understanding the difference in N contents with Venus and the evolution of surface environments of extrasolar rocky planets.

\appendix

\section{N solubility model}
\label{sec:appendix_SN}

\begin{figure}
    \centering
    \includegraphics[width=\linewidth]{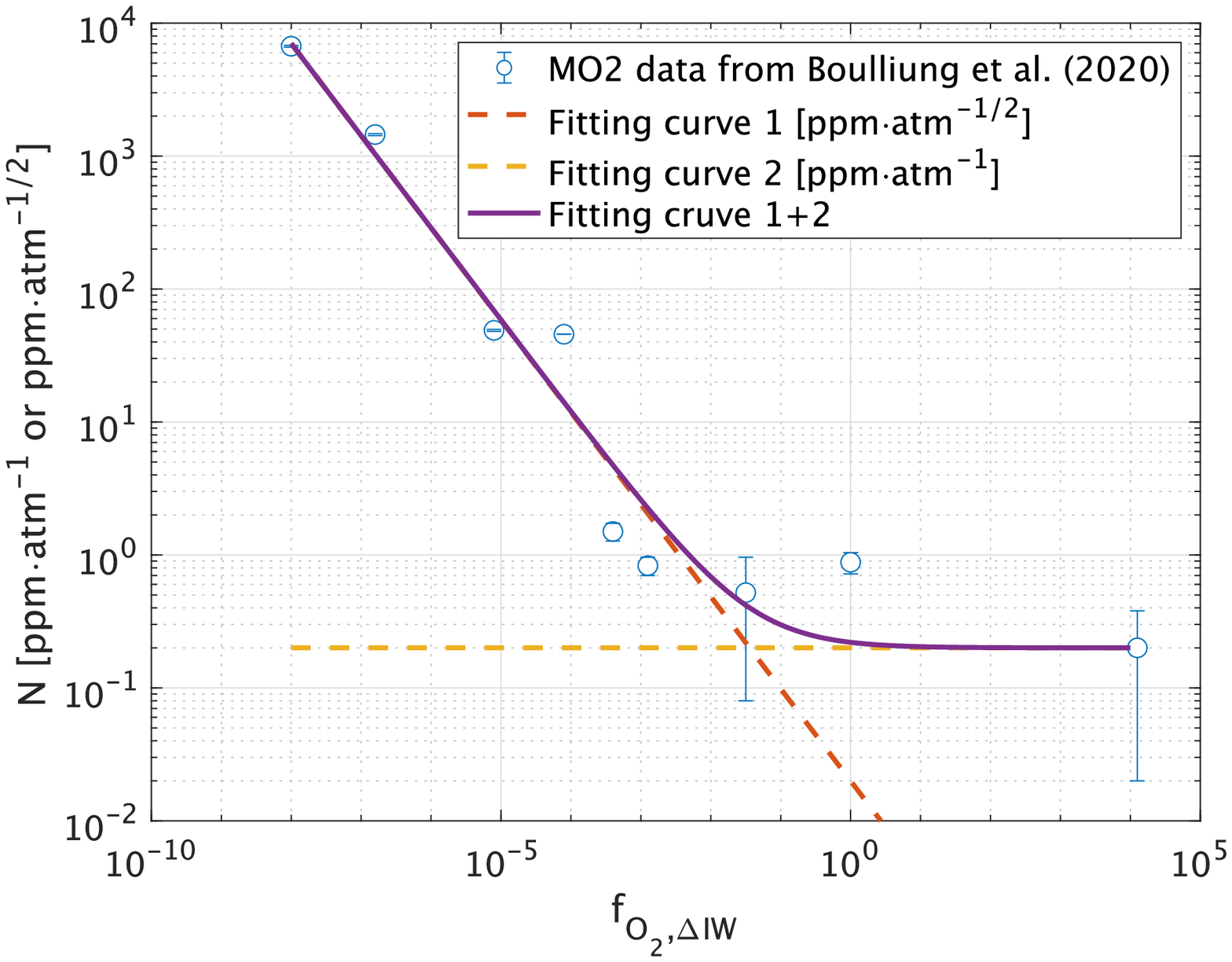}
    \caption{N solubility model used in this study (Equation \ref{eq:SN}, purple curve). Two terms in the fitting formula are also shown (fitting curves 1 and 2, red and yellow lines). Original data are from \citeA{Boulliung+2020} (Blue points).}
    \label{fig:SN}
\end{figure}

We developed a model for N solubility as a function of the oxygen fugacity and the N partial pressure. Solubility data measured by \citeA{Boulliung+2020} for the terrestrial magma ocean-like composition (MO2) were fitted by the sum of two terms following \citeA{Libourel+2003} as (Figure \ref{fig:SN}),
\begin{equation}
    S_{\rm N} {\rm [ppm \cdot atm^{-1}]} = 10^{-0.6934{\log_{10}f_{\rm O_2,\Delta{\rm IW}}} - 1.7002} \cdot \biggl( \frac{p_{\rm N_2}}{\rm 1\ atm} \biggr)^{-\frac{1}{2}} + 0.2. \label{eq:SN}
\end{equation}
The first and second terms denote chemical and physical solubilities dominant under reducing and oxidizing conditions, respectively \cite{Libourel+2003,Bernadou+2021}. The first term was given with the least squares method for $\log_{10}f_{\rm O_2,\Delta{\rm IW}} < -1.5$ data, assuming $p_{\rm N_2}^{1/2}$ dependence \cite{Boulliung+2020}. The second term was given to much the data point for the oxidized limit ($\log_{10}f_{\rm O_2,\Delta{\rm IW}} = +4$).

\red{\section{Derivation of Equation \ref{eq:Miatm}}
\label{sec:appendix_Miatm}

The equation for the solubility equilibrium between the atmosphere and the magma ocean (Equation \ref{eq:Miatm}) is derived from those for hydrostatic equilibrium, the ideal gas, and the definition of solubility, which are given by,
\begin{equation}
    \frac{dp}{dz} = - \rho_{\rm atm} g, \label{eq:static}
\end{equation}
\begin{equation}
    p_i = n_i k_{\rm B} T = \rho_i k_{\rm B} T / m_i, \label{eq:ideal}
\end{equation}
and,
\begin{equation}
    p_{i} = \frac{x_i}{S_i}, \label{eq:solbility}
\end{equation}
where $z$, $\rho_{\rm atm}$, $n_i$ and, $x_i$ are the height from the surface, the density of atmospheric gas, the number density of species $i$, and the mass fraction of species $i$ in the magma ocean, respectively. From Equations \ref{eq:static} and \ref{eq:ideal}, we obtain,
\begin{equation}
    \frac{d}{dz} \biggl( \frac{n}{n_i} p_i  \biggr) = - \frac{n}{n_i} \frac{\bar{m}}{m_i} \rho_i g. \label{eq:dpidz}
\end{equation}
Assuming the mixing ratio $n_i/n$ to be constant in the atmosphere, Equation \ref{eq:dpidz} can be integrated analytically to lead,
\begin{equation}
    p_i (z=0) = \frac{\bar{m}}{m_i} \frac{M_{{\rm atm},i}}{A} g. \label{eq:pi}
\end{equation}
We can rewrite $M_{{\rm mo},i}$ using Equations \ref{eq:solbility} and \ref{eq:pi} and obtain,
\begin{equation}
    M_{{\rm mo},i} = x_i M_{\rm mo} = \frac{\bar{m}}{m_i} \frac{M_{{\rm atm},i}}{A} g \cdot S_i \cdot M_{\rm mo}. \label{eq:Mmoi}
\end{equation}
Finally, substituting Equation \ref{eq:Mmoi} into Equation \ref{eq:Miatmmo} leads to Equation \ref{eq:Miatm}.
}

\section{Supplementary results of the cycling model}

{Figure \ref{fig:Ncycle1_WO_MccMocCONSTRAINT} shows the result of the nominal model runs without the constraints on the crustal N masses and isotopic ratios. Figures \ref{fig:Ncycle2} and \ref{fig:Ncycle3} show the variants of the nominal model, where we assumed the onset time of plate tectonics to be 4.5 Ga (Figure \ref{fig:Ncycle2}) and the initiation of N subduction at 2.5 Ga (Figure \ref{fig:Ncycle3}), respectively. }

\begin{landscape}

\begin{figure}
    \centering
    \includegraphics[width=\linewidth]{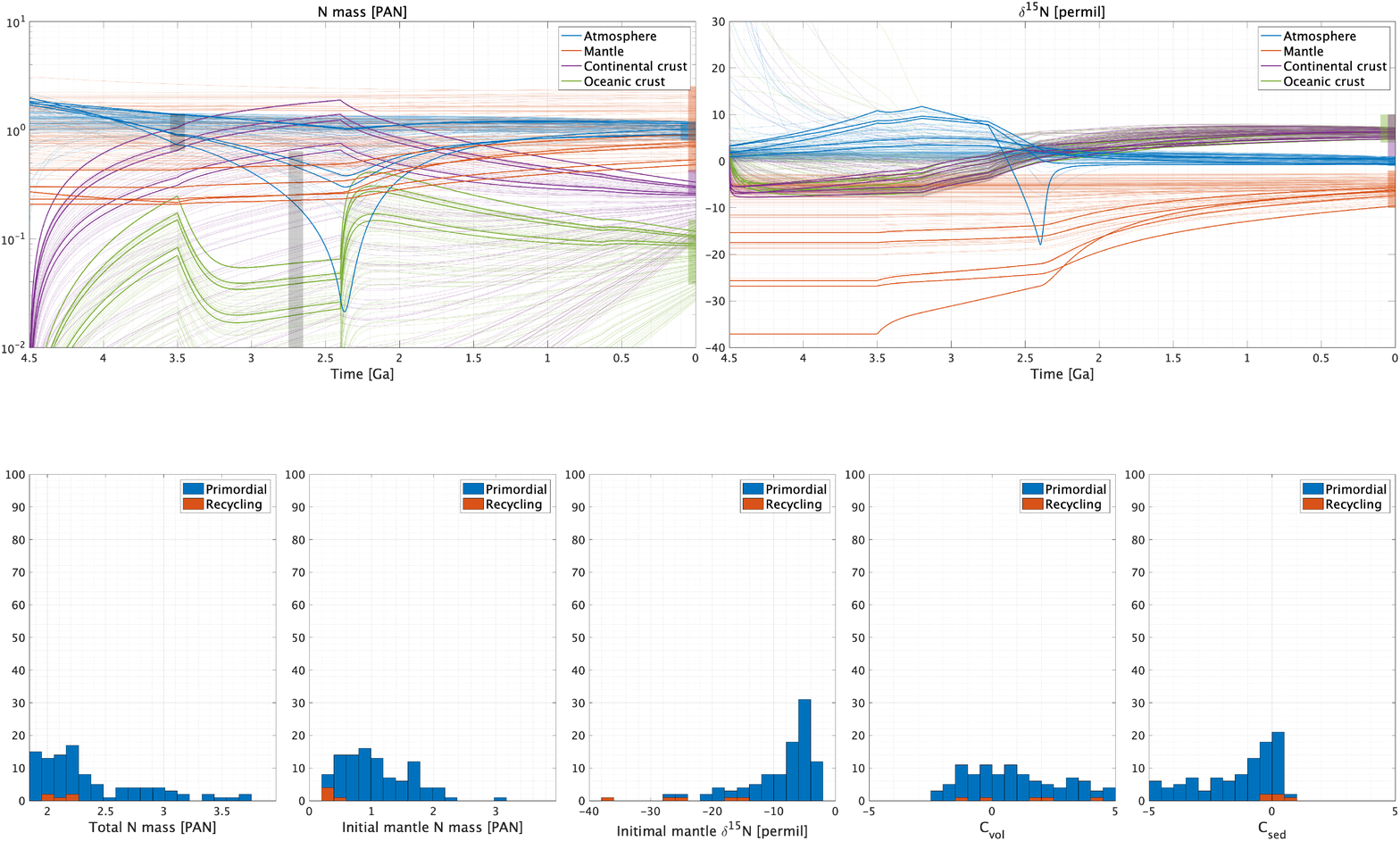}
    \caption{Cycling model results for {the case where we omitted the constraints on the N content and isotopic ratios of the modern continental and oceanic crust, while the other settings are the same as in the nominal case (Figure \ref{fig:Ncycle1})}. See Figure \ref{fig:Ncycle1} caption for details.}
    \label{fig:Ncycle1_WO_MccMocCONSTRAINT}
\end{figure}

\begin{figure}
    \centering
    \includegraphics[width=\linewidth]{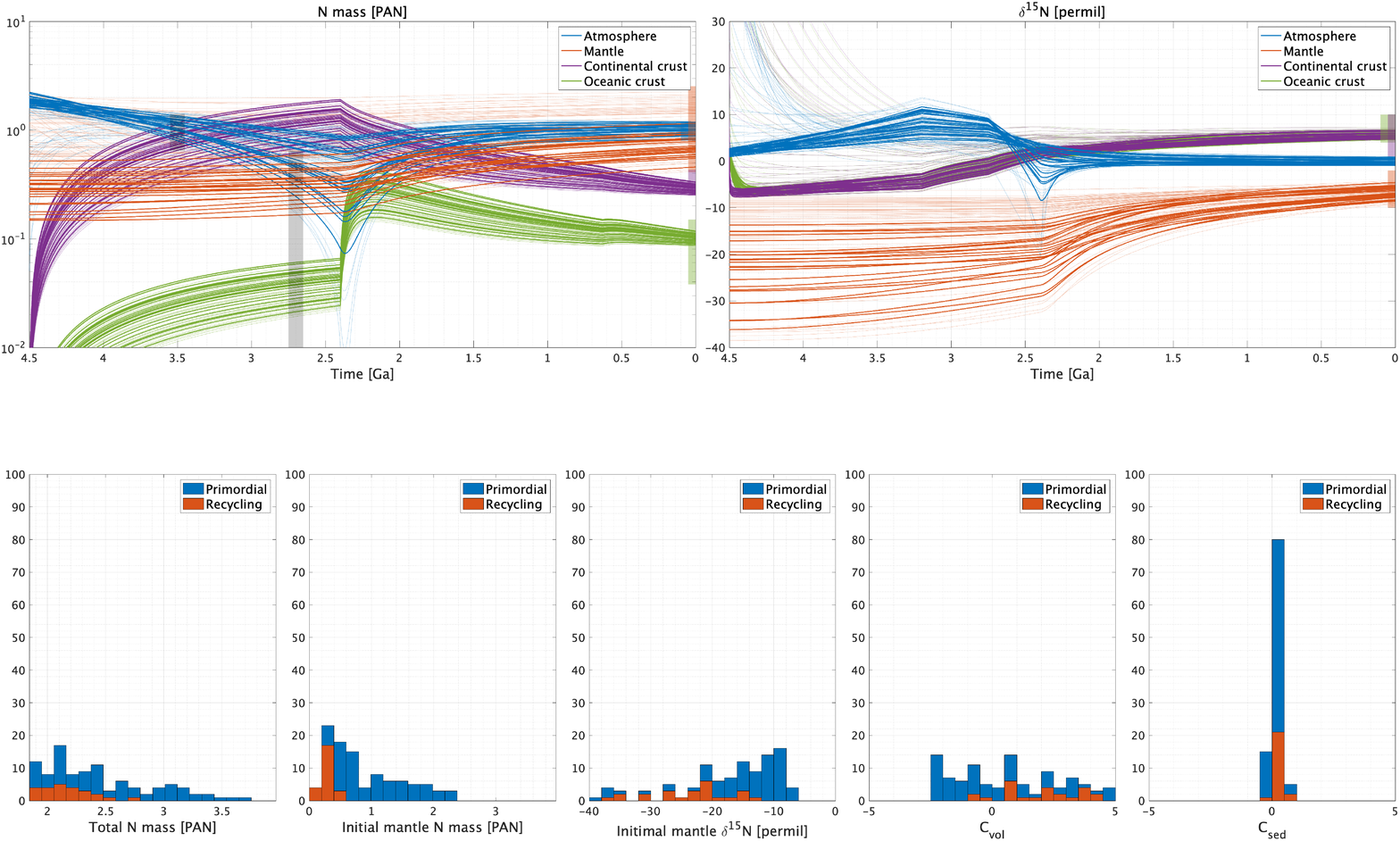}
    \caption{Cycling model results for {the case where the plate tectonics started at 4.5 Ga, while the other settings are the same as in the nominal case (Figure \ref{fig:Ncycle1})}. See Figure \ref{fig:Ncycle1} caption for details.}
    \label{fig:Ncycle2}
\end{figure}

\begin{figure}
    \centering
    \includegraphics[width=\linewidth]{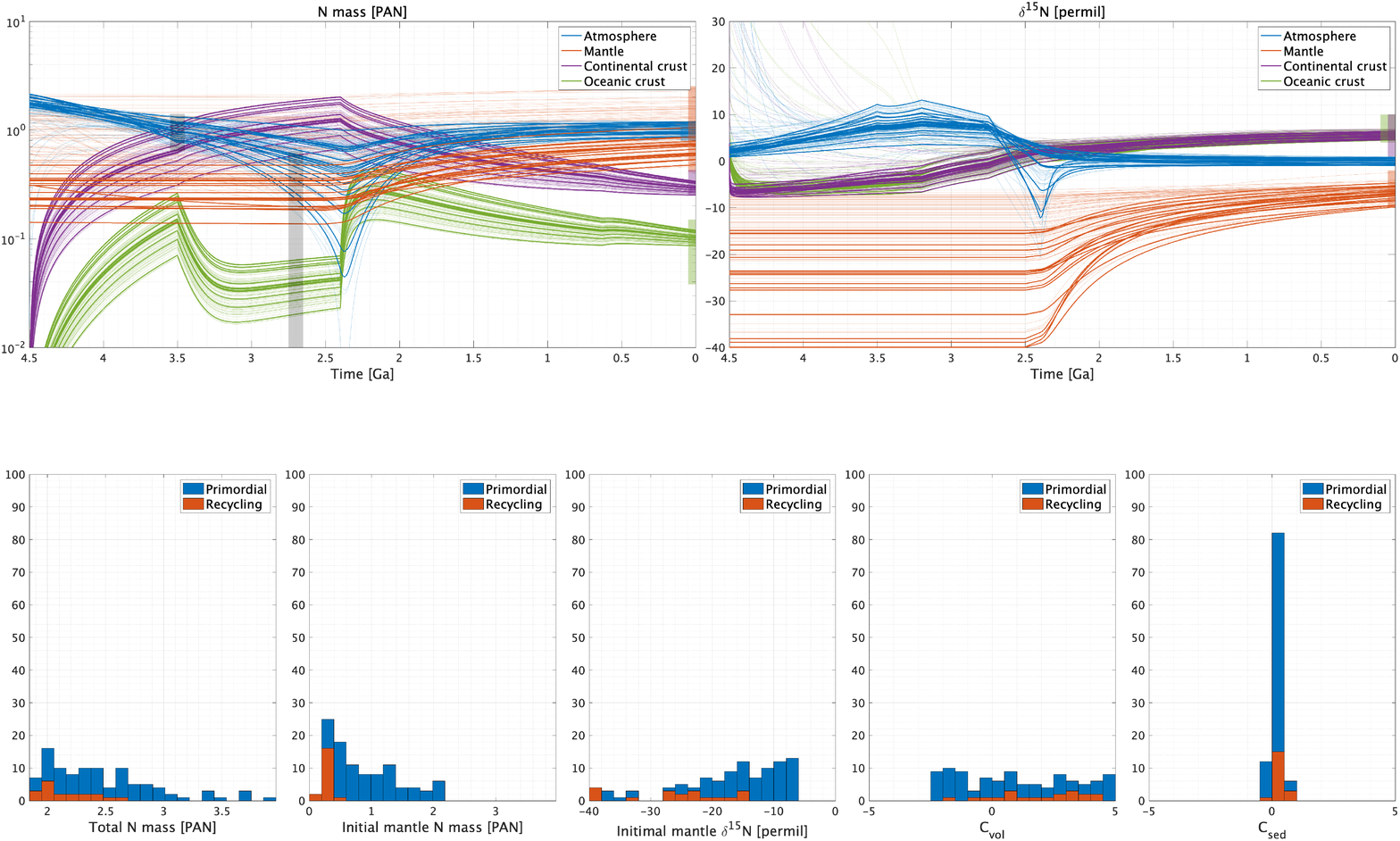}
    \caption{Cycling model results for {the case where N subduction started 2.5 Ga, while the other settings are the same as in the nominal case (Figure \ref{fig:Ncycle1})}. See Figure \ref{fig:Ncycle1} caption for details.}
    \label{fig:Ncycle3}
\end{figure}

\end{landscape}

\section{Atmospheric escape models}
\label{sec:appendix_escape}

The isotopic fractionation factor, defined as, 
\begin{equation}
    f \equiv \frac{F_{\rm ^{15}N}/F_{\rm ^{14}N}}{[{\rm ^{15}N}]/[{\rm ^{14}N}]}, 
\end{equation}
{was} estimated for EUV-driven hydrodynamic escape and solar wind-induced ion pickup, where $F_i$ is the escape flux. Then we estimated the amounts of N removed by the escape processes $M_{\rm lost}$ to enrich the remnant atmosphere (1 PAN) in $^{15}$N by $\Delta {\rm ^{15}N} = +6$\textperthousand \ (Figure \ref{fig:escape}), using the Rayleigh dissolution equation, 
\begin{equation}
    M_{\rm lost} = \biggl( \frac{1}{1-10^{-3}\times \Delta {\rm ^{15}N}} \biggr)^\frac{1}{1-f} -1 \ [{\rm PAN}].
\end{equation}

Escaping hydrogen drags $^{14}$N$^{14}$N and $^{15}$N$^{14}$N mass-dependently, and causes isotopic fractionation. The fractionation factor for hydrodynamic escape {was} estimated using a model of \citeA{Lammer+2020}. In the model, the radii of visible- and EUV-photospheres are estimated with their Equations 5, 11, and 12. Then escape fluxes of hydrogen and other minor species are computed with their Equations 10, 14, and 15. For simplicity, we assumed that the model proto-Earth {had} a H$_2$ plus N$_2$ atmosphere, whose mass is 10$^{-3}$ times Earth's mass. The mixing ratio of N$_2$ {was} assumed to be 10$^{-3}$ considering a primordial dense H$_2$ atmosphere. We assume skin and upper atmospheric temperatures to be 255 K and 400 K \cite{Watson+1981}, respectively. The binary diffusion coefficient between H and N$_2$ was taken from \citeA{Zahnle+1986}. EUV luminosity of the young Sun {was} taken to be a parameter, and is expected to be 30--500 times the present-day value during the first 100 Myrs, depending on its rotation rate \cite{Tu+2015}.

Ion pickup by the solar wind removes N from the exobase level, where heavier $^{15}$N is depleted due to diffusive separation. The fractionation factor {was} estimated by adapting an isothermal, hydrostatic structure for both $^{14}$N$^{14}$N and $^{15}$N$^{14}$N, given as,
\begin{equation}
    f = \exp{\biggl[ \frac{G M_{\rm E} \Delta m}{k_b T_{\rm exo} r_{\rm homo}}{\biggl( \frac{r_{\rm homo}}{r_{\rm exo}} - 1 \biggr)}  \biggr]}, 
\end{equation}
where $G$ is the gravitational constant, $k_b$ is the Boltzmann constant, $\Delta m$ is the mass difference between $^{14}$N$^{14}$N and $^{15}$N$^{14}$N, $T_{\rm exo}$ is the exospheric temperature, and $r_{\rm homo}$ and $r_{\rm exo}$ is the radii of homopause and exobase, respectively. We {assumed} $T_{\rm exo} = 10^4\ {\rm K}$, $r_{\rm homo} = R_{\rm E} + 100\ {\rm km}$, and $r_{\rm exo} = 12.7 R_{\rm E}$ \cite{Tian+2008,Lichtenegger+2010}.

\section{Impact erosion model}
\label{sec:appendix_impacts}

Atmospheric loss by impact-induced atmospheric erosion during late accretion was calculated by using the model of \citeA{Sakuraba+2019} (Figure \ref{fig:impacts}). The model setting is the same for the early Earth model of \citeA{Sakuraba+2019} except for the impactor composition and the initial N content, for the latter we assumed the estimate from our partitioning model for the oxidized magma-ocean case (7.7 PAN to 150 PAN).

Here we summarize the main feature of the impact erosion model \cite<see>[for details]{Sakuraba+2019}. The impact erosion model computes the loss and supply of volatile elements by statistically-averaged impacts. Assuming the carbonate-silicate cycle, we {fixed} the background CO$_2$ partial pressure at 1 bar. A trace amount of water vapor (0.017 bar) {was} included assuming the saturated vapor pressure. We assumed the surface temperature to be 288 K. The results depend on the assumed temperature only weakly \cite{Sakuraba+2019,Sakuraba+2021}.

We assumed two cases for the N content of impactors: enstatite chondrite-like impactors containing 400 ppm N$_2$ \cite{hirschmann2016, Bergin+2015} and \textquotedblleft dry\textquotedblright \ objects containing no N. The former is consistent with the N isotopic signatures of Earth and enstatite chondrites \cite{Dalou+2019,Piani+2020}. The latter is an extreme case to maximize the eroded N mass case. {We did not compute the case of carbonaceous chondrite-like impactors, but they are richer in N than enstatite chondrites \cite<e.g.,>[]{Grady+Wright2003} so that the net effect to remove N from the atmosphere is weaker \cite{Sakuraba+2019,Sakuraba+2021}.} Isotopic signatures of siderophile elements in the mantle also point to the types of late accretion impactors containing a non-negligible amount of N, such as enstatite or carbonaceous chondtites \cite{Dauphas2017,Fischer+2020}. 

The assumed size distribution of the impactors is $\mathrm dN (D)/\mathrm dD \propto D^{-q}$ with $q = 3$, where $N(D)$ is the number of objects of diameter smaller than $D$. The minimum and maximum sizes are assumed to be $D = 10^{-1.5}$ km and $10^3$ km, respectively. {A steeper size distribution (larger $q$) leads to more efficient impact erosion.} The assumed size distribution is similar to that of the asteroid belt {\cite{Bottke+2005}}, {and is likely an upper limit for $q$ \cite<see,>[for discussion]{Sakuraba+2021}}.

{The cumulative mass of late accretion is constrained by highly siderophile element concentrations in Earth's mantle \cite{Chou1978,Marchi+2018}. We assumed 0.5\% of Earth's mass as a minimum estimate, but calculated up to 1\% to take a larger estimate into account.}

\acknowledgments
Data and codes to generate figures in this paper are available at \url{https://doi.org/10.6084/m9.figshare.17128421.v1}. \red{We thank Bernard Marty and an anonymous reviewer for constructive comments.} This work was supported by JSPS KAKENHI Grant number 15H05832, 17H01175, 17H06457, 18K13601, 18K13602, 19H01960, 19H05072, 19K15671, 20K04126, 20KK0080, 21H04514, 20H04608, and 21K13976.


%
%

\bibliography{references.bib}

\begin{thebibliography}{}

\bibitem [\protect \citeauthoryear {%
Ader%
\ \protect \BOthers {.}}{%
Ader%
\ \protect \BOthers {.}}{%
{\protect \APACyear {2016}}%
}]{%
Ader+2016}
\APACinsertmetastar {%
Ader+2016}%
\begin{APACrefauthors}%
Ader, M.%
, Thomazo, C.%
, Sansjofre, P.%
, Busigny, V.%
, Papineau, D.%
, Laffont, R.%
\BDBL {}Halverson, G\BPBI P.%
\end{APACrefauthors}%
\unskip\
\newblock
\APACrefYearMonthDay{2016}{}{}.
\newblock
{\BBOQ}\APACrefatitle {Interpretation of the nitrogen isotopic composition of
  {P}recambrian sedimentary rocks: {A}ssumptions and perspectives}
  {Interpretation of the nitrogen isotopic composition of {P}recambrian
  sedimentary rocks: {A}ssumptions and perspectives}.{\BBCQ}
\newblock
\APACjournalVolNumPages{Chemical Geology}{429}{}{93--110}.
\PrintBackRefs{\CurrentBib}

\bibitem [\protect \citeauthoryear {%
Armstrong%
, Frost%
, McCammon%
, Rubie%
\BCBL {}\ \BBA {} Ballaran%
}{%
Armstrong%
\ \protect \BOthers {.}}{%
{\protect \APACyear {2019}}%
}]{%
Armstrong+2019}
\APACinsertmetastar {%
Armstrong+2019}%
\begin{APACrefauthors}%
Armstrong, K.%
, Frost, D\BPBI J.%
, McCammon, C\BPBI A.%
, Rubie, D\BPBI C.%
\BCBL {}\ \BBA {} Ballaran, T\BPBI B.%
\end{APACrefauthors}%
\unskip\
\newblock
\APACrefYearMonthDay{2019}{}{}.
\newblock
{\BBOQ}\APACrefatitle {Deep magma ocean formation set the oxidation state of
  {E}arth’s mantle} {Deep magma ocean formation set the oxidation state of
  {E}arth’s mantle}.{\BBCQ}
\newblock
\APACjournalVolNumPages{Science}{365}{6456}{903--906}.
\PrintBackRefs{\CurrentBib}

\bibitem [\protect \citeauthoryear {%
Aulbach%
\ \BBA {} Stagno%
}{%
Aulbach%
\ \BBA {} Stagno%
}{%
{\protect \APACyear {2016}}%
}]{%
Aulbach+Stagno2016}
\APACinsertmetastar {%
Aulbach+Stagno2016}%
\begin{APACrefauthors}%
Aulbach, S.%
\BCBT {}\ \BBA {} Stagno, V.%
\end{APACrefauthors}%
\unskip\
\newblock
\APACrefYearMonthDay{2016}{}{}.
\newblock
{\BBOQ}\APACrefatitle {Evidence for a reducing {A}rchean ambient mantle and its
  effects on the carbon cycle} {Evidence for a reducing {A}rchean ambient
  mantle and its effects on the carbon cycle}.{\BBCQ}
\newblock
\APACjournalVolNumPages{Geology}{44}{9}{751--754}.
\PrintBackRefs{\CurrentBib}

\bibitem [\protect \citeauthoryear {%
Aulbach%
\ \protect \BOthers {.}}{%
Aulbach%
\ \protect \BOthers {.}}{%
{\protect \APACyear {2019}}%
}]{%
Aulbach+2019}
\APACinsertmetastar {%
Aulbach+2019}%
\begin{APACrefauthors}%
Aulbach, S.%
, Woodland, A\BPBI B.%
, Stern, R\BPBI A.%
, Vasilyev, P.%
, Heaman, L\BPBI M.%
\BCBL {}\ \BBA {} Viljoen, K.%
\end{APACrefauthors}%
\unskip\
\newblock
\APACrefYearMonthDay{2019}{}{}.
\newblock
{\BBOQ}\APACrefatitle {Evidence for a dominantly reducing {A}rchaean ambient
  mantle from two redox proxies, and low oxygen fugacity of deeply subducted
  oceanic crust} {Evidence for a dominantly reducing {A}rchaean ambient mantle
  from two redox proxies, and low oxygen fugacity of deeply subducted oceanic
  crust}.{\BBCQ}
\newblock
\APACjournalVolNumPages{Scientific Reports}{9}{1}{1--11}.
\PrintBackRefs{\CurrentBib}

\bibitem [\protect \citeauthoryear {%
Avice%
\ \BBA {} Marty%
}{%
Avice%
\ \BBA {} Marty%
}{%
{\protect \APACyear {2020}}%
}]{%
Avice+Marty2020}
\APACinsertmetastar {%
Avice+Marty2020}%
\begin{APACrefauthors}%
Avice, G.%
\BCBT {}\ \BBA {} Marty, B.%
\end{APACrefauthors}%
\unskip\
\newblock
\APACrefYearMonthDay{2020}{}{}.
\newblock
{\BBOQ}\APACrefatitle {Perspectives on atmospheric evolution from noble gas and
  nitrogen isotopes on {E}arth, {M}ars \& {V}enus} {Perspectives on atmospheric
  evolution from noble gas and nitrogen isotopes on {E}arth, {M}ars \&
  {V}enus}.{\BBCQ}
\newblock
\APACjournalVolNumPages{Space Science Reviews}{216}{3}{1--18}.
\PrintBackRefs{\CurrentBib}

\bibitem [\protect \citeauthoryear {%
Avice%
, Marty%
\BCBL {}\ \BBA {} Burgess%
}{%
Avice%
\ \protect \BOthers {.}}{%
{\protect \APACyear {2017}}%
}]{%
Avice+2017}
\APACinsertmetastar {%
Avice+2017}%
\begin{APACrefauthors}%
Avice, G.%
, Marty, B.%
\BCBL {}\ \BBA {} Burgess, R.%
\end{APACrefauthors}%
\unskip\
\newblock
\APACrefYearMonthDay{2017}{}{}.
\newblock
{\BBOQ}\APACrefatitle {The origin and degassing history of the {E}arth's
  atmosphere revealed by {A}rchean xenon} {The origin and degassing history of
  the {E}arth's atmosphere revealed by {A}rchean xenon}.{\BBCQ}
\newblock
\APACjournalVolNumPages{Nature communications}{8}{1}{1--9}.
\PrintBackRefs{\CurrentBib}

\bibitem [\protect \citeauthoryear {%
Avice%
\ \protect \BOthers {.}}{%
Avice%
\ \protect \BOthers {.}}{%
{\protect \APACyear {2018}}%
}]{%
Avice+2018}
\APACinsertmetastar {%
Avice+2018}%
\begin{APACrefauthors}%
Avice, G.%
, Marty, B.%
, Burgess, R.%
, Hofmann, A.%
, Philippot, P.%
, Zahnle, K.%
\BCBL {}\ \BBA {} Zakharov, D.%
\end{APACrefauthors}%
\unskip\
\newblock
\APACrefYearMonthDay{2018}{}{}.
\newblock
{\BBOQ}\APACrefatitle {Evolution of atmospheric xenon and other noble gases
  inferred from {A}rchean to {P}aleoproterozoic rocks} {Evolution of
  atmospheric xenon and other noble gases inferred from {A}rchean to
  {P}aleoproterozoic rocks}.{\BBCQ}
\newblock
\APACjournalVolNumPages{Geochimica et Cosmochimica Acta}{232}{}{82--100}.
\PrintBackRefs{\CurrentBib}

\bibitem [\protect \citeauthoryear {%
Badro%
, Brodholt%
, Piet%
, Siebert%
\BCBL {}\ \BBA {} Ryerson%
}{%
Badro%
\ \protect \BOthers {.}}{%
{\protect \APACyear {2015}}%
}]{%
Badro+2015}
\APACinsertmetastar {%
Badro+2015}%
\begin{APACrefauthors}%
Badro, J.%
, Brodholt, J\BPBI P.%
, Piet, H.%
, Siebert, J.%
\BCBL {}\ \BBA {} Ryerson, F\BPBI J.%
\end{APACrefauthors}%
\unskip\
\newblock
\APACrefYearMonthDay{2015}{}{}.
\newblock
{\BBOQ}\APACrefatitle {Core formation and core composition from coupled
  geochemical and geophysical constraints} {Core formation and core composition
  from coupled geochemical and geophysical constraints}.{\BBCQ}
\newblock
\APACjournalVolNumPages{Proceedings of the National Academy of
  Sciences}{112}{40}{12310--12314}.
\PrintBackRefs{\CurrentBib}

\bibitem [\protect \citeauthoryear {%
Bajgain%
, Mookherjee%
, Dasgupta%
, Ghosh%
\BCBL {}\ \BBA {} Karki%
}{%
Bajgain%
\ \protect \BOthers {.}}{%
{\protect \APACyear {2019}}%
}]{%
Bajgain+2019}
\APACinsertmetastar {%
Bajgain+2019}%
\begin{APACrefauthors}%
Bajgain, S\BPBI K.%
, Mookherjee, M.%
, Dasgupta, R.%
, Ghosh, D\BPBI B.%
\BCBL {}\ \BBA {} Karki, B\BPBI B.%
\end{APACrefauthors}%
\unskip\
\newblock
\APACrefYearMonthDay{2019}{}{}.
\newblock
{\BBOQ}\APACrefatitle {Nitrogen content in the {E}arth's outer core} {Nitrogen
  content in the {E}arth's outer core}.{\BBCQ}
\newblock
\APACjournalVolNumPages{Geophysical Research Letters}{46}{1}{89--98}.
\PrintBackRefs{\CurrentBib}

\bibitem [\protect \citeauthoryear {%
Barry%
\ \BBA {} Hilton%
}{%
Barry%
\ \BBA {} Hilton%
}{%
{\protect \APACyear {2016}}%
}]{%
Barry+Hilton2016}
\APACinsertmetastar {%
Barry+Hilton2016}%
\begin{APACrefauthors}%
Barry, P.%
\BCBT {}\ \BBA {} Hilton, D.%
\end{APACrefauthors}%
\unskip\
\newblock
\APACrefYearMonthDay{2016}{}{}.
\newblock
{\BBOQ}\APACrefatitle {Release of subducted sedimentary nitrogen throughout
  {E}arth’s mantle} {Release of subducted sedimentary nitrogen throughout
  {E}arth’s mantle}.{\BBCQ}
\newblock
\APACjournalVolNumPages{Geochemical Perspectives Letters}{2}{}{138--147}.
\PrintBackRefs{\CurrentBib}

\bibitem [\protect \citeauthoryear {%
Bekaert%
\ \protect \BOthers {.}}{%
Bekaert%
\ \protect \BOthers {.}}{%
{\protect \APACyear {2018}}%
}]{%
Bekaert+2018}
\APACinsertmetastar {%
Bekaert+2018}%
\begin{APACrefauthors}%
Bekaert, D\BPBI V.%
, Broadley, M\BPBI W.%
, Delarue, F.%
, Avice, G.%
, Robert, F.%
\BCBL {}\ \BBA {} Marty, B.%
\end{APACrefauthors}%
\unskip\
\newblock
\APACrefYearMonthDay{2018}{}{}.
\newblock
{\BBOQ}\APACrefatitle {Archean kerogen as a new tracer of atmospheric
  evolution: Implications for dating the widespread nature of early life}
  {Archean kerogen as a new tracer of atmospheric evolution: Implications for
  dating the widespread nature of early life}.{\BBCQ}
\newblock
\APACjournalVolNumPages{Science advances}{4}{2}{eaar2091}.
\PrintBackRefs{\CurrentBib}

\bibitem [\protect \citeauthoryear {%
Bekaert%
\ \protect \BOthers {.}}{%
Bekaert%
\ \protect \BOthers {.}}{%
{\protect \APACyear {2020}}%
}]{%
Bekaert2020xenon}
\APACinsertmetastar {%
Bekaert2020xenon}%
\begin{APACrefauthors}%
Bekaert, D\BPBI V.%
, Broadley, M\BPBI W.%
, Delarue, F.%
, Druzhinina, Z.%
, Paris, G.%
, Robert, F.%
\BDBL {}Marty, B.%
\end{APACrefauthors}%
\unskip\
\newblock
\APACrefYearMonthDay{2020}{}{}.
\newblock
{\BBOQ}\APACrefatitle {Xenon isotopes in {A}rchean and {P}roterozoic insoluble
  organic matter: A robust indicator of syngenecity?} {Xenon isotopes in
  {A}rchean and {P}roterozoic insoluble organic matter: A robust indicator of
  syngenecity?}{\BBCQ}
\newblock
\APACjournalVolNumPages{Precambrian Research}{336}{}{105505}.
\PrintBackRefs{\CurrentBib}

\bibitem [\protect \citeauthoryear {%
Bekaert%
\ \protect \BOthers {.}}{%
Bekaert%
\ \protect \BOthers {.}}{%
{\protect \APACyear {2021}}%
}]{%
Bekaert+2020}
\APACinsertmetastar {%
Bekaert+2020}%
\begin{APACrefauthors}%
Bekaert, D\BPBI V.%
, Turner, S\BPBI J.%
, Broadley, M\BPBI W.%
, Barnes, J\BPBI D.%
, Halld{\'o}rsson, S\BPBI A.%
, Labidi, J.%
\BDBL {}Barry, P\BPBI H.%
\end{APACrefauthors}%
\unskip\
\newblock
\APACrefYearMonthDay{2021}{}{}.
\newblock
{\BBOQ}\APACrefatitle {Subduction-Driven Volatile Recycling: A Global Mass
  Balance} {Subduction-driven volatile recycling: A global mass
  balance}.{\BBCQ}
\newblock
\APACjournalVolNumPages{Annual Review of Earth and Planetary
  Sciences}{49}{}{37--70}.
\PrintBackRefs{\CurrentBib}

\bibitem [\protect \citeauthoryear {%
Bergin%
, Blake%
, Ciesla%
, Hirschmann%
\BCBL {}\ \BBA {} Li%
}{%
Bergin%
\ \protect \BOthers {.}}{%
{\protect \APACyear {2015}}%
}]{%
Bergin+2015}
\APACinsertmetastar {%
Bergin+2015}%
\begin{APACrefauthors}%
Bergin, E\BPBI A.%
, Blake, G\BPBI A.%
, Ciesla, F.%
, Hirschmann, M\BPBI M.%
\BCBL {}\ \BBA {} Li, J.%
\end{APACrefauthors}%
\unskip\
\newblock
\APACrefYearMonthDay{2015}{}{}.
\newblock
{\BBOQ}\APACrefatitle {Tracing the ingredients for a habitable {E}arth from
  interstellar space through planet formation} {Tracing the ingredients for a
  habitable {E}arth from interstellar space through planet formation}.{\BBCQ}
\newblock
\APACjournalVolNumPages{Proceedings of the National Academy of
  Sciences}{112}{29}{8965--8970}.
\PrintBackRefs{\CurrentBib}

\bibitem [\protect \citeauthoryear {%
Bernadou%
, Gaillard%
, F{\"u}ri%
, Marrocchi%
\BCBL {}\ \BBA {} Slodczyk%
}{%
Bernadou%
\ \protect \BOthers {.}}{%
{\protect \APACyear {2021}}%
}]{%
Bernadou+2021}
\APACinsertmetastar {%
Bernadou+2021}%
\begin{APACrefauthors}%
Bernadou, F.%
, Gaillard, F.%
, F{\"u}ri, E.%
, Marrocchi, Y.%
\BCBL {}\ \BBA {} Slodczyk, A.%
\end{APACrefauthors}%
\unskip\
\newblock
\APACrefYearMonthDay{2021}{}{}.
\newblock
{\BBOQ}\APACrefatitle {Nitrogen solubility in basaltic silicate melt -
  {I}mplications for degassing processes} {Nitrogen solubility in basaltic
  silicate melt - {I}mplications for degassing processes}.{\BBCQ}
\newblock
\APACjournalVolNumPages{Chemical Geology}{573}{}{120192}.
\PrintBackRefs{\CurrentBib}

\bibitem [\protect \citeauthoryear {%
Berner%
}{%
Berner%
}{%
{\protect \APACyear {1982}}%
}]{%
Berner1982}
\APACinsertmetastar {%
Berner1982}%
\begin{APACrefauthors}%
Berner, R\BPBI A.%
\end{APACrefauthors}%
\unskip\
\newblock
\APACrefYearMonthDay{1982}{}{}.
\newblock
{\BBOQ}\APACrefatitle {Burial of organic carbon and pyrite sulfur in the modern
  ocean: its geochemical and environmental significance} {Burial of organic
  carbon and pyrite sulfur in the modern ocean: its geochemical and
  environmental significance}.{\BBCQ}
\newblock
\APACjournalVolNumPages{Am. J. Sci.;(United States)}{282}{}{}.
\PrintBackRefs{\CurrentBib}

\bibitem [\protect \citeauthoryear {%
Berner%
}{%
Berner%
}{%
{\protect \APACyear {2006}}%
}]{%
Berner2006}
\APACinsertmetastar {%
Berner2006}%
\begin{APACrefauthors}%
Berner, R\BPBI A.%
\end{APACrefauthors}%
\unskip\
\newblock
\APACrefYearMonthDay{2006}{}{}.
\newblock
{\BBOQ}\APACrefatitle {Geological nitrogen cycle and atmospheric {N}$_2$ over
  Phanerozoic time} {Geological nitrogen cycle and atmospheric {N}$_2$ over
  phanerozoic time}.{\BBCQ}
\newblock
\APACjournalVolNumPages{Geology}{34}{5}{413--415}.
\PrintBackRefs{\CurrentBib}

\bibitem [\protect \citeauthoryear {%
Bottke%
\ \protect \BOthers {.}}{%
Bottke%
\ \protect \BOthers {.}}{%
{\protect \APACyear {2005}}%
}]{%
Bottke+2005}
\APACinsertmetastar {%
Bottke+2005}%
\begin{APACrefauthors}%
Bottke, W\BPBI F.%
, Durda, D\BPBI D.%
, Nesvorn{\`y}, D.%
, Jedicke, R.%
, Morbidelli, A.%
, Vokrouhlick{\`y}, D.%
\BCBL {}\ \BBA {} Levison, H.%
\end{APACrefauthors}%
\unskip\
\newblock
\APACrefYearMonthDay{2005}{}{}.
\newblock
{\BBOQ}\APACrefatitle {The fossilized size distribution of the main asteroid
  belt} {The fossilized size distribution of the main asteroid belt}.{\BBCQ}
\newblock
\APACjournalVolNumPages{Icarus}{175}{1}{111--140}.
\PrintBackRefs{\CurrentBib}

\bibitem [\protect \citeauthoryear {%
Bottke%
, Walker%
, Day%
, Nesvorny%
\BCBL {}\ \BBA {} Elkins-Tanton%
}{%
Bottke%
\ \protect \BOthers {.}}{%
{\protect \APACyear {2010}}%
}]{%
Bottke+2010}
\APACinsertmetastar {%
Bottke+2010}%
\begin{APACrefauthors}%
Bottke, W\BPBI F.%
, Walker, R\BPBI J.%
, Day, J\BPBI M.%
, Nesvorny, D.%
\BCBL {}\ \BBA {} Elkins-Tanton, L.%
\end{APACrefauthors}%
\unskip\
\newblock
\APACrefYearMonthDay{2010}{}{}.
\newblock
{\BBOQ}\APACrefatitle {Stochastic late accretion to {E}arth, the {M}oon, and
  {M}ars} {Stochastic late accretion to {E}arth, the {M}oon, and
  {M}ars}.{\BBCQ}
\newblock
\APACjournalVolNumPages{Science}{330}{6010}{1527--1530}.
\PrintBackRefs{\CurrentBib}

\bibitem [\protect \citeauthoryear {%
Boulliung%
\ \protect \BOthers {.}}{%
Boulliung%
\ \protect \BOthers {.}}{%
{\protect \APACyear {2020}}%
}]{%
Boulliung+2020}
\APACinsertmetastar {%
Boulliung+2020}%
\begin{APACrefauthors}%
Boulliung, J.%
, F{\"u}ri, E.%
, Dalou, C.%
, Tissandier, L.%
, Zimmermann, L.%
\BCBL {}\ \BBA {} Marrocchi, Y.%
\end{APACrefauthors}%
\unskip\
\newblock
\APACrefYearMonthDay{2020}{}{}.
\newblock
{\BBOQ}\APACrefatitle {Oxygen fugacity and melt composition controls on
  nitrogen solubility in silicate melts} {Oxygen fugacity and melt composition
  controls on nitrogen solubility in silicate melts}.{\BBCQ}
\newblock
\APACjournalVolNumPages{Geochimica et Cosmochimica Acta}{284}{}{120--133}.
\PrintBackRefs{\CurrentBib}

\bibitem [\protect \citeauthoryear {%
Bradley%
}{%
Bradley%
}{%
{\protect \APACyear {2008}}%
}]{%
Bradley+2008}
\APACinsertmetastar {%
Bradley+2008}%
\begin{APACrefauthors}%
Bradley, D\BPBI C.%
\end{APACrefauthors}%
\unskip\
\newblock
\APACrefYearMonthDay{2008}{}{}.
\newblock
{\BBOQ}\APACrefatitle {Passive margins through {E}arth history} {Passive
  margins through {E}arth history}.{\BBCQ}
\newblock
\APACjournalVolNumPages{Earth-Science Reviews}{91}{1-4}{1--26}.
\PrintBackRefs{\CurrentBib}

\bibitem [\protect \citeauthoryear {%
Busigny%
\ \BBA {} Bebout%
}{%
Busigny%
\ \BBA {} Bebout%
}{%
{\protect \APACyear {2013}}%
}]{%
Busigny+Bebout2013}
\APACinsertmetastar {%
Busigny+Bebout2013}%
\begin{APACrefauthors}%
Busigny, V.%
\BCBT {}\ \BBA {} Bebout, G\BPBI E.%
\end{APACrefauthors}%
\unskip\
\newblock
\APACrefYearMonthDay{2013}{}{}.
\newblock
{\BBOQ}\APACrefatitle {Nitrogen in the silicate {E}arth: Speciation and
  isotopic behavior during mineral--fluid interactions} {Nitrogen in the
  silicate {E}arth: Speciation and isotopic behavior during mineral--fluid
  interactions}.{\BBCQ}
\newblock
\APACjournalVolNumPages{Elements}{9}{5}{353--358}.
\PrintBackRefs{\CurrentBib}

\bibitem [\protect \citeauthoryear {%
Busigny%
, Cartigny%
\BCBL {}\ \BBA {} Philippot%
}{%
Busigny%
\ \protect \BOthers {.}}{%
{\protect \APACyear {2011}}%
}]{%
Busigny+2011}
\APACinsertmetastar {%
Busigny+2011}%
\begin{APACrefauthors}%
Busigny, V.%
, Cartigny, P.%
\BCBL {}\ \BBA {} Philippot, P.%
\end{APACrefauthors}%
\unskip\
\newblock
\APACrefYearMonthDay{2011}{}{}.
\newblock
{\BBOQ}\APACrefatitle {Nitrogen isotopes in ophiolitic metagabbros: A
  re-evaluation of modern nitrogen fluxes in subduction zones and implication
  for the early {E}arth atmosphere} {Nitrogen isotopes in ophiolitic
  metagabbros: A re-evaluation of modern nitrogen fluxes in subduction zones
  and implication for the early {E}arth atmosphere}.{\BBCQ}
\newblock
\APACjournalVolNumPages{Geochimica et Cosmochimica Acta}{75}{23}{7502--7521}.
\PrintBackRefs{\CurrentBib}

\bibitem [\protect \citeauthoryear {%
Cabral%
\ \protect \BOthers {.}}{%
Cabral%
\ \protect \BOthers {.}}{%
{\protect \APACyear {2013}}%
}]{%
Cabral+2013}
\APACinsertmetastar {%
Cabral+2013}%
\begin{APACrefauthors}%
Cabral, R\BPBI A.%
, Jackson, M\BPBI G.%
, Rose-Koga, E\BPBI F.%
, Koga, K\BPBI T.%
, Whitehouse, M\BPBI J.%
, Antonelli, M\BPBI A.%
\BDBL {}Hauri, E\BPBI H.%
\end{APACrefauthors}%
\unskip\
\newblock
\APACrefYearMonthDay{2013}{}{}.
\newblock
{\BBOQ}\APACrefatitle {Anomalous sulphur isotopes in plume lavas reveal deep
  mantle storage of {A}rchaean crust} {Anomalous sulphur isotopes in plume
  lavas reveal deep mantle storage of {A}rchaean crust}.{\BBCQ}
\newblock
\APACjournalVolNumPages{Nature}{496}{7446}{490--493}.
\PrintBackRefs{\CurrentBib}

\bibitem [\protect \citeauthoryear {%
Cartigny%
\ \BBA {} Marty%
}{%
Cartigny%
\ \BBA {} Marty%
}{%
{\protect \APACyear {2013}}%
}]{%
Cartigny+Marty2013}
\APACinsertmetastar {%
Cartigny+Marty2013}%
\begin{APACrefauthors}%
Cartigny, P.%
\BCBT {}\ \BBA {} Marty, B.%
\end{APACrefauthors}%
\unskip\
\newblock
\APACrefYearMonthDay{2013}{}{}.
\newblock
{\BBOQ}\APACrefatitle {Nitrogen isotopes and mantle geodynamics: The emergence
  of life and the atmosphere--crust--mantle connection} {Nitrogen isotopes and
  mantle geodynamics: The emergence of life and the atmosphere--crust--mantle
  connection}.{\BBCQ}
\newblock
\APACjournalVolNumPages{Elements}{9}{5}{359--366}.
\PrintBackRefs{\CurrentBib}

\bibitem [\protect \citeauthoryear {%
Catling%
\ \BBA {} Zahnle%
}{%
Catling%
\ \BBA {} Zahnle%
}{%
{\protect \APACyear {2020}}%
}]{%
Catling+Zahnle2020}
\APACinsertmetastar {%
Catling+Zahnle2020}%
\begin{APACrefauthors}%
Catling, D\BPBI C.%
\BCBT {}\ \BBA {} Zahnle, K\BPBI J.%
\end{APACrefauthors}%
\unskip\
\newblock
\APACrefYearMonthDay{2020}{}{}.
\newblock
{\BBOQ}\APACrefatitle {The {A}rchean atmosphere} {The {A}rchean
  atmosphere}.{\BBCQ}
\newblock
\APACjournalVolNumPages{Science Advances}{6}{9}{eaax1420}.
\PrintBackRefs{\CurrentBib}

\bibitem [\protect \citeauthoryear {%
Chou%
}{%
Chou%
}{%
{\protect \APACyear {1978}}%
}]{%
Chou1978}
\APACinsertmetastar {%
Chou1978}%
\begin{APACrefauthors}%
Chou, C\BHBI L.%
\end{APACrefauthors}%
\unskip\
\newblock
\APACrefYearMonthDay{1978}{}{}.
\newblock
{\BBOQ}\APACrefatitle {Fractionation of siderophile elements in the {E}arth's
  upper mantle} {Fractionation of siderophile elements in the {E}arth's upper
  mantle}.{\BBCQ}
\newblock
\BIn{} \APACrefbtitle {Lunar and Planetary Science Conference Proceedings}
  {Lunar and planetary science conference proceedings}\ (\BVOL~9).
\PrintBackRefs{\CurrentBib}

\bibitem [\protect \citeauthoryear {%
Condie%
, Pisarevsky%
, Korenaga%
\BCBL {}\ \BBA {} Gardoll%
}{%
Condie%
\ \protect \BOthers {.}}{%
{\protect \APACyear {2015}}%
}]{%
Condie+2015}
\APACinsertmetastar {%
Condie+2015}%
\begin{APACrefauthors}%
Condie, K.%
, Pisarevsky, S\BPBI A.%
, Korenaga, J.%
\BCBL {}\ \BBA {} Gardoll, S.%
\end{APACrefauthors}%
\unskip\
\newblock
\APACrefYearMonthDay{2015}{}{}.
\newblock
{\BBOQ}\APACrefatitle {Is the rate of supercontinent assembly changing with
  time?} {Is the rate of supercontinent assembly changing with time?}{\BBCQ}
\newblock
\APACjournalVolNumPages{Precambrian Research}{259}{}{278--289}.
\PrintBackRefs{\CurrentBib}

\bibitem [\protect \citeauthoryear {%
Dalou%
\ \protect \BOthers {.}}{%
Dalou%
\ \protect \BOthers {.}}{%
{\protect \APACyear {2019}}%
}]{%
Dalou+2019}
\APACinsertmetastar {%
Dalou+2019}%
\begin{APACrefauthors}%
Dalou, C.%
, F{\"u}ri, E.%
, Deligny, C.%
, Piani, L.%
, Caumon, M\BHBI C.%
, Laumonier, M.%
\BDBL {}Ed{\'e}n, M.%
\end{APACrefauthors}%
\unskip\
\newblock
\APACrefYearMonthDay{2019}{}{}.
\newblock
{\BBOQ}\APACrefatitle {Redox control on nitrogen isotope fractionation during
  planetary core formation} {Redox control on nitrogen isotope fractionation
  during planetary core formation}.{\BBCQ}
\newblock
\APACjournalVolNumPages{Proceedings of the National Academy of
  Sciences}{116}{29}{14485--14494}.
\PrintBackRefs{\CurrentBib}

\bibitem [\protect \citeauthoryear {%
Dalou%
, Hirschmann%
, von~der Handt%
, Mosenfelder%
\BCBL {}\ \BBA {} Armstrong%
}{%
Dalou%
\ \protect \BOthers {.}}{%
{\protect \APACyear {2017}}%
}]{%
Dalou+2017}
\APACinsertmetastar {%
Dalou+2017}%
\begin{APACrefauthors}%
Dalou, C.%
, Hirschmann, M\BPBI M.%
, von~der Handt, A.%
, Mosenfelder, J.%
\BCBL {}\ \BBA {} Armstrong, L\BPBI S.%
\end{APACrefauthors}%
\unskip\
\newblock
\APACrefYearMonthDay{2017}{}{}.
\newblock
{\BBOQ}\APACrefatitle {Nitrogen and carbon fractionation during core--mantle
  differentiation at shallow depth} {Nitrogen and carbon fractionation during
  core--mantle differentiation at shallow depth}.{\BBCQ}
\newblock
\APACjournalVolNumPages{Earth and Planetary Science Letters}{458}{}{141--151}.
\PrintBackRefs{\CurrentBib}

\bibitem [\protect \citeauthoryear {%
Dauphas%
}{%
Dauphas%
}{%
{\protect \APACyear {2017}}%
}]{%
Dauphas2017}
\APACinsertmetastar {%
Dauphas2017}%
\begin{APACrefauthors}%
Dauphas, N.%
\end{APACrefauthors}%
\unskip\
\newblock
\APACrefYearMonthDay{2017}{}{}.
\newblock
{\BBOQ}\APACrefatitle {The isotopic nature of the {E}arth’s accreting
  material through time} {The isotopic nature of the {E}arth’s accreting
  material through time}.{\BBCQ}
\newblock
\APACjournalVolNumPages{Nature}{541}{7638}{521--524}.
\PrintBackRefs{\CurrentBib}

\bibitem [\protect \citeauthoryear {%
Dottin~III%
, Labidi%
, Jackson%
, Woodhead%
\BCBL {}\ \BBA {} Farquhar%
}{%
Dottin~III%
\ \protect \BOthers {.}}{%
{\protect \APACyear {2020}}%
}]{%
Dottin+2020}
\APACinsertmetastar {%
Dottin+2020}%
\begin{APACrefauthors}%
Dottin~III, J.%
, Labidi, J.%
, Jackson, M.%
, Woodhead, J.%
\BCBL {}\ \BBA {} Farquhar, J.%
\end{APACrefauthors}%
\unskip\
\newblock
\APACrefYearMonthDay{2020}{}{}.
\newblock
{\BBOQ}\APACrefatitle {Isotopic evidence for multiple recycled sulfur
  reservoirs in the {M}angaia mantle plume} {Isotopic evidence for multiple
  recycled sulfur reservoirs in the {M}angaia mantle plume}.{\BBCQ}
\newblock
\APACjournalVolNumPages{Geochemistry, Geophysics,
  Geosystems}{21}{10}{e2020GC009081}.
\PrintBackRefs{\CurrentBib}

\bibitem [\protect \citeauthoryear {%
Dziewonski%
\ \BBA {} Anderson%
}{%
Dziewonski%
\ \BBA {} Anderson%
}{%
{\protect \APACyear {1981}}%
}]{%
Dziewonski+Anderson1981}
\APACinsertmetastar {%
Dziewonski+Anderson1981}%
\begin{APACrefauthors}%
Dziewonski, A\BPBI M.%
\BCBT {}\ \BBA {} Anderson, D\BPBI L.%
\end{APACrefauthors}%
\unskip\
\newblock
\APACrefYearMonthDay{1981}{}{}.
\newblock
{\BBOQ}\APACrefatitle {Preliminary reference {E}arth model} {Preliminary
  reference {E}arth model}.{\BBCQ}
\newblock
\APACjournalVolNumPages{Physics of the Rarth and Planetary
  Interiors}{25}{4}{297--356}.
\PrintBackRefs{\CurrentBib}

\bibitem [\protect \citeauthoryear {%
D’Incecco%
\ \protect \BOthers {.}}{%
D’Incecco%
\ \protect \BOthers {.}}{%
{\protect \APACyear {2021}}%
}]{%
DIncecco2021}
\APACinsertmetastar {%
DIncecco2021}%
\begin{APACrefauthors}%
D’Incecco, P.%
, Filiberto, J.%
, L{\'o}pez, I.%
, Gorinov, D.%
, Komatsu, G.%
, Martynov, A.%
\BCBL {}\ \BBA {} Pisarenko, P.%
\end{APACrefauthors}%
\unskip\
\newblock
\APACrefYearMonthDay{2021}{}{}.
\newblock
{\BBOQ}\APACrefatitle {The Young Volcanic Rises on {V}enus: a Key Scientific
  Target for Future Orbital and in-situ Measurements on {V}enus} {The young
  volcanic rises on {V}enus: a key scientific target for future orbital and
  in-situ measurements on {V}enus}.{\BBCQ}
\newblock
\APACjournalVolNumPages{Solar System Research}{55}{4}{315--323}.
\PrintBackRefs{\CurrentBib}

\bibitem [\protect \citeauthoryear {%
Fischer-G{\"o}dde%
\ \protect \BOthers {.}}{%
Fischer-G{\"o}dde%
\ \protect \BOthers {.}}{%
{\protect \APACyear {2020}}%
}]{%
Fischer+2020}
\APACinsertmetastar {%
Fischer+2020}%
\begin{APACrefauthors}%
Fischer-G{\"o}dde, M.%
, Elfers, B\BHBI M.%
, M{\"u}nker, C.%
, Szilas, K.%
, Maier, W\BPBI D.%
, Messling, N.%
\BDBL {}Smithies, H.%
\end{APACrefauthors}%
\unskip\
\newblock
\APACrefYearMonthDay{2020}{}{}.
\newblock
{\BBOQ}\APACrefatitle {Ruthenium isotope vestige of {E}arth’s pre-late-veneer
  mantle preserved in {A}rchaean rocks} {Ruthenium isotope vestige of
  {E}arth’s pre-late-veneer mantle preserved in {A}rchaean rocks}.{\BBCQ}
\newblock
\APACjournalVolNumPages{Nature}{579}{7798}{240--244}.
\PrintBackRefs{\CurrentBib}

\bibitem [\protect \citeauthoryear {%
Fischer-G{\"o}dde%
\ \BBA {} Kleine%
}{%
Fischer-G{\"o}dde%
\ \BBA {} Kleine%
}{%
{\protect \APACyear {2017}}%
}]{%
Fischer+Kleine2017}
\APACinsertmetastar {%
Fischer+Kleine2017}%
\begin{APACrefauthors}%
Fischer-G{\"o}dde, M.%
\BCBT {}\ \BBA {} Kleine, T.%
\end{APACrefauthors}%
\unskip\
\newblock
\APACrefYearMonthDay{2017}{}{}.
\newblock
{\BBOQ}\APACrefatitle {Ruthenium isotopic evidence for an inner {S}olar
  {S}ystem origin of the late veneer} {Ruthenium isotopic evidence for an inner
  {S}olar {S}ystem origin of the late veneer}.{\BBCQ}
\newblock
\APACjournalVolNumPages{Nature}{541}{7638}{525--527}.
\PrintBackRefs{\CurrentBib}

\bibitem [\protect \citeauthoryear {%
Gaillard%
\ \protect \BOthers {.}}{%
Gaillard%
\ \protect \BOthers {.}}{%
{\protect \APACyear {2022}}%
}]{%
Gaillard+2022}
\APACinsertmetastar {%
Gaillard+2022}%
\begin{APACrefauthors}%
Gaillard, F.%
, Bernadou, F.%
, Roskosz, M.%
, Bouhifd, M\BPBI A.%
, Marrocchi, Y.%
, Iacono-Marziano, G.%
\BDBL {}Rogerie, G.%
\end{APACrefauthors}%
\unskip\
\newblock
\APACrefYearMonthDay{2022}{}{}.
\newblock
{\BBOQ}\APACrefatitle {Redox controls during magma ocean degassing} {Redox
  controls during magma ocean degassing}.{\BBCQ}
\newblock
\APACjournalVolNumPages{Earth and Planetary Science Letters}{577}{}{117255}.
\PrintBackRefs{\CurrentBib}

\bibitem [\protect \citeauthoryear {%
Garvin%
, Buick%
, Anbar%
, Arnold%
\BCBL {}\ \BBA {} Kaufman%
}{%
Garvin%
\ \protect \BOthers {.}}{%
{\protect \APACyear {2009}}%
}]{%
Garvin+2009}
\APACinsertmetastar {%
Garvin+2009}%
\begin{APACrefauthors}%
Garvin, J.%
, Buick, R.%
, Anbar, A\BPBI D.%
, Arnold, G\BPBI L.%
\BCBL {}\ \BBA {} Kaufman, A\BPBI J.%
\end{APACrefauthors}%
\unskip\
\newblock
\APACrefYearMonthDay{2009}{}{}.
\newblock
{\BBOQ}\APACrefatitle {Isotopic evidence for an aerobic nitrogen cycle in the
  latest {A}rchean} {Isotopic evidence for an aerobic nitrogen cycle in the
  latest {A}rchean}.{\BBCQ}
\newblock
\APACjournalVolNumPages{Science}{323}{5917}{1045--1048}.
\PrintBackRefs{\CurrentBib}

\bibitem [\protect \citeauthoryear {%
Goldblatt%
\ \protect \BOthers {.}}{%
Goldblatt%
\ \protect \BOthers {.}}{%
{\protect \APACyear {2009}}%
}]{%
Goldblatt+2009}
\APACinsertmetastar {%
Goldblatt+2009}%
\begin{APACrefauthors}%
Goldblatt, C.%
, Claire, M\BPBI W.%
, Lenton, T\BPBI M.%
, Matthews, A\BPBI J.%
, Watson, A\BPBI J.%
\BCBL {}\ \BBA {} Zahnle, K\BPBI J.%
\end{APACrefauthors}%
\unskip\
\newblock
\APACrefYearMonthDay{2009}{}{}.
\newblock
{\BBOQ}\APACrefatitle {Nitrogen-enhanced greenhouse warming on early {E}arth}
  {Nitrogen-enhanced greenhouse warming on early {E}arth}.{\BBCQ}
\newblock
\APACjournalVolNumPages{Nature Geoscience}{2}{12}{891--896}.
\PrintBackRefs{\CurrentBib}

\bibitem [\protect \citeauthoryear {%
Grady%
\ \BBA {} Wright%
}{%
Grady%
\ \BBA {} Wright%
}{%
{\protect \APACyear {2003}}%
}]{%
Grady+Wright2003}
\APACinsertmetastar {%
Grady+Wright2003}%
\begin{APACrefauthors}%
Grady, M\BPBI M.%
\BCBT {}\ \BBA {} Wright, I\BPBI P.%
\end{APACrefauthors}%
\unskip\
\newblock
\APACrefYearMonthDay{2003}{}{}.
\newblock
{\BBOQ}\APACrefatitle {Elemental and isotopic abundances of carbon and nitrogen
  in meteorites} {Elemental and isotopic abundances of carbon and nitrogen in
  meteorites}.{\BBCQ}
\newblock
\APACjournalVolNumPages{Space Science Reviews}{106}{1}{231--248}.
\PrintBackRefs{\CurrentBib}

\bibitem [\protect \citeauthoryear {%
Grewal%
\ \protect \BOthers {.}}{%
Grewal%
\ \protect \BOthers {.}}{%
{\protect \APACyear {2019}}%
}]{%
Grewal+2019}
\APACinsertmetastar {%
Grewal+2019}%
\begin{APACrefauthors}%
Grewal, D\BPBI S.%
, Dasgupta, R.%
, Holmes, A\BPBI K.%
, Costin, G.%
, Li, Y.%
\BCBL {}\ \BBA {} Tsuno, K.%
\end{APACrefauthors}%
\unskip\
\newblock
\APACrefYearMonthDay{2019}{}{}.
\newblock
{\BBOQ}\APACrefatitle {The fate of nitrogen during core-mantle separation on
  {E}arth} {The fate of nitrogen during core-mantle separation on
  {E}arth}.{\BBCQ}
\newblock
\APACjournalVolNumPages{Geochimica et Cosmochimica Acta}{251}{}{87--115}.
\PrintBackRefs{\CurrentBib}

\bibitem [\protect \citeauthoryear {%
Grewal%
, Dasgupta%
, Hough%
\BCBL {}\ \BBA {} Farnell%
}{%
Grewal%
\ \protect \BOthers {.}}{%
{\protect \APACyear {2021}}%
}]{%
Grewal+2021}
\APACinsertmetastar {%
Grewal+2021}%
\begin{APACrefauthors}%
Grewal, D\BPBI S.%
, Dasgupta, R.%
, Hough, T.%
\BCBL {}\ \BBA {} Farnell, A.%
\end{APACrefauthors}%
\unskip\
\newblock
\APACrefYearMonthDay{2021}{}{}.
\newblock
{\BBOQ}\APACrefatitle {Rates of protoplanetary accretion and differentiation
  set nitrogen budget of rocky planets} {Rates of protoplanetary accretion and
  differentiation set nitrogen budget of rocky planets}.{\BBCQ}
\newblock
\APACjournalVolNumPages{Nature Geoscience}{14}{6}{369--376}.
\PrintBackRefs{\CurrentBib}

\bibitem [\protect \citeauthoryear {%
Grott%
\ \protect \BOthers {.}}{%
Grott%
\ \protect \BOthers {.}}{%
{\protect \APACyear {2013}}%
}]{%
Grott+2013}
\APACinsertmetastar {%
Grott+2013}%
\begin{APACrefauthors}%
Grott, M.%
, Baratoux, D.%
, Hauber, E.%
, Sautter, V.%
, Mustard, J.%
, Gasnault, O.%
\BDBL {}others%
\end{APACrefauthors}%
\unskip\
\newblock
\APACrefYearMonthDay{2013}{}{}.
\newblock
{\BBOQ}\APACrefatitle {Long-term evolution of the {M}artian crust-mantle
  system} {Long-term evolution of the {M}artian crust-mantle system}.{\BBCQ}
\newblock
\APACjournalVolNumPages{Space Science Reviews}{174}{1-4}{49--111}.
\PrintBackRefs{\CurrentBib}

\bibitem [\protect \citeauthoryear {%
Guo%
\ \BBA {} Korenaga%
}{%
Guo%
\ \BBA {} Korenaga%
}{%
{\protect \APACyear {2020}}%
}]{%
Guo+Korenaga2020}
\APACinsertmetastar {%
Guo+Korenaga2020}%
\begin{APACrefauthors}%
Guo, M.%
\BCBT {}\ \BBA {} Korenaga, J.%
\end{APACrefauthors}%
\unskip\
\newblock
\APACrefYearMonthDay{2020}{}{}.
\newblock
{\BBOQ}\APACrefatitle {Argon constraints on the early growth of felsic
  continental crust} {Argon constraints on the early growth of felsic
  continental crust}.{\BBCQ}
\newblock
\APACjournalVolNumPages{Science advances}{6}{21}{eaaz6234}.
\PrintBackRefs{\CurrentBib}

\bibitem [\protect \citeauthoryear {%
Haendel%
, M{\"u}hle%
, Nitzsche%
, Stiehl%
\BCBL {}\ \BBA {} Wand%
}{%
Haendel%
\ \protect \BOthers {.}}{%
{\protect \APACyear {1986}}%
}]{%
Haendel+1986}
\APACinsertmetastar {%
Haendel+1986}%
\begin{APACrefauthors}%
Haendel, D.%
, M{\"u}hle, K.%
, Nitzsche, H\BHBI M.%
, Stiehl, G.%
\BCBL {}\ \BBA {} Wand, U.%
\end{APACrefauthors}%
\unskip\
\newblock
\APACrefYearMonthDay{1986}{}{}.
\newblock
{\BBOQ}\APACrefatitle {Isotopic variations of the fixed nitrogen in metamorphic
  rocks} {Isotopic variations of the fixed nitrogen in metamorphic
  rocks}.{\BBCQ}
\newblock
\APACjournalVolNumPages{Geochimica et cosmochimica Acta}{50}{5}{749--758}.
\PrintBackRefs{\CurrentBib}

\bibitem [\protect \citeauthoryear {%
Halama%
, Bebout%
, John%
\BCBL {}\ \BBA {} Scambelluri%
}{%
Halama%
\ \protect \BOthers {.}}{%
{\protect \APACyear {2014}}%
}]{%
Halama+2014}
\APACinsertmetastar {%
Halama+2014}%
\begin{APACrefauthors}%
Halama, R.%
, Bebout, G\BPBI E.%
, John, T.%
\BCBL {}\ \BBA {} Scambelluri, M.%
\end{APACrefauthors}%
\unskip\
\newblock
\APACrefYearMonthDay{2014}{}{}.
\newblock
{\BBOQ}\APACrefatitle {Nitrogen recycling in subducted mantle rocks and
  implications for the global nitrogen cycle} {Nitrogen recycling in subducted
  mantle rocks and implications for the global nitrogen cycle}.{\BBCQ}
\newblock
\APACjournalVolNumPages{International Journal of Earth
  Sciences}{103}{7}{2081--2099}.
\PrintBackRefs{\CurrentBib}

\bibitem [\protect \citeauthoryear {%
Halliday%
}{%
Halliday%
}{%
{\protect \APACyear {2013}}%
}]{%
Halliday2013}
\APACinsertmetastar {%
Halliday2013}%
\begin{APACrefauthors}%
Halliday, A\BPBI N.%
\end{APACrefauthors}%
\unskip\
\newblock
\APACrefYearMonthDay{2013}{}{}.
\newblock
{\BBOQ}\APACrefatitle {The origins of volatiles in the terrestrial planets}
  {The origins of volatiles in the terrestrial planets}.{\BBCQ}
\newblock
\APACjournalVolNumPages{Geochimica et Cosmochimica Acta}{105}{}{146--171}.
\PrintBackRefs{\CurrentBib}

\bibitem [\protect \citeauthoryear {%
Hamano%
, Abe%
\BCBL {}\ \BBA {} Genda%
}{%
Hamano%
\ \protect \BOthers {.}}{%
{\protect \APACyear {2013}}%
}]{%
Hamano+2013}
\APACinsertmetastar {%
Hamano+2013}%
\begin{APACrefauthors}%
Hamano, K.%
, Abe, Y.%
\BCBL {}\ \BBA {} Genda, H.%
\end{APACrefauthors}%
\unskip\
\newblock
\APACrefYearMonthDay{2013}{}{}.
\newblock
{\BBOQ}\APACrefatitle {Emergence of two types of terrestrial planet on
  solidification of magma ocean} {Emergence of two types of terrestrial planet
  on solidification of magma ocean}.{\BBCQ}
\newblock
\APACjournalVolNumPages{Nature}{497}{7451}{607--610}.
\PrintBackRefs{\CurrentBib}

\bibitem [\protect \citeauthoryear {%
Heber%
, Brooker%
, Kelley%
\BCBL {}\ \BBA {} Wood%
}{%
Heber%
\ \protect \BOthers {.}}{%
{\protect \APACyear {2007}}%
}]{%
Heber+2007}
\APACinsertmetastar {%
Heber+2007}%
\begin{APACrefauthors}%
Heber, V\BPBI S.%
, Brooker, R\BPBI A.%
, Kelley, S\BPBI P.%
\BCBL {}\ \BBA {} Wood, B\BPBI J.%
\end{APACrefauthors}%
\unskip\
\newblock
\APACrefYearMonthDay{2007}{}{}.
\newblock
{\BBOQ}\APACrefatitle {Crystal--melt partitioning of noble gases (helium, neon,
  argon, krypton, and xenon) for olivine and clinopyroxene} {Crystal--melt
  partitioning of noble gases (helium, neon, argon, krypton, and xenon) for
  olivine and clinopyroxene}.{\BBCQ}
\newblock
\APACjournalVolNumPages{Geochimica et Cosmochimica Acta}{71}{4}{1041--1061}.
\PrintBackRefs{\CurrentBib}

\bibitem [\protect \citeauthoryear {%
Hernlund%
, McNamara%
\BCBL {}\ \BBA {} Schubert%
}{%
Hernlund%
\ \protect \BOthers {.}}{%
{\protect \APACyear {2015}}%
}]{%
Hernlund+2015}
\APACinsertmetastar {%
Hernlund+2015}%
\begin{APACrefauthors}%
Hernlund, J.%
, McNamara, A.%
\BCBL {}\ \BBA {} Schubert, G.%
\end{APACrefauthors}%
\unskip\
\newblock
\APACrefYearMonthDay{2015}{}{}.
\newblock
{\BBOQ}\APACrefatitle {7.11: {T}he core--mantle boundary region} {7.11: {T}he
  core--mantle boundary region}.{\BBCQ}
\newblock
\APACjournalVolNumPages{Treatise on geophysics}{2}{7}{461--519}.
\PrintBackRefs{\CurrentBib}

\bibitem [\protect \citeauthoryear {%
Hier-Majumder%
\ \BBA {} Hirschmann%
}{%
Hier-Majumder%
\ \BBA {} Hirschmann%
}{%
{\protect \APACyear {2017}}%
}]{%
Hier-Majumder+Hirschmann2017}
\APACinsertmetastar {%
Hier-Majumder+Hirschmann2017}%
\begin{APACrefauthors}%
Hier-Majumder, S.%
\BCBT {}\ \BBA {} Hirschmann, M\BPBI M.%
\end{APACrefauthors}%
\unskip\
\newblock
\APACrefYearMonthDay{2017}{}{}.
\newblock
{\BBOQ}\APACrefatitle {The origin of volatiles in the {E}arth's mantle} {The
  origin of volatiles in the {E}arth's mantle}.{\BBCQ}
\newblock
\APACjournalVolNumPages{Geochemistry, Geophysics,
  Geosystems}{18}{8}{3078--3092}.
\PrintBackRefs{\CurrentBib}

\bibitem [\protect \citeauthoryear {%
Hirschmann%
}{%
Hirschmann%
}{%
{\protect \APACyear {2016}}%
}]{%
hirschmann2016}
\APACinsertmetastar {%
hirschmann2016}%
\begin{APACrefauthors}%
Hirschmann, M\BPBI M.%
\end{APACrefauthors}%
\unskip\
\newblock
\APACrefYearMonthDay{2016}{}{}.
\newblock
{\BBOQ}\APACrefatitle {Constraints on the early delivery and fractionation of
  Earth’s major volatiles from {C/H}, {C/N}, and {C/S} ratios} {Constraints
  on the early delivery and fractionation of earth’s major volatiles from
  {C/H}, {C/N}, and {C/S} ratios}.{\BBCQ}
\newblock
\APACjournalVolNumPages{American Mineralogist}{101}{3}{540--553}.
\PrintBackRefs{\CurrentBib}

\bibitem [\protect \citeauthoryear {%
Houlton%
, Morford%
\BCBL {}\ \BBA {} Dahlgren%
}{%
Houlton%
\ \protect \BOthers {.}}{%
{\protect \APACyear {2018}}%
}]{%
Houlton+2018}
\APACinsertmetastar {%
Houlton+2018}%
\begin{APACrefauthors}%
Houlton, B.%
, Morford, S.%
\BCBL {}\ \BBA {} Dahlgren, R.%
\end{APACrefauthors}%
\unskip\
\newblock
\APACrefYearMonthDay{2018}{}{}.
\newblock
{\BBOQ}\APACrefatitle {Convergent evidence for widespread rock nitrogen sources
  in {E}arth's surface environment} {Convergent evidence for widespread rock
  nitrogen sources in {E}arth's surface environment}.{\BBCQ}
\newblock
\APACjournalVolNumPages{Science}{360}{6384}{58--62}.
\PrintBackRefs{\CurrentBib}

\bibitem [\protect \citeauthoryear {%
Hu%
\ \BBA {} Diaz%
}{%
Hu%
\ \BBA {} Diaz%
}{%
{\protect \APACyear {2019}}%
}]{%
Hu+Diaz2019}
\APACinsertmetastar {%
Hu+Diaz2019}%
\begin{APACrefauthors}%
Hu, R.%
\BCBT {}\ \BBA {} Diaz, H\BPBI D.%
\end{APACrefauthors}%
\unskip\
\newblock
\APACrefYearMonthDay{2019}{}{}.
\newblock
{\BBOQ}\APACrefatitle {Stability of Nitrogen in Planetary Atmospheres in
  Contact with Liquid Water} {Stability of nitrogen in planetary atmospheres in
  contact with liquid water}.{\BBCQ}
\newblock
\APACjournalVolNumPages{The Astrophysical Journal}{886}{2}{126}.
\PrintBackRefs{\CurrentBib}

\bibitem [\protect \citeauthoryear {%
Iacono-Marziano%
, Paonita%
, Rizzo%
, Scaillet%
\BCBL {}\ \BBA {} Gaillard%
}{%
Iacono-Marziano%
\ \protect \BOthers {.}}{%
{\protect \APACyear {2010}}%
}]{%
Iacono+2010}
\APACinsertmetastar {%
Iacono+2010}%
\begin{APACrefauthors}%
Iacono-Marziano, G.%
, Paonita, A.%
, Rizzo, A.%
, Scaillet, B.%
\BCBL {}\ \BBA {} Gaillard, F.%
\end{APACrefauthors}%
\unskip\
\newblock
\APACrefYearMonthDay{2010}{}{}.
\newblock
{\BBOQ}\APACrefatitle {Noble gas solubilities in silicate melts: {N}ew
  experimental results and a comprehensive model of the effects of liquid
  composition, temperature and pressure} {Noble gas solubilities in silicate
  melts: {N}ew experimental results and a comprehensive model of the effects of
  liquid composition, temperature and pressure}.{\BBCQ}
\newblock
\APACjournalVolNumPages{Chemical Geology}{279}{3-4}{145--157}.
\PrintBackRefs{\CurrentBib}

\bibitem [\protect \citeauthoryear {%
Johnson%
\ \BBA {} Goldblatt%
}{%
Johnson%
\ \BBA {} Goldblatt%
}{%
{\protect \APACyear {2015}}%
}]{%
Johnson+Goldblatt2015}
\APACinsertmetastar {%
Johnson+Goldblatt2015}%
\begin{APACrefauthors}%
Johnson, B\BPBI W.%
\BCBT {}\ \BBA {} Goldblatt, C.%
\end{APACrefauthors}%
\unskip\
\newblock
\APACrefYearMonthDay{2015}{}{}.
\newblock
{\BBOQ}\APACrefatitle {The nitrogen budget of {E}arth} {The nitrogen budget of
  {E}arth}.{\BBCQ}
\newblock
\APACjournalVolNumPages{Earth-Science Reviews}{148}{}{150--173}.
\PrintBackRefs{\CurrentBib}

\bibitem [\protect \citeauthoryear {%
Johnson%
\ \BBA {} Goldblatt%
}{%
Johnson%
\ \BBA {} Goldblatt%
}{%
{\protect \APACyear {2017}}%
}]{%
Johnson+Goldblatt2017}
\APACinsertmetastar {%
Johnson+Goldblatt2017}%
\begin{APACrefauthors}%
Johnson, B\BPBI W.%
\BCBT {}\ \BBA {} Goldblatt, C.%
\end{APACrefauthors}%
\unskip\
\newblock
\APACrefYearMonthDay{2017}{}{}.
\newblock
{\BBOQ}\APACrefatitle {A secular increase in continental crust nitrogen during
  the {P}recambrian} {A secular increase in continental crust nitrogen during
  the {P}recambrian}.{\BBCQ}
\newblock
\APACjournalVolNumPages{Geochemical Perspectives Letters}{4}{}{24--28}.
\PrintBackRefs{\CurrentBib}

\bibitem [\protect \citeauthoryear {%
Johnson%
\ \BBA {} Goldblatt%
}{%
Johnson%
\ \BBA {} Goldblatt%
}{%
{\protect \APACyear {2018}}%
}]{%
Johnson+Goldblatt2018}
\APACinsertmetastar {%
Johnson+Goldblatt2018}%
\begin{APACrefauthors}%
Johnson, B\BPBI W.%
\BCBT {}\ \BBA {} Goldblatt, C.%
\end{APACrefauthors}%
\unskip\
\newblock
\APACrefYearMonthDay{2018}{}{}.
\newblock
{\BBOQ}\APACrefatitle {Earth{N}: {A} new {E}arth system nitrogen model}
  {Earth{N}: {A} new {E}arth system nitrogen model}.{\BBCQ}
\newblock
\APACjournalVolNumPages{Geochemistry, Geophysics,
  Geosystems}{19}{8}{2516--2542}.
\PrintBackRefs{\CurrentBib}

\bibitem [\protect \citeauthoryear {%
Keller%
\ \BBA {} Schoene%
}{%
Keller%
\ \BBA {} Schoene%
}{%
{\protect \APACyear {2018}}%
}]{%
Keller+Schoene2018}
\APACinsertmetastar {%
Keller+Schoene2018}%
\begin{APACrefauthors}%
Keller, B.%
\BCBT {}\ \BBA {} Schoene, B.%
\end{APACrefauthors}%
\unskip\
\newblock
\APACrefYearMonthDay{2018}{}{}.
\newblock
{\BBOQ}\APACrefatitle {Plate tectonics and continental basaltic geochemistry
  throughout {E}arth history} {Plate tectonics and continental basaltic
  geochemistry throughout {E}arth history}.{\BBCQ}
\newblock
\APACjournalVolNumPages{Earth and Planetary Science Letters}{481}{}{290--304}.
\PrintBackRefs{\CurrentBib}

\bibitem [\protect \citeauthoryear {%
Korenaga%
}{%
Korenaga%
}{%
{\protect \APACyear {2003}}%
}]{%
Korenaga2003}
\APACinsertmetastar {%
Korenaga2003}%
\begin{APACrefauthors}%
Korenaga, J.%
\end{APACrefauthors}%
\unskip\
\newblock
\APACrefYearMonthDay{2003}{}{}.
\newblock
{\BBOQ}\APACrefatitle {Energetics of mantle convection and the fate of fossil
  heat} {Energetics of mantle convection and the fate of fossil heat}.{\BBCQ}
\newblock
\APACjournalVolNumPages{Geophysical Research Letters}{30}{8}{}.
\PrintBackRefs{\CurrentBib}

\bibitem [\protect \citeauthoryear {%
Korenaga%
}{%
Korenaga%
}{%
{\protect \APACyear {2013}}%
}]{%
Korenaga+2013}
\APACinsertmetastar {%
Korenaga+2013}%
\begin{APACrefauthors}%
Korenaga, J.%
\end{APACrefauthors}%
\unskip\
\newblock
\APACrefYearMonthDay{2013}{}{}.
\newblock
{\BBOQ}\APACrefatitle {Initiation and evolution of plate tectonics on {E}arth:
  theories and observations} {Initiation and evolution of plate tectonics on
  {E}arth: theories and observations}.{\BBCQ}
\newblock
\APACjournalVolNumPages{Annual review of earth and planetary
  sciences}{41}{}{117--151}.
\PrintBackRefs{\CurrentBib}

\bibitem [\protect \citeauthoryear {%
Korenaga%
}{%
Korenaga%
}{%
{\protect \APACyear {2018}}%
}]{%
Korenaga2018}
\APACinsertmetastar {%
Korenaga2018}%
\begin{APACrefauthors}%
Korenaga, J.%
\end{APACrefauthors}%
\unskip\
\newblock
\APACrefYearMonthDay{2018}{}{}.
\newblock
{\BBOQ}\APACrefatitle {Crustal evolution and mantle dynamics through {E}arth
  history} {Crustal evolution and mantle dynamics through {E}arth
  history}.{\BBCQ}
\newblock
\APACjournalVolNumPages{Philosophical Transactions of the Royal Society A:
  Mathematical, Physical and Engineering Sciences}{376}{2132}{20170408}.
\PrintBackRefs{\CurrentBib}

\bibitem [\protect \citeauthoryear {%
Krissansen-Totton%
, Fortney%
\BCBL {}\ \BBA {} Nimmo%
}{%
Krissansen-Totton%
\ \protect \BOthers {.}}{%
{\protect \APACyear {2021}}%
}]{%
Krissansen+2021}
\APACinsertmetastar {%
Krissansen+2021}%
\begin{APACrefauthors}%
Krissansen-Totton, J.%
, Fortney, J\BPBI J.%
\BCBL {}\ \BBA {} Nimmo, F.%
\end{APACrefauthors}%
\unskip\
\newblock
\APACrefYearMonthDay{2021}{}{}.
\newblock
{\BBOQ}\APACrefatitle {Was {V}enus Ever Habitable? {C}onstraints from a Coupled
  Interior--Atmosphere--Redox Evolution Model} {Was {V}enus ever habitable?
  {C}onstraints from a coupled interior--atmosphere--redox evolution
  model}.{\BBCQ}
\newblock
\APACjournalVolNumPages{The Planetary Science Journal}{2}{5}{216}.
\PrintBackRefs{\CurrentBib}

\bibitem [\protect \citeauthoryear {%
Kuga%
\ \protect \BOthers {.}}{%
Kuga%
\ \protect \BOthers {.}}{%
{\protect \APACyear {2014}}%
}]{%
Kuga+2014}
\APACinsertmetastar {%
Kuga+2014}%
\begin{APACrefauthors}%
Kuga, M.%
, Carrasco, N.%
, Marty, B.%
, Marrocchi, Y.%
, Bernard, S.%
, Rigaudier, T.%
\BDBL {}Tissandier, L.%
\end{APACrefauthors}%
\unskip\
\newblock
\APACrefYearMonthDay{2014}{}{}.
\newblock
{\BBOQ}\APACrefatitle {Nitrogen isotopic fractionation during abiotic synthesis
  of organic solid particles} {Nitrogen isotopic fractionation during abiotic
  synthesis of organic solid particles}.{\BBCQ}
\newblock
\APACjournalVolNumPages{Earth and Planetary Science Letters}{393}{}{2--13}.
\PrintBackRefs{\CurrentBib}

\bibitem [\protect \citeauthoryear {%
Kurokawa%
, Foriel%
, Laneuville%
, Houser%
\BCBL {}\ \BBA {} Usui%
}{%
Kurokawa%
, Foriel%
\BCBL {}\ \protect \BOthers {.}}{%
{\protect \APACyear {2018}}%
}]{%
Kurokawa+2018EPSL}
\APACinsertmetastar {%
Kurokawa+2018EPSL}%
\begin{APACrefauthors}%
Kurokawa, H.%
, Foriel, J.%
, Laneuville, M.%
, Houser, C.%
\BCBL {}\ \BBA {} Usui, T.%
\end{APACrefauthors}%
\unskip\
\newblock
\APACrefYearMonthDay{2018}{}{}.
\newblock
{\BBOQ}\APACrefatitle {Subduction and atmospheric escape of {E}arth's seawater
  constrained by hydrogen isotopes} {Subduction and atmospheric escape of
  {E}arth's seawater constrained by hydrogen isotopes}.{\BBCQ}
\newblock
\APACjournalVolNumPages{Earth and Planetary Science Letters}{497}{}{149--160}.
\PrintBackRefs{\CurrentBib}

\bibitem [\protect \citeauthoryear {%
Kurokawa%
, Kurosawa%
\BCBL {}\ \BBA {} Usui%
}{%
Kurokawa%
, Kurosawa%
\BCBL {}\ \BBA {} Usui%
}{%
{\protect \APACyear {2018}}%
}]{%
Kurokawa+2018}
\APACinsertmetastar {%
Kurokawa+2018}%
\begin{APACrefauthors}%
Kurokawa, H.%
, Kurosawa, K.%
\BCBL {}\ \BBA {} Usui, T.%
\end{APACrefauthors}%
\unskip\
\newblock
\APACrefYearMonthDay{2018}{}{}.
\newblock
{\BBOQ}\APACrefatitle {A lower limit of atmospheric pressure on early {M}ars
  inferred from nitrogen and argon isotopic compositions} {A lower limit of
  atmospheric pressure on early {M}ars inferred from nitrogen and argon
  isotopic compositions}.{\BBCQ}
\newblock
\APACjournalVolNumPages{Icarus}{299}{}{443--459}.
\PrintBackRefs{\CurrentBib}

\bibitem [\protect \citeauthoryear {%
Kurokawa%
\ \protect \BOthers {.}}{%
Kurokawa%
\ \protect \BOthers {.}}{%
{\protect \APACyear {2021}}%
}]{%
Kurokawa+2021}
\APACinsertmetastar {%
Kurokawa+2021}%
\begin{APACrefauthors}%
Kurokawa, H.%
, Miura, Y\BPBI N.%
, Sugita, S.%
, Cho, Y.%
, Leblanc, F.%
, Terada, N.%
\BCBL {}\ \BBA {} Nakagawa, H.%
\end{APACrefauthors}%
\unskip\
\newblock
\APACrefYearMonthDay{2021}{}{}.
\newblock
{\BBOQ}\APACrefatitle {Mars' atmospheric neon suggests volatile-rich primitive
  mantle} {Mars' atmospheric neon suggests volatile-rich primitive
  mantle}.{\BBCQ}
\newblock
\APACjournalVolNumPages{Icarus}{370}{}{114685}.
\PrintBackRefs{\CurrentBib}

\bibitem [\protect \citeauthoryear {%
Labidi%
\ \protect \BOthers {.}}{%
Labidi%
\ \protect \BOthers {.}}{%
{\protect \APACyear {2020}}%
}]{%
Labidi+2020}
\APACinsertmetastar {%
Labidi+2020}%
\begin{APACrefauthors}%
Labidi, J.%
, Barry, P.%
, Bekaert, D.%
, Broadley, M.%
, Marty, B.%
, Giunta, T.%
\BDBL {}others%
\end{APACrefauthors}%
\unskip\
\newblock
\APACrefYearMonthDay{2020}{}{}.
\newblock
{\BBOQ}\APACrefatitle {Hydrothermal $^{15}${N}$^{15}${N} abundances constrain
  the origins of mantle nitrogen} {Hydrothermal $^{15}${N}$^{15}${N} abundances
  constrain the origins of mantle nitrogen}.{\BBCQ}
\newblock
\APACjournalVolNumPages{Nature}{580}{7803}{367--371}.
\PrintBackRefs{\CurrentBib}

\bibitem [\protect \citeauthoryear {%
Labrosse%
, Hernlund%
\BCBL {}\ \BBA {} Coltice%
}{%
Labrosse%
\ \protect \BOthers {.}}{%
{\protect \APACyear {2007}}%
}]{%
Labrosse+2007}
\APACinsertmetastar {%
Labrosse+2007}%
\begin{APACrefauthors}%
Labrosse, S.%
, Hernlund, J.%
\BCBL {}\ \BBA {} Coltice, N.%
\end{APACrefauthors}%
\unskip\
\newblock
\APACrefYearMonthDay{2007}{}{}.
\newblock
{\BBOQ}\APACrefatitle {A crystallizing dense magma ocean at the base of the
  {E}arth's mantle} {A crystallizing dense magma ocean at the base of the
  {E}arth's mantle}.{\BBCQ}
\newblock
\APACjournalVolNumPages{Nature}{450}{7171}{866--869}.
\PrintBackRefs{\CurrentBib}

\bibitem [\protect \citeauthoryear {%
Lammer%
, Leitzinger%
\BCBL {}\ \protect \BOthers {.}}{%
Lammer%
, Leitzinger%
\BCBL {}\ \protect \BOthers {.}}{%
{\protect \APACyear {2020}}%
}]{%
Lammer+2020}
\APACinsertmetastar {%
Lammer+2020}%
\begin{APACrefauthors}%
Lammer, H.%
, Leitzinger, M.%
, Scherf, M.%
, Odert, P.%
, Burger, C.%
, Kubyshkina, D.%
\BDBL {}others%
\end{APACrefauthors}%
\unskip\
\newblock
\APACrefYearMonthDay{2020}{}{}.
\newblock
{\BBOQ}\APACrefatitle {Constraining the early evolution of {V}enus and {E}arth
  through atmospheric {A}r, {N}e isotope and bulk {K}/{U} ratios} {Constraining
  the early evolution of {V}enus and {E}arth through atmospheric {A}r, {N}e
  isotope and bulk {K}/{U} ratios}.{\BBCQ}
\newblock
\APACjournalVolNumPages{Icarus}{339}{}{113551}.
\PrintBackRefs{\CurrentBib}

\bibitem [\protect \citeauthoryear {%
Lammer%
, Scherf%
\BCBL {}\ \protect \BOthers {.}}{%
Lammer%
, Scherf%
\BCBL {}\ \protect \BOthers {.}}{%
{\protect \APACyear {2020}}%
}]{%
Lammer+2020SSR}
\APACinsertmetastar {%
Lammer+2020SSR}%
\begin{APACrefauthors}%
Lammer, H.%
, Scherf, M.%
, Kurokawa, H.%
, Ueno, Y.%
, Burger, C.%
, Maindl, T.%
\BDBL {}others%
\end{APACrefauthors}%
\unskip\
\newblock
\APACrefYearMonthDay{2020}{}{}.
\newblock
{\BBOQ}\APACrefatitle {Loss and Fractionation of Noble Gas Isotopes and
  Moderately Volatile Elements from Planetary Embryos and Early {V}enus,
  {E}arth and {M}ars} {Loss and fractionation of noble gas isotopes and
  moderately volatile elements from planetary embryos and early {V}enus,
  {E}arth and {M}ars}.{\BBCQ}
\newblock
\APACjournalVolNumPages{Space Science Reviews}{216}{4}{1--50}.
\PrintBackRefs{\CurrentBib}

\bibitem [\protect \citeauthoryear {%
Laneuville%
, Kameya%
\BCBL {}\ \BBA {} Cleaves%
}{%
Laneuville%
\ \protect \BOthers {.}}{%
{\protect \APACyear {2018}}%
}]{%
Laneuville+2018}
\APACinsertmetastar {%
Laneuville+2018}%
\begin{APACrefauthors}%
Laneuville, M.%
, Kameya, M.%
\BCBL {}\ \BBA {} Cleaves, H\BPBI J.%
\end{APACrefauthors}%
\unskip\
\newblock
\APACrefYearMonthDay{2018}{}{}.
\newblock
{\BBOQ}\APACrefatitle {Earth without life: {a} systems model of a global
  abiotic nitrogen cycle} {Earth without life: {a} systems model of a global
  abiotic nitrogen cycle}.{\BBCQ}
\newblock
\APACjournalVolNumPages{Astrobiology}{18}{7}{897--914}.
\PrintBackRefs{\CurrentBib}

\bibitem [\protect \citeauthoryear {%
Li%
, Bebout%
\BCBL {}\ \BBA {} Idleman%
}{%
Li%
\ \protect \BOthers {.}}{%
{\protect \APACyear {2007}}%
}]{%
Li+2007}
\APACinsertmetastar {%
Li+2007}%
\begin{APACrefauthors}%
Li, L.%
, Bebout, G\BPBI E.%
\BCBL {}\ \BBA {} Idleman, B\BPBI D.%
\end{APACrefauthors}%
\unskip\
\newblock
\APACrefYearMonthDay{2007}{}{}.
\newblock
{\BBOQ}\APACrefatitle {Nitrogen concentration and $\delta^{15}${N} of altered
  oceanic crust obtained on {ODP} {L}egs 129 and 185: Insights into
  alteration-related nitrogen enrichment and the nitrogen subduction budget}
  {Nitrogen concentration and $\delta^{15}${N} of altered oceanic crust
  obtained on {ODP} {L}egs 129 and 185: Insights into alteration-related
  nitrogen enrichment and the nitrogen subduction budget}.{\BBCQ}
\newblock
\APACjournalVolNumPages{Geochimica et Cosmochimica Acta}{71}{9}{2344--2360}.
\PrintBackRefs{\CurrentBib}

\bibitem [\protect \citeauthoryear {%
Libourel%
, Marty%
\BCBL {}\ \BBA {} Humbert%
}{%
Libourel%
\ \protect \BOthers {.}}{%
{\protect \APACyear {2003}}%
}]{%
Libourel+2003}
\APACinsertmetastar {%
Libourel+2003}%
\begin{APACrefauthors}%
Libourel, G.%
, Marty, B.%
\BCBL {}\ \BBA {} Humbert, F.%
\end{APACrefauthors}%
\unskip\
\newblock
\APACrefYearMonthDay{2003}{}{}.
\newblock
{\BBOQ}\APACrefatitle {Nitrogen solubility in basaltic melt. {P}art {I}. Effect
  of oxygen fugacity} {Nitrogen solubility in basaltic melt. {P}art {I}. effect
  of oxygen fugacity}.{\BBCQ}
\newblock
\APACjournalVolNumPages{Geochimica et Cosmochimica Acta}{67}{21}{4123--4135}.
\PrintBackRefs{\CurrentBib}

\bibitem [\protect \citeauthoryear {%
Lichtenegger%
\ \protect \BOthers {.}}{%
Lichtenegger%
\ \protect \BOthers {.}}{%
{\protect \APACyear {2010}}%
}]{%
Lichtenegger+2010}
\APACinsertmetastar {%
Lichtenegger+2010}%
\begin{APACrefauthors}%
Lichtenegger, H.%
, Lammer, H.%
, Grie{\ss}meier, J\BHBI M.%
, Kulikov, Y\BPBI N.%
, von Paris, P.%
, Hausleitner, W.%
\BDBL {}Rauer, H.%
\end{APACrefauthors}%
\unskip\
\newblock
\APACrefYearMonthDay{2010}{}{}.
\newblock
{\BBOQ}\APACrefatitle {Aeronomical evidence for higher {CO}$_2$ levels during
  {E}arth’s {H}adean epoch} {Aeronomical evidence for higher {CO}$_2$ levels
  during {E}arth’s {H}adean epoch}.{\BBCQ}
\newblock
\APACjournalVolNumPages{Icarus}{210}{1}{1--7}.
\PrintBackRefs{\CurrentBib}

\bibitem [\protect \citeauthoryear {%
Lim%
, Bonati%
\BCBL {}\ \BBA {} Hernlund%
}{%
Lim%
\ \protect \BOthers {.}}{%
{\protect \APACyear {2021}}%
}]{%
Lim+2021}
\APACinsertmetastar {%
Lim+2021}%
\begin{APACrefauthors}%
Lim, K\BPBI W.%
, Bonati, I.%
\BCBL {}\ \BBA {} Hernlund, J\BPBI W.%
\end{APACrefauthors}%
\unskip\
\newblock
\APACrefYearMonthDay{2021}{}{}.
\newblock
{\BBOQ}\APACrefatitle {A Hybrid Mechanism for Enhanced Core-Mantle Boundary
  Chemical Interaction} {A hybrid mechanism for enhanced core-mantle boundary
  chemical interaction}.{\BBCQ}
\newblock
\APACjournalVolNumPages{Geophysical Research Letters}{48}{23}{e2021GL094456}.
\PrintBackRefs{\CurrentBib}

\bibitem [\protect \citeauthoryear {%
Lyons%
, Reinhard%
\BCBL {}\ \BBA {} Planavsky%
}{%
Lyons%
\ \protect \BOthers {.}}{%
{\protect \APACyear {2014}}%
}]{%
Lyons+2014}
\APACinsertmetastar {%
Lyons+2014}%
\begin{APACrefauthors}%
Lyons, T\BPBI W.%
, Reinhard, C\BPBI T.%
\BCBL {}\ \BBA {} Planavsky, N\BPBI J.%
\end{APACrefauthors}%
\unskip\
\newblock
\APACrefYearMonthDay{2014}{}{}.
\newblock
{\BBOQ}\APACrefatitle {The rise of oxygen in {E}arth's early ocean and
  atmosphere} {The rise of oxygen in {E}arth's early ocean and
  atmosphere}.{\BBCQ}
\newblock
\APACjournalVolNumPages{Nature}{506}{7488}{307--315}.
\PrintBackRefs{\CurrentBib}

\bibitem [\protect \citeauthoryear {%
Mallik%
, Li%
\BCBL {}\ \BBA {} Wiedenbeck%
}{%
Mallik%
\ \protect \BOthers {.}}{%
{\protect \APACyear {2018}}%
}]{%
Mallik+2018}
\APACinsertmetastar {%
Mallik+2018}%
\begin{APACrefauthors}%
Mallik, A.%
, Li, Y.%
\BCBL {}\ \BBA {} Wiedenbeck, M.%
\end{APACrefauthors}%
\unskip\
\newblock
\APACrefYearMonthDay{2018}{}{}.
\newblock
{\BBOQ}\APACrefatitle {Nitrogen evolution within the {E}arth's
  atmosphere--mantle system assessed by recycling in subduction zones}
  {Nitrogen evolution within the {E}arth's atmosphere--mantle system assessed
  by recycling in subduction zones}.{\BBCQ}
\newblock
\APACjournalVolNumPages{Earth and Planetary Science Letters}{482}{}{556--566}.
\PrintBackRefs{\CurrentBib}

\bibitem [\protect \citeauthoryear {%
Marchi%
, Canup%
\BCBL {}\ \BBA {} Walker%
}{%
Marchi%
\ \protect \BOthers {.}}{%
{\protect \APACyear {2018}}%
}]{%
Marchi+2018}
\APACinsertmetastar {%
Marchi+2018}%
\begin{APACrefauthors}%
Marchi, S.%
, Canup, R.%
\BCBL {}\ \BBA {} Walker, R.%
\end{APACrefauthors}%
\unskip\
\newblock
\APACrefYearMonthDay{2018}{}{}.
\newblock
{\BBOQ}\APACrefatitle {Heterogeneous delivery of silicate and metal to the
  {E}arth by large planetesimals} {Heterogeneous delivery of silicate and metal
  to the {E}arth by large planetesimals}.{\BBCQ}
\newblock
\APACjournalVolNumPages{Nature geoscience}{11}{1}{77--81}.
\PrintBackRefs{\CurrentBib}

\bibitem [\protect \citeauthoryear {%
Marty%
}{%
Marty%
}{%
{\protect \APACyear {1995}}%
}]{%
Marty1995}
\APACinsertmetastar {%
Marty1995}%
\begin{APACrefauthors}%
Marty, B.%
\end{APACrefauthors}%
\unskip\
\newblock
\APACrefYearMonthDay{1995}{}{}.
\newblock
{\BBOQ}\APACrefatitle {Nitrogen content of the mantle inferred from
  {N}$_2$--{A}r correlation in oceanic basalts} {Nitrogen content of the mantle
  inferred from {N}$_2$--{A}r correlation in oceanic basalts}.{\BBCQ}
\newblock
\APACjournalVolNumPages{Nature}{377}{6547}{326--329}.
\PrintBackRefs{\CurrentBib}

\bibitem [\protect \citeauthoryear {%
Marty%
\ \protect \BOthers {.}}{%
Marty%
\ \protect \BOthers {.}}{%
{\protect \APACyear {2017}}%
}]{%
Marty+2017}
\APACinsertmetastar {%
Marty+2017}%
\begin{APACrefauthors}%
Marty, B.%
, Altwegg, K.%
, Balsiger, H.%
, Bar-Nun, A.%
, Bekaert, D\BPBI V.%
, Berthelier, J\BHBI J.%
\BDBL {}others%
\end{APACrefauthors}%
\unskip\
\newblock
\APACrefYearMonthDay{2017}{}{}.
\newblock
{\BBOQ}\APACrefatitle {Xenon isotopes in 67{P}/{C}huryumov-{G}erasimenko show
  that comets contributed to {E}arth's atmosphere} {Xenon isotopes in
  67{P}/{C}huryumov-{G}erasimenko show that comets contributed to {E}arth's
  atmosphere}.{\BBCQ}
\newblock
\APACjournalVolNumPages{Science}{356}{6342}{1069--1072}.
\PrintBackRefs{\CurrentBib}

\bibitem [\protect \citeauthoryear {%
Marty%
\ \protect \BOthers {.}}{%
Marty%
\ \protect \BOthers {.}}{%
{\protect \APACyear {2016}}%
}]{%
Marty+2016}
\APACinsertmetastar {%
Marty+2016}%
\begin{APACrefauthors}%
Marty, B.%
, Avice, G.%
, Sano, Y.%
, Altwegg, K.%
, Balsiger, H.%
, H{\"a}ssig, M.%
\BDBL {}Rubin, M.%
\end{APACrefauthors}%
\unskip\
\newblock
\APACrefYearMonthDay{2016}{}{}.
\newblock
{\BBOQ}\APACrefatitle {Origins of volatile elements ({H}, {C}, {N}, noble
  gases) on {E}arth and {M}ars in light of recent results from the {ROSETTA}
  cometary mission} {Origins of volatile elements ({H}, {C}, {N}, noble gases)
  on {E}arth and {M}ars in light of recent results from the {ROSETTA} cometary
  mission}.{\BBCQ}
\newblock
\APACjournalVolNumPages{Earth and Planetary Science Letters}{441}{}{91--102}.
\PrintBackRefs{\CurrentBib}

\bibitem [\protect \citeauthoryear {%
Marty%
, Bekaert%
, Broadley%
\BCBL {}\ \BBA {} Jaupart%
}{%
Marty%
\ \protect \BOthers {.}}{%
{\protect \APACyear {2019}}%
}]{%
Marty+2019}
\APACinsertmetastar {%
Marty+2019}%
\begin{APACrefauthors}%
Marty, B.%
, Bekaert, D\BPBI V.%
, Broadley, M\BPBI W.%
\BCBL {}\ \BBA {} Jaupart, C.%
\end{APACrefauthors}%
\unskip\
\newblock
\APACrefYearMonthDay{2019}{}{}.
\newblock
{\BBOQ}\APACrefatitle {Geochemical evidence for high volatile fluxes from the
  mantle at the end of the {A}rchaean} {Geochemical evidence for high volatile
  fluxes from the mantle at the end of the {A}rchaean}.{\BBCQ}
\newblock
\APACjournalVolNumPages{Nature}{575}{7783}{485--488}.
\PrintBackRefs{\CurrentBib}

\bibitem [\protect \citeauthoryear {%
Marty%
\ \BBA {} Dauphas%
}{%
Marty%
\ \BBA {} Dauphas%
}{%
{\protect \APACyear {2003}}%
}]{%
Marty+Dauphas2003}
\APACinsertmetastar {%
Marty+Dauphas2003}%
\begin{APACrefauthors}%
Marty, B.%
\BCBT {}\ \BBA {} Dauphas, N.%
\end{APACrefauthors}%
\unskip\
\newblock
\APACrefYearMonthDay{2003}{}{}.
\newblock
{\BBOQ}\APACrefatitle {The nitrogen record of crust--mantle interaction and
  mantle convection from {A}rchean to present} {The nitrogen record of
  crust--mantle interaction and mantle convection from {A}rchean to
  present}.{\BBCQ}
\newblock
\APACjournalVolNumPages{Earth and Planetary Science
  Letters}{206}{3-4}{397--410}.
\PrintBackRefs{\CurrentBib}

\bibitem [\protect \citeauthoryear {%
Marty%
, Zimmermann%
, Pujol%
, Burgess%
\BCBL {}\ \BBA {} Philippot%
}{%
Marty%
\ \protect \BOthers {.}}{%
{\protect \APACyear {2013}}%
}]{%
Marty+2013}
\APACinsertmetastar {%
Marty+2013}%
\begin{APACrefauthors}%
Marty, B.%
, Zimmermann, L.%
, Pujol, M.%
, Burgess, R.%
\BCBL {}\ \BBA {} Philippot, P.%
\end{APACrefauthors}%
\unskip\
\newblock
\APACrefYearMonthDay{2013}{}{}.
\newblock
{\BBOQ}\APACrefatitle {Nitrogen isotopic composition and density of the
  {A}rchean atmosphere} {Nitrogen isotopic composition and density of the
  {A}rchean atmosphere}.{\BBCQ}
\newblock
\APACjournalVolNumPages{Science}{342}{6154}{101--104}.
\PrintBackRefs{\CurrentBib}

\bibitem [\protect \citeauthoryear {%
McDonough%
\ \BBA {} Sun%
}{%
McDonough%
\ \BBA {} Sun%
}{%
{\protect \APACyear {1995}}%
}]{%
McDonough+Sun1995}
\APACinsertmetastar {%
McDonough+Sun1995}%
\begin{APACrefauthors}%
McDonough, W\BPBI F.%
\BCBT {}\ \BBA {} Sun, S\BHBI S.%
\end{APACrefauthors}%
\unskip\
\newblock
\APACrefYearMonthDay{1995}{}{}.
\newblock
{\BBOQ}\APACrefatitle {The composition of the {E}arth} {The composition of the
  {E}arth}.{\BBCQ}
\newblock
\APACjournalVolNumPages{Chemical Geology}{120}{3-4}{223--253}.
\PrintBackRefs{\CurrentBib}

\bibitem [\protect \citeauthoryear {%
Mikhail%
\ \BBA {} Sverjensky%
}{%
Mikhail%
\ \BBA {} Sverjensky%
}{%
{\protect \APACyear {2014}}%
}]{%
Mikhail+Sverjensky2014}
\APACinsertmetastar {%
Mikhail+Sverjensky2014}%
\begin{APACrefauthors}%
Mikhail, S.%
\BCBT {}\ \BBA {} Sverjensky, D\BPBI A.%
\end{APACrefauthors}%
\unskip\
\newblock
\APACrefYearMonthDay{2014}{}{}.
\newblock
{\BBOQ}\APACrefatitle {Nitrogen speciation in upper mantle fluids and the
  origin of {E}arth's nitrogen-rich atmosphere} {Nitrogen speciation in upper
  mantle fluids and the origin of {E}arth's nitrogen-rich atmosphere}.{\BBCQ}
\newblock
\APACjournalVolNumPages{Nature Geoscience}{7}{11}{816--819}.
\PrintBackRefs{\CurrentBib}

\bibitem [\protect \citeauthoryear {%
Mitchell%
\ \protect \BOthers {.}}{%
Mitchell%
\ \protect \BOthers {.}}{%
{\protect \APACyear {2010}}%
}]{%
Mitchell+2010}
\APACinsertmetastar {%
Mitchell+2010}%
\begin{APACrefauthors}%
Mitchell, E\BPBI C.%
, Fischer, T\BPBI P.%
, Hilton, D\BPBI R.%
, Hauri, E\BPBI H.%
, Shaw, A\BPBI M.%
, de Moor, J\BPBI M.%
\BDBL {}Kazahaya, K.%
\end{APACrefauthors}%
\unskip\
\newblock
\APACrefYearMonthDay{2010}{}{}.
\newblock
{\BBOQ}\APACrefatitle {Nitrogen sources and recycling at subduction zones:
  Insights from the {I}zu-{B}onin-{M}ariana arc} {Nitrogen sources and
  recycling at subduction zones: Insights from the {I}zu-{B}onin-{M}ariana
  arc}.{\BBCQ}
\newblock
\APACjournalVolNumPages{Geochemistry, Geophysics, Geosystems}{11}{2}{}.
\PrintBackRefs{\CurrentBib}

\bibitem [\protect \citeauthoryear {%
Monteux%
, Andrault%
\BCBL {}\ \BBA {} Samuel%
}{%
Monteux%
\ \protect \BOthers {.}}{%
{\protect \APACyear {2016}}%
}]{%
Monteux+2016}
\APACinsertmetastar {%
Monteux+2016}%
\begin{APACrefauthors}%
Monteux, J.%
, Andrault, D.%
\BCBL {}\ \BBA {} Samuel, H.%
\end{APACrefauthors}%
\unskip\
\newblock
\APACrefYearMonthDay{2016}{}{}.
\newblock
{\BBOQ}\APACrefatitle {On the cooling of a deep terrestrial magma ocean} {On
  the cooling of a deep terrestrial magma ocean}.{\BBCQ}
\newblock
\APACjournalVolNumPages{Earth and Planetary Science Letters}{448}{}{140--149}.
\PrintBackRefs{\CurrentBib}

\bibitem [\protect \citeauthoryear {%
Moore%
}{%
Moore%
}{%
{\protect \APACyear {1977}}%
}]{%
Moore+1977}
\APACinsertmetastar {%
Moore+1977}%
\begin{APACrefauthors}%
Moore, H.%
\end{APACrefauthors}%
\unskip\
\newblock
\APACrefYearMonthDay{1977}{}{}.
\newblock
{\BBOQ}\APACrefatitle {The isotopic composition of ammonia, nitrogen dioxide
  and nitrate in the atmosphere} {The isotopic composition of ammonia, nitrogen
  dioxide and nitrate in the atmosphere}.{\BBCQ}
\newblock
\APACjournalVolNumPages{Atmospheric Environment}{11}{12}{1239--1243}.
\PrintBackRefs{\CurrentBib}

\bibitem [\protect \citeauthoryear {%
Navarro-Gonz{\'a}lez%
, McKay%
\BCBL {}\ \BBA {} Mvondo%
}{%
Navarro-Gonz{\'a}lez%
\ \protect \BOthers {.}}{%
{\protect \APACyear {2001}}%
}]{%
Navarro+2001}
\APACinsertmetastar {%
Navarro+2001}%
\begin{APACrefauthors}%
Navarro-Gonz{\'a}lez, R.%
, McKay, C\BPBI P.%
\BCBL {}\ \BBA {} Mvondo, D\BPBI N.%
\end{APACrefauthors}%
\unskip\
\newblock
\APACrefYearMonthDay{2001}{}{}.
\newblock
{\BBOQ}\APACrefatitle {A possible nitrogen crisis for {A}rchaean life due to
  reduced nitrogen fixation by lightning} {A possible nitrogen crisis for
  {A}rchaean life due to reduced nitrogen fixation by lightning}.{\BBCQ}
\newblock
\APACjournalVolNumPages{Nature}{412}{6842}{61--64}.
\PrintBackRefs{\CurrentBib}

\bibitem [\protect \citeauthoryear {%
Nicklas%
, Puchtel%
\BCBL {}\ \BBA {} Ash%
}{%
Nicklas%
\ \protect \BOthers {.}}{%
{\protect \APACyear {2018}}%
}]{%
Nicklas+2018}
\APACinsertmetastar {%
Nicklas+2018}%
\begin{APACrefauthors}%
Nicklas, R\BPBI W.%
, Puchtel, I\BPBI S.%
\BCBL {}\ \BBA {} Ash, R\BPBI D.%
\end{APACrefauthors}%
\unskip\
\newblock
\APACrefYearMonthDay{2018}{}{}.
\newblock
{\BBOQ}\APACrefatitle {Redox state of the {A}rchean mantle: evidence from V
  partitioning in 3.5--2.4 Ga komatiites} {Redox state of the {A}rchean mantle:
  evidence from v partitioning in 3.5--2.4 ga komatiites}.{\BBCQ}
\newblock
\APACjournalVolNumPages{Geochimica et Cosmochimica Acta}{222}{}{447--466}.
\PrintBackRefs{\CurrentBib}

\bibitem [\protect \citeauthoryear {%
Nicklas%
\ \protect \BOthers {.}}{%
Nicklas%
\ \protect \BOthers {.}}{%
{\protect \APACyear {2019}}%
}]{%
Nicklas+2019}
\APACinsertmetastar {%
Nicklas+2019}%
\begin{APACrefauthors}%
Nicklas, R\BPBI W.%
, Puchtel, I\BPBI S.%
, Ash, R\BPBI D.%
, Piccoli, P\BPBI M.%
, Hanski, E.%
, Nisbet, E\BPBI G.%
\BDBL {}Anbar, A\BPBI D.%
\end{APACrefauthors}%
\unskip\
\newblock
\APACrefYearMonthDay{2019}{}{}.
\newblock
{\BBOQ}\APACrefatitle {Secular mantle oxidation across the
  {A}rchean-{P}roterozoic boundary: {E}vidence from {V} partitioning in
  komatiites and picrites} {Secular mantle oxidation across the
  {A}rchean-{P}roterozoic boundary: {E}vidence from {V} partitioning in
  komatiites and picrites}.{\BBCQ}
\newblock
\APACjournalVolNumPages{Geochimica et Cosmochimica Acta}{250}{}{49--75}.
\PrintBackRefs{\CurrentBib}

\bibitem [\protect \citeauthoryear {%
Nikolaou%
\ \protect \BOthers {.}}{%
Nikolaou%
\ \protect \BOthers {.}}{%
{\protect \APACyear {2019}}%
}]{%
Nikolaou+2019}
\APACinsertmetastar {%
Nikolaou+2019}%
\begin{APACrefauthors}%
Nikolaou, A.%
, Katyal, N.%
, Tosi, N.%
, Godolt, M.%
, Grenfell, J\BPBI L.%
\BCBL {}\ \BBA {} Rauer, H.%
\end{APACrefauthors}%
\unskip\
\newblock
\APACrefYearMonthDay{2019}{}{}.
\newblock
{\BBOQ}\APACrefatitle {What factors affect the duration and outgassing of the
  terrestrial magma ocean?} {What factors affect the duration and outgassing of
  the terrestrial magma ocean?}{\BBCQ}
\newblock
\APACjournalVolNumPages{The Astrophysical Journal}{875}{1}{11}.
\PrintBackRefs{\CurrentBib}

\bibitem [\protect \citeauthoryear {%
Ozima%
\ \BBA {} Podosek%
}{%
Ozima%
\ \BBA {} Podosek%
}{%
{\protect \APACyear {2002}}%
}]{%
Ozima+Podosek2002}
\APACinsertmetastar {%
Ozima+Podosek2002}%
\begin{APACrefauthors}%
Ozima, M.%
\BCBT {}\ \BBA {} Podosek, F\BPBI A.%
\end{APACrefauthors}%
\unskip\
\newblock
\APACrefYear{2002}.
\newblock
\APACrefbtitle {Noble gas geochemistry} {Noble gas geochemistry}.
\newblock
\APACaddressPublisher{}{Cambridge University Press}.
\PrintBackRefs{\CurrentBib}

\bibitem [\protect \citeauthoryear {%
Palot%
, Cartigny%
, Harris%
, Kaminsky%
\BCBL {}\ \BBA {} Stachel%
}{%
Palot%
\ \protect \BOthers {.}}{%
{\protect \APACyear {2012}}%
}]{%
Palot+2012}
\APACinsertmetastar {%
Palot+2012}%
\begin{APACrefauthors}%
Palot, M.%
, Cartigny, P.%
, Harris, J.%
, Kaminsky, F.%
\BCBL {}\ \BBA {} Stachel, T.%
\end{APACrefauthors}%
\unskip\
\newblock
\APACrefYearMonthDay{2012}{}{}.
\newblock
{\BBOQ}\APACrefatitle {Evidence for deep mantle convection and primordial
  heterogeneity from nitrogen and carbon stable isotopes in diamond} {Evidence
  for deep mantle convection and primordial heterogeneity from nitrogen and
  carbon stable isotopes in diamond}.{\BBCQ}
\newblock
\APACjournalVolNumPages{Earth and Planetary Science Letters}{357}{}{179--193}.
\PrintBackRefs{\CurrentBib}

\bibitem [\protect \citeauthoryear {%
Parai%
\ \BBA {} Mukhopadhyay%
}{%
Parai%
\ \BBA {} Mukhopadhyay%
}{%
{\protect \APACyear {2018}}%
}]{%
Parai+Mukhopadhyay2018}
\APACinsertmetastar {%
Parai+Mukhopadhyay2018}%
\begin{APACrefauthors}%
Parai, R.%
\BCBT {}\ \BBA {} Mukhopadhyay, S.%
\end{APACrefauthors}%
\unskip\
\newblock
\APACrefYearMonthDay{2018}{}{}.
\newblock
{\BBOQ}\APACrefatitle {Xenon isotopic constraints on the history of volatile
  recycling into the mantle} {Xenon isotopic constraints on the history of
  volatile recycling into the mantle}.{\BBCQ}
\newblock
\APACjournalVolNumPages{Nature}{560}{7717}{223--227}.
\PrintBackRefs{\CurrentBib}

\bibitem [\protect \citeauthoryear {%
Pehrsson%
, Eglington%
, Evans%
, Huston%
\BCBL {}\ \BBA {} Reddy%
}{%
Pehrsson%
\ \protect \BOthers {.}}{%
{\protect \APACyear {2016}}%
}]{%
Pehrsson+2016}
\APACinsertmetastar {%
Pehrsson+2016}%
\begin{APACrefauthors}%
Pehrsson, S\BPBI J.%
, Eglington, B\BPBI M.%
, Evans, D\BPBI A.%
, Huston, D.%
\BCBL {}\ \BBA {} Reddy, S\BPBI M.%
\end{APACrefauthors}%
\unskip\
\newblock
\APACrefYearMonthDay{2016}{}{}.
\newblock
{\BBOQ}\APACrefatitle {Metallogeny and its link to orogenic style during the
  Nuna supercontinent cycle} {Metallogeny and its link to orogenic style during
  the nuna supercontinent cycle}.{\BBCQ}
\newblock
\APACjournalVolNumPages{Geological Society, London, Special
  Publications}{424}{1}{83--94}.
\PrintBackRefs{\CurrentBib}

\bibitem [\protect \citeauthoryear {%
Pepin%
}{%
Pepin%
}{%
{\protect \APACyear {1991}}%
}]{%
Pepin1991}
\APACinsertmetastar {%
Pepin1991}%
\begin{APACrefauthors}%
Pepin, R\BPBI O.%
\end{APACrefauthors}%
\unskip\
\newblock
\APACrefYearMonthDay{1991}{}{}.
\newblock
{\BBOQ}\APACrefatitle {On the origin and early evolution of terrestrial planet
  atmospheres and meteoritic volatiles} {On the origin and early evolution of
  terrestrial planet atmospheres and meteoritic volatiles}.{\BBCQ}
\newblock
\APACjournalVolNumPages{Icarus}{92}{1}{2--79}.
\PrintBackRefs{\CurrentBib}

\bibitem [\protect \citeauthoryear {%
Piani%
\ \protect \BOthers {.}}{%
Piani%
\ \protect \BOthers {.}}{%
{\protect \APACyear {2020}}%
}]{%
Piani+2020}
\APACinsertmetastar {%
Piani+2020}%
\begin{APACrefauthors}%
Piani, L.%
, Marrocchi, Y.%
, Rigaudier, T.%
, Vacher, L\BPBI G.%
, Thomassin, D.%
\BCBL {}\ \BBA {} Marty, B.%
\end{APACrefauthors}%
\unskip\
\newblock
\APACrefYearMonthDay{2020}{}{}.
\newblock
{\BBOQ}\APACrefatitle {Earth’s water may have been inherited from material
  similar to enstatite chondrite meteorites} {Earth’s water may have been
  inherited from material similar to enstatite chondrite meteorites}.{\BBCQ}
\newblock
\APACjournalVolNumPages{Science}{369}{6507}{1110--1113}.
\PrintBackRefs{\CurrentBib}

\bibitem [\protect \citeauthoryear {%
Pujol%
, Marty%
\BCBL {}\ \BBA {} Burgess%
}{%
Pujol%
\ \protect \BOthers {.}}{%
{\protect \APACyear {2011}}%
}]{%
Pujol+2011}
\APACinsertmetastar {%
Pujol+2011}%
\begin{APACrefauthors}%
Pujol, M.%
, Marty, B.%
\BCBL {}\ \BBA {} Burgess, R.%
\end{APACrefauthors}%
\unskip\
\newblock
\APACrefYearMonthDay{2011}{}{}.
\newblock
{\BBOQ}\APACrefatitle {Chondritic-like xenon trapped in Archean rocks: A
  possible signature of the ancient atmosphere} {Chondritic-like xenon trapped
  in archean rocks: A possible signature of the ancient atmosphere}.{\BBCQ}
\newblock
\APACjournalVolNumPages{Earth and Planetary Science
  Letters}{308}{3-4}{298--306}.
\PrintBackRefs{\CurrentBib}

\bibitem [\protect \citeauthoryear {%
Pujol%
, Marty%
, Burgess%
, Turner%
\BCBL {}\ \BBA {} Philippot%
}{%
Pujol%
\ \protect \BOthers {.}}{%
{\protect \APACyear {2013}}%
}]{%
Pujol+2013}
\APACinsertmetastar {%
Pujol+2013}%
\begin{APACrefauthors}%
Pujol, M.%
, Marty, B.%
, Burgess, R.%
, Turner, G.%
\BCBL {}\ \BBA {} Philippot, P.%
\end{APACrefauthors}%
\unskip\
\newblock
\APACrefYearMonthDay{2013}{}{}.
\newblock
{\BBOQ}\APACrefatitle {Argon isotopic composition of {A}rchaean atmosphere
  probes early {E}arth geodynamics} {Argon isotopic composition of {A}rchaean
  atmosphere probes early {E}arth geodynamics}.{\BBCQ}
\newblock
\APACjournalVolNumPages{Nature}{498}{7452}{87--90}.
\PrintBackRefs{\CurrentBib}

\bibitem [\protect \citeauthoryear {%
Regier%
\ \protect \BOthers {.}}{%
Regier%
\ \protect \BOthers {.}}{%
{\protect \APACyear {2020}}%
}]{%
Regier2020}
\APACinsertmetastar {%
Regier2020}%
\begin{APACrefauthors}%
Regier, M.%
, Pearson, D.%
, Stachel, T.%
, Luth, R.%
, Stern, R.%
\BCBL {}\ \BBA {} Harris, J.%
\end{APACrefauthors}%
\unskip\
\newblock
\APACrefYearMonthDay{2020}{}{}.
\newblock
{\BBOQ}\APACrefatitle {The lithospheric-to-lower-mantle carbon cycle recorded
  in superdeep diamonds} {The lithospheric-to-lower-mantle carbon cycle
  recorded in superdeep diamonds}.{\BBCQ}
\newblock
\APACjournalVolNumPages{Nature}{585}{}{234-239}.
\newblock
\begin{APACrefDOI} \doi{10.1038/s41586-020-2676-z} \end{APACrefDOI}
\PrintBackRefs{\CurrentBib}

\bibitem [\protect \citeauthoryear {%
Rudnick%
, Gao%
, Holland%
, Turekian%
\BCBL {}\ \protect \BOthers {.}}{%
Rudnick%
\ \protect \BOthers {.}}{%
{\protect \APACyear {2003}}%
}]{%
Rudnick+Gao2003}
\APACinsertmetastar {%
Rudnick+Gao2003}%
\begin{APACrefauthors}%
Rudnick, R.%
, Gao, S.%
, Holland, H.%
, Turekian, K.%
\BCBL {}\ \BOthersPeriod {.}\end{APACrefauthors}%
\unskip\
\newblock
\APACrefYearMonthDay{2003}{}{}.
\newblock
{\BBOQ}\APACrefatitle {Composition of the continental crust} {Composition of
  the continental crust}.{\BBCQ}
\newblock
\APACjournalVolNumPages{The Crust}{3}{}{1--64}.
\PrintBackRefs{\CurrentBib}

\bibitem [\protect \citeauthoryear {%
Sakuraba%
, Kurokawa%
\BCBL {}\ \BBA {} Genda%
}{%
Sakuraba%
\ \protect \BOthers {.}}{%
{\protect \APACyear {2019}}%
}]{%
Sakuraba+2019}
\APACinsertmetastar {%
Sakuraba+2019}%
\begin{APACrefauthors}%
Sakuraba, H.%
, Kurokawa, H.%
\BCBL {}\ \BBA {} Genda, H.%
\end{APACrefauthors}%
\unskip\
\newblock
\APACrefYearMonthDay{2019}{}{}.
\newblock
{\BBOQ}\APACrefatitle {Impact degassing and atmospheric erosion on {V}enus,
  {E}arth, and {M}ars during the late accretion} {Impact degassing and
  atmospheric erosion on {V}enus, {E}arth, and {M}ars during the late
  accretion}.{\BBCQ}
\newblock
\APACjournalVolNumPages{Icarus}{317}{}{48--58}.
\PrintBackRefs{\CurrentBib}

\bibitem [\protect \citeauthoryear {%
Sakuraba%
, Kurokawa%
, Genda%
\BCBL {}\ \BBA {} Ohta%
}{%
Sakuraba%
\ \protect \BOthers {.}}{%
{\protect \APACyear {2021}}%
}]{%
Sakuraba+2021}
\APACinsertmetastar {%
Sakuraba+2021}%
\begin{APACrefauthors}%
Sakuraba, H.%
, Kurokawa, H.%
, Genda, H.%
\BCBL {}\ \BBA {} Ohta, K.%
\end{APACrefauthors}%
\unskip\
\newblock
\APACrefYearMonthDay{2021}{}{}.
\newblock
{\BBOQ}\APACrefatitle {Numerous chondritic impactors and oxidized magma ocean
  set {E}arth's volatile depletion} {Numerous chondritic impactors and oxidized
  magma ocean set {E}arth's volatile depletion}.{\BBCQ}
\newblock
\APACjournalVolNumPages{Scientific Reports}{TBD}{}{TBD}.
\PrintBackRefs{\CurrentBib}

\bibitem [\protect \citeauthoryear {%
Salvador%
\ \protect \BOthers {.}}{%
Salvador%
\ \protect \BOthers {.}}{%
{\protect \APACyear {2017}}%
}]{%
Salvador+2017}
\APACinsertmetastar {%
Salvador+2017}%
\begin{APACrefauthors}%
Salvador, A.%
, Massol, H.%
, Davaille, A.%
, Marcq, E.%
, Sarda, P.%
\BCBL {}\ \BBA {} Chassefi{\`e}re, E.%
\end{APACrefauthors}%
\unskip\
\newblock
\APACrefYearMonthDay{2017}{}{}.
\newblock
{\BBOQ}\APACrefatitle {The relative influence of {H}$_2${O} and {CO}$_2$ on the
  primitive surface conditions and evolution of rocky planets} {The relative
  influence of {H}$_2${O} and {CO}$_2$ on the primitive surface conditions and
  evolution of rocky planets}.{\BBCQ}
\newblock
\APACjournalVolNumPages{Journal of Geophysical Research:
  Planets}{122}{7}{1458--1486}.
\PrintBackRefs{\CurrentBib}

\bibitem [\protect \citeauthoryear {%
Saxena%
\ \protect \BOthers {.}}{%
Saxena%
\ \protect \BOthers {.}}{%
{\protect \APACyear {2019}}%
}]{%
Saxena+2019}
\APACinsertmetastar {%
Saxena+2019}%
\begin{APACrefauthors}%
Saxena, P.%
, Killen, R\BPBI M.%
, Airapetian, V.%
, Petro, N\BPBI E.%
, Curran, N\BPBI M.%
\BCBL {}\ \BBA {} Mandell, A\BPBI M.%
\end{APACrefauthors}%
\unskip\
\newblock
\APACrefYearMonthDay{2019}{}{}.
\newblock
{\BBOQ}\APACrefatitle {Was the {S}un a slow rotator? {S}odium and potassium
  constraints from the lunar regolith} {Was the {S}un a slow rotator? {S}odium
  and potassium constraints from the lunar regolith}.{\BBCQ}
\newblock
\APACjournalVolNumPages{The Astrophysical Journal Letters}{876}{1}{L16}.
\PrintBackRefs{\CurrentBib}

\bibitem [\protect \citeauthoryear {%
Schaefer%
\ \BBA {} Fegley~Jr%
}{%
Schaefer%
\ \BBA {} Fegley~Jr%
}{%
{\protect \APACyear {2017}}%
}]{%
Schaefer+Fegley2017}
\APACinsertmetastar {%
Schaefer+Fegley2017}%
\begin{APACrefauthors}%
Schaefer, L.%
\BCBT {}\ \BBA {} Fegley~Jr, B.%
\end{APACrefauthors}%
\unskip\
\newblock
\APACrefYearMonthDay{2017}{}{}.
\newblock
{\BBOQ}\APACrefatitle {Redox states of initial atmospheres outgassed on rocky
  planets and planetesimals} {Redox states of initial atmospheres outgassed on
  rocky planets and planetesimals}.{\BBCQ}
\newblock
\APACjournalVolNumPages{The Astrophysical Journal}{843}{2}{120}.
\PrintBackRefs{\CurrentBib}

\bibitem [\protect \citeauthoryear {%
Schwieterman%
, Robinson%
, Meadows%
, Misra%
\BCBL {}\ \BBA {} Domagal-Goldman%
}{%
Schwieterman%
\ \protect \BOthers {.}}{%
{\protect \APACyear {2015}}%
}]{%
Schwieterman+2015}
\APACinsertmetastar {%
Schwieterman+2015}%
\begin{APACrefauthors}%
Schwieterman, E\BPBI W.%
, Robinson, T\BPBI D.%
, Meadows, V\BPBI S.%
, Misra, A.%
\BCBL {}\ \BBA {} Domagal-Goldman, S.%
\end{APACrefauthors}%
\unskip\
\newblock
\APACrefYearMonthDay{2015}{}{}.
\newblock
{\BBOQ}\APACrefatitle {Detecting and constraining {N}$_2$ abundances in
  planetary atmospheres using collisional pairs} {Detecting and constraining
  {N}$_2$ abundances in planetary atmospheres using collisional pairs}.{\BBCQ}
\newblock
\APACjournalVolNumPages{The Astrophysical Journal}{810}{1}{57}.
\PrintBackRefs{\CurrentBib}

\bibitem [\protect \citeauthoryear {%
Shen%
, Pinti%
\BCBL {}\ \BBA {} Hashizume%
}{%
Shen%
\ \protect \BOthers {.}}{%
{\protect \APACyear {2006}}%
}]{%
Shen+2006}
\APACinsertmetastar {%
Shen+2006}%
\begin{APACrefauthors}%
Shen, Y.%
, Pinti, D\BPBI L.%
\BCBL {}\ \BBA {} Hashizume, K.%
\end{APACrefauthors}%
\unskip\
\newblock
\APACrefYearMonthDay{2006}{}{}.
\newblock
{\BBOQ}\APACrefatitle {Biogeochemical cycles of sulfur and nitrogen in the
  Archean ocean and atmosphere} {Biogeochemical cycles of sulfur and nitrogen
  in the archean ocean and atmosphere}.{\BBCQ}
\newblock
\APACjournalVolNumPages{Archean Geodynamics and Environments}{164}{}{305--320}.
\PrintBackRefs{\CurrentBib}

\bibitem [\protect \citeauthoryear {%
Solomatov%
}{%
Solomatov%
}{%
{\protect \APACyear {2015}}%
}]{%
Solomatov2015}
\APACinsertmetastar {%
Solomatov2015}%
\begin{APACrefauthors}%
Solomatov, V.%
\end{APACrefauthors}%
\unskip\
\newblock
\APACrefYearMonthDay{2015}{}{}.
\newblock
{\BBOQ}\APACrefatitle {Magma oceans and primordial mantle differentiation}
  {Magma oceans and primordial mantle differentiation}.{\BBCQ}
\newblock
\BIn{} \APACrefbtitle {Treatise on Geophysics: Second Edition} {Treatise on
  geophysics: Second edition}\ (\BPGS\ 81--104).
\PrintBackRefs{\CurrentBib}

\bibitem [\protect \citeauthoryear {%
Som%
\ \protect \BOthers {.}}{%
Som%
\ \protect \BOthers {.}}{%
{\protect \APACyear {2016}}%
}]{%
Som+2016}
\APACinsertmetastar {%
Som+2016}%
\begin{APACrefauthors}%
Som, S\BPBI M.%
, Buick, R.%
, Hagadorn, J\BPBI W.%
, Blake, T\BPBI S.%
, Perreault, J\BPBI M.%
, Harnmeijer, J\BPBI P.%
\BCBL {}\ \BBA {} Catling, D\BPBI C.%
\end{APACrefauthors}%
\unskip\
\newblock
\APACrefYearMonthDay{2016}{}{}.
\newblock
{\BBOQ}\APACrefatitle {Earth's air pressure 2.7 billion years ago constrained
  to less than half of modern levels} {Earth's air pressure 2.7 billion years
  ago constrained to less than half of modern levels}.{\BBCQ}
\newblock
\APACjournalVolNumPages{Nature Geoscience}{9}{6}{448--451}.
\PrintBackRefs{\CurrentBib}

\bibitem [\protect \citeauthoryear {%
Som%
, Catling%
, Harnmeijer%
, Polivka%
\BCBL {}\ \BBA {} Buick%
}{%
Som%
\ \protect \BOthers {.}}{%
{\protect \APACyear {2012}}%
}]{%
Som+2012}
\APACinsertmetastar {%
Som+2012}%
\begin{APACrefauthors}%
Som, S\BPBI M.%
, Catling, D\BPBI C.%
, Harnmeijer, J\BPBI P.%
, Polivka, P\BPBI M.%
\BCBL {}\ \BBA {} Buick, R.%
\end{APACrefauthors}%
\unskip\
\newblock
\APACrefYearMonthDay{2012}{}{}.
\newblock
{\BBOQ}\APACrefatitle {Air density 2.7 billion years ago limited to less than
  twice modern levels by fossil raindrop imprints} {Air density 2.7 billion
  years ago limited to less than twice modern levels by fossil raindrop
  imprints}.{\BBCQ}
\newblock
\APACjournalVolNumPages{Nature}{484}{7394}{359--362}.
\PrintBackRefs{\CurrentBib}

\bibitem [\protect \citeauthoryear {%
Sossi%
\ \protect \BOthers {.}}{%
Sossi%
\ \protect \BOthers {.}}{%
{\protect \APACyear {2020}}%
}]{%
Sossi+2020}
\APACinsertmetastar {%
Sossi+2020}%
\begin{APACrefauthors}%
Sossi, P\BPBI A.%
, Burnham, A\BPBI D.%
, Badro, J.%
, Lanzirotti, A.%
, Newville, M.%
\BCBL {}\ \BBA {} O’neill, H\BPBI S\BPBI C.%
\end{APACrefauthors}%
\unskip\
\newblock
\APACrefYearMonthDay{2020}{}{}.
\newblock
{\BBOQ}\APACrefatitle {Redox state of {E}arth’s magma ocean and its
  {V}enus-like early atmosphere} {Redox state of {E}arth’s magma ocean and
  its {V}enus-like early atmosphere}.{\BBCQ}
\newblock
\APACjournalVolNumPages{Science advances}{6}{48}{eabd1387}.
\PrintBackRefs{\CurrentBib}

\bibitem [\protect \citeauthoryear {%
Speelmanns%
, Schmidt%
\BCBL {}\ \BBA {} Liebske%
}{%
Speelmanns%
\ \protect \BOthers {.}}{%
{\protect \APACyear {2018}}%
}]{%
Speelmanns+2018}
\APACinsertmetastar {%
Speelmanns+2018}%
\begin{APACrefauthors}%
Speelmanns, I\BPBI M.%
, Schmidt, M\BPBI W.%
\BCBL {}\ \BBA {} Liebske, C.%
\end{APACrefauthors}%
\unskip\
\newblock
\APACrefYearMonthDay{2018}{}{}.
\newblock
{\BBOQ}\APACrefatitle {Nitrogen solubility in core materials} {Nitrogen
  solubility in core materials}.{\BBCQ}
\newblock
\APACjournalVolNumPages{Geophysical Research Letters}{45}{15}{7434--7443}.
\PrintBackRefs{\CurrentBib}

\bibitem [\protect \citeauthoryear {%
St{\"u}eken%
}{%
St{\"u}eken%
}{%
{\protect \APACyear {2016}}%
}]{%
Stueken2016AB}
\APACinsertmetastar {%
Stueken2016AB}%
\begin{APACrefauthors}%
St{\"u}eken, E\BPBI E.%
\end{APACrefauthors}%
\unskip\
\newblock
\APACrefYearMonthDay{2016}{}{}.
\newblock
{\BBOQ}\APACrefatitle {Nitrogen in ancient mud: a biosignature?} {Nitrogen in
  ancient mud: a biosignature?}{\BBCQ}
\newblock
\APACjournalVolNumPages{Astrobiology}{16}{9}{730--735}.
\PrintBackRefs{\CurrentBib}

\bibitem [\protect \citeauthoryear {%
St{\"u}eken%
, Boocock%
, Szilas%
, Mikhail%
\BCBL {}\ \BBA {} Gardiner%
}{%
St{\"u}eken%
\ \protect \BOthers {.}}{%
{\protect \APACyear {2021}}%
}]{%
Stueken+2021}
\APACinsertmetastar {%
Stueken+2021}%
\begin{APACrefauthors}%
St{\"u}eken, E\BPBI E.%
, Boocock, T.%
, Szilas, K.%
, Mikhail, S.%
\BCBL {}\ \BBA {} Gardiner, N\BPBI J.%
\end{APACrefauthors}%
\unskip\
\newblock
\APACrefYearMonthDay{2021}{}{}.
\newblock
{\BBOQ}\APACrefatitle {Reconstructing Nitrogen Sources to {E}arth's Earliest
  Biosphere at 3.7 Ga} {Reconstructing nitrogen sources to {E}arth's earliest
  biosphere at 3.7 ga}.{\BBCQ}
\newblock
\APACjournalVolNumPages{Frontiers in Earth Science}{9}{}{286}.
\PrintBackRefs{\CurrentBib}

\bibitem [\protect \citeauthoryear {%
St{\"u}eken%
, Buick%
, Guy%
\BCBL {}\ \BBA {} Koehler%
}{%
St{\"u}eken%
\ \protect \BOthers {.}}{%
{\protect \APACyear {2015}}%
}]{%
Stueken+2015}
\APACinsertmetastar {%
Stueken+2015}%
\begin{APACrefauthors}%
St{\"u}eken, E\BPBI E.%
, Buick, R.%
, Guy, B\BPBI M.%
\BCBL {}\ \BBA {} Koehler, M\BPBI C.%
\end{APACrefauthors}%
\unskip\
\newblock
\APACrefYearMonthDay{2015}{}{}.
\newblock
{\BBOQ}\APACrefatitle {Isotopic evidence for biological nitrogen fixation by
  molybdenum-nitrogenase from 3.2 {G}yr} {Isotopic evidence for biological
  nitrogen fixation by molybdenum-nitrogenase from 3.2 {G}yr}.{\BBCQ}
\newblock
\APACjournalVolNumPages{Nature}{520}{7549}{666--669}.
\PrintBackRefs{\CurrentBib}

\bibitem [\protect \citeauthoryear {%
St{\"u}eken%
, Kipp%
, Koehler%
\BCBL {}\ \BBA {} Buick%
}{%
St{\"u}eken%
, Kipp%
, Koehler%
\BCBL {}\ \BBA {} Buick%
}{%
{\protect \APACyear {2016}}%
}]{%
Stueken+2016}
\APACinsertmetastar {%
Stueken+2016}%
\begin{APACrefauthors}%
St{\"u}eken, E\BPBI E.%
, Kipp, M\BPBI A.%
, Koehler, M\BPBI C.%
\BCBL {}\ \BBA {} Buick, R.%
\end{APACrefauthors}%
\unskip\
\newblock
\APACrefYearMonthDay{2016}{}{}.
\newblock
{\BBOQ}\APACrefatitle {The evolution of {E}arth's biogeochemical nitrogen
  cycle} {The evolution of {E}arth's biogeochemical nitrogen cycle}.{\BBCQ}
\newblock
\APACjournalVolNumPages{Earth-Science Reviews}{160}{}{220--239}.
\PrintBackRefs{\CurrentBib}

\bibitem [\protect \citeauthoryear {%
St{\"u}eken%
, Kipp%
, Koehler%
, Schwieterman%
\BCBL {}\ \protect \BOthers {.}}{%
St{\"u}eken%
, Kipp%
, Koehler%
, Schwieterman%
\BCBL {}\ \protect \BOthers {.}}{%
{\protect \APACyear {2016}}%
}]{%
Stueken+2016ABb}
\APACinsertmetastar {%
Stueken+2016ABb}%
\begin{APACrefauthors}%
St{\"u}eken, E\BPBI E.%
, Kipp, M\BPBI A.%
, Koehler, M\BPBI C.%
, Schwieterman, E\BPBI W.%
, Johnson, B.%
\BCBL {}\ \BBA {} Buick, R.%
\end{APACrefauthors}%
\unskip\
\newblock
\APACrefYearMonthDay{2016}{}{}.
\newblock
{\BBOQ}\APACrefatitle {Modeling $p${N}$_2$ through geological time:
  {I}mplications for planetary climates and atmospheric biosignatures}
  {Modeling $p${N}$_2$ through geological time: {I}mplications for planetary
  climates and atmospheric biosignatures}.{\BBCQ}
\newblock
\APACjournalVolNumPages{Astrobiology}{16}{12}{949--963}.
\PrintBackRefs{\CurrentBib}

\bibitem [\protect \citeauthoryear {%
Tian%
, Kasting%
, Liu%
\BCBL {}\ \BBA {} Roble%
}{%
Tian%
\ \protect \BOthers {.}}{%
{\protect \APACyear {2008}}%
}]{%
Tian+2008}
\APACinsertmetastar {%
Tian+2008}%
\begin{APACrefauthors}%
Tian, F.%
, Kasting, J\BPBI F.%
, Liu, H\BHBI L.%
\BCBL {}\ \BBA {} Roble, R\BPBI G.%
\end{APACrefauthors}%
\unskip\
\newblock
\APACrefYearMonthDay{2008}{}{}.
\newblock
{\BBOQ}\APACrefatitle {Hydrodynamic planetary thermosphere model: 1. {R}esponse
  of the {E}arth's thermosphere to extreme solar {EUV} conditions and the
  significance of adiabatic cooling} {Hydrodynamic planetary thermosphere
  model: 1. {R}esponse of the {E}arth's thermosphere to extreme solar {EUV}
  conditions and the significance of adiabatic cooling}.{\BBCQ}
\newblock
\APACjournalVolNumPages{Journal of Geophysical Research:
  Planets}{113}{}{E05008}.
\PrintBackRefs{\CurrentBib}

\bibitem [\protect \citeauthoryear {%
Tolstikhin%
\ \BBA {} Marty%
}{%
Tolstikhin%
\ \BBA {} Marty%
}{%
{\protect \APACyear {1998}}%
}]{%
Tolstikhin+Marty1998}
\APACinsertmetastar {%
Tolstikhin+Marty1998}%
\begin{APACrefauthors}%
Tolstikhin, I.%
\BCBT {}\ \BBA {} Marty, B.%
\end{APACrefauthors}%
\unskip\
\newblock
\APACrefYearMonthDay{1998}{}{}.
\newblock
{\BBOQ}\APACrefatitle {The evolution of terrestrial volatiles: a view from
  helium, neon, argon and nitrogen isotope modelling} {The evolution of
  terrestrial volatiles: a view from helium, neon, argon and nitrogen isotope
  modelling}.{\BBCQ}
\newblock
\APACjournalVolNumPages{Chemical Geology}{147}{1-2}{27--52}.
\PrintBackRefs{\CurrentBib}

\bibitem [\protect \citeauthoryear {%
Tu%
, Johnstone%
, G{\"u}del%
\BCBL {}\ \BBA {} Lammer%
}{%
Tu%
\ \protect \BOthers {.}}{%
{\protect \APACyear {2015}}%
}]{%
Tu+2015}
\APACinsertmetastar {%
Tu+2015}%
\begin{APACrefauthors}%
Tu, L.%
, Johnstone, C\BPBI P.%
, G{\"u}del, M.%
\BCBL {}\ \BBA {} Lammer, H.%
\end{APACrefauthors}%
\unskip\
\newblock
\APACrefYearMonthDay{2015}{}{}.
\newblock
{\BBOQ}\APACrefatitle {The extreme ultraviolet and {X}-ray {S}un in time:
  high-energy evolutionary tracks of a solar-like star} {The extreme
  ultraviolet and {X}-ray {S}un in time: high-energy evolutionary tracks of a
  solar-like star}.{\BBCQ}
\newblock
\APACjournalVolNumPages{Astronomy \& Astrophysics}{577}{}{L3}.
\PrintBackRefs{\CurrentBib}

\bibitem [\protect \citeauthoryear {%
Turbet%
\ \protect \BOthers {.}}{%
Turbet%
\ \protect \BOthers {.}}{%
{\protect \APACyear {2021}}%
}]{%
Turbet+2021}
\APACinsertmetastar {%
Turbet+2021}%
\begin{APACrefauthors}%
Turbet, M.%
, Bolmont, E.%
, Chaverot, G.%
, Ehrenreich, D.%
, Leconte, J.%
\BCBL {}\ \BBA {} Marcq, E.%
\end{APACrefauthors}%
\unskip\
\newblock
\APACrefYearMonthDay{2021}{}{}.
\newblock
{\BBOQ}\APACrefatitle {Day--night cloud asymmetry prevents early oceans on
  {V}enus but not on {E}arth} {Day--night cloud asymmetry prevents early oceans
  on {V}enus but not on {E}arth}.{\BBCQ}
\newblock
\APACjournalVolNumPages{Nature}{598}{7880}{276--280}.
\PrintBackRefs{\CurrentBib}

\bibitem [\protect \citeauthoryear {%
Vladilo%
\ \protect \BOthers {.}}{%
Vladilo%
\ \protect \BOthers {.}}{%
{\protect \APACyear {2013}}%
}]{%
Vladilo+2013}
\APACinsertmetastar {%
Vladilo+2013}%
\begin{APACrefauthors}%
Vladilo, G.%
, Murante, G.%
, Silva, L.%
, Provenzale, A.%
, Ferri, G.%
\BCBL {}\ \BBA {} Ragazzini, G.%
\end{APACrefauthors}%
\unskip\
\newblock
\APACrefYearMonthDay{2013}{}{}.
\newblock
{\BBOQ}\APACrefatitle {The habitable zone of Earth-like planets with different
  levels of atmospheric pressure} {The habitable zone of earth-like planets
  with different levels of atmospheric pressure}.{\BBCQ}
\newblock
\APACjournalVolNumPages{The Astrophysical Journal}{767}{1}{65}.
\PrintBackRefs{\CurrentBib}

\bibitem [\protect \citeauthoryear {%
Watson%
, Donahue%
\BCBL {}\ \BBA {} Walker%
}{%
Watson%
\ \protect \BOthers {.}}{%
{\protect \APACyear {1981}}%
}]{%
Watson+1981}
\APACinsertmetastar {%
Watson+1981}%
\begin{APACrefauthors}%
Watson, A\BPBI J.%
, Donahue, T\BPBI M.%
\BCBL {}\ \BBA {} Walker, J\BPBI C.%
\end{APACrefauthors}%
\unskip\
\newblock
\APACrefYearMonthDay{1981}{}{}.
\newblock
{\BBOQ}\APACrefatitle {The dynamics of a rapidly escaping atmosphere:
  {A}pplications to the evolution of {E}arth and {V}enus} {The dynamics of a
  rapidly escaping atmosphere: {A}pplications to the evolution of {E}arth and
  {V}enus}.{\BBCQ}
\newblock
\APACjournalVolNumPages{Icarus}{48}{}{150-166}.
\PrintBackRefs{\CurrentBib}

\bibitem [\protect \citeauthoryear {%
Way%
\ \BBA {} Del~Genio%
}{%
Way%
\ \BBA {} Del~Genio%
}{%
{\protect \APACyear {2020}}%
}]{%
Way+2020}
\APACinsertmetastar {%
Way+2020}%
\begin{APACrefauthors}%
Way, M\BPBI J.%
\BCBT {}\ \BBA {} Del~Genio, A\BPBI D.%
\end{APACrefauthors}%
\unskip\
\newblock
\APACrefYearMonthDay{2020}{}{}.
\newblock
{\BBOQ}\APACrefatitle {Venusian habitable climate scenarios: {M}odeling {V}enus
  through time and applications to slowly rotating {V}enus-like exoplanets}
  {Venusian habitable climate scenarios: {M}odeling {V}enus through time and
  applications to slowly rotating {V}enus-like exoplanets}.{\BBCQ}
\newblock
\APACjournalVolNumPages{Journal of Geophysical Research:
  Planets}{125}{5}{e2019JE006276}.
\PrintBackRefs{\CurrentBib}

\bibitem [\protect \citeauthoryear {%
Weiss%
\ \protect \BOthers {.}}{%
Weiss%
\ \protect \BOthers {.}}{%
{\protect \APACyear {2016}}%
}]{%
Weiss+2016}
\APACinsertmetastar {%
Weiss+2016}%
\begin{APACrefauthors}%
Weiss, M\BPBI C.%
, Sousa, F\BPBI L.%
, Mrnjavac, N.%
, Neukirchen, S.%
, Roettger, M.%
, Nelson-Sathi, S.%
\BCBL {}\ \BBA {} Martin, W\BPBI F.%
\end{APACrefauthors}%
\unskip\
\newblock
\APACrefYearMonthDay{2016}{}{}.
\newblock
{\BBOQ}\APACrefatitle {The physiology and habitat of the last universal common
  ancestor} {The physiology and habitat of the last universal common
  ancestor}.{\BBCQ}
\newblock
\APACjournalVolNumPages{Nature microbiology}{1}{9}{1--8}.
\PrintBackRefs{\CurrentBib}

\bibitem [\protect \citeauthoryear {%
Wordsworth%
}{%
Wordsworth%
}{%
{\protect \APACyear {2016}}%
}]{%
Wordsworth2016}
\APACinsertmetastar {%
Wordsworth2016}%
\begin{APACrefauthors}%
Wordsworth, R.%
\end{APACrefauthors}%
\unskip\
\newblock
\APACrefYearMonthDay{2016}{}{}.
\newblock
{\BBOQ}\APACrefatitle {Atmospheric nitrogen evolution on {E}arth and {V}enus}
  {Atmospheric nitrogen evolution on {E}arth and {V}enus}.{\BBCQ}
\newblock
\APACjournalVolNumPages{Earth and Planetary Science Letters}{447}{}{103--111}.
\PrintBackRefs{\CurrentBib}

\bibitem [\protect \citeauthoryear {%
Wordsworth%
\ \BBA {} Pierrehumbert%
}{%
Wordsworth%
\ \BBA {} Pierrehumbert%
}{%
{\protect \APACyear {2014}}%
}]{%
Wordsworth+Pierrehumbert2014}
\APACinsertmetastar {%
Wordsworth+Pierrehumbert2014}%
\begin{APACrefauthors}%
Wordsworth, R.%
\BCBT {}\ \BBA {} Pierrehumbert, R.%
\end{APACrefauthors}%
\unskip\
\newblock
\APACrefYearMonthDay{2014}{}{}.
\newblock
{\BBOQ}\APACrefatitle {Abiotic oxygen-dominated atmospheres on terrestrial
  habitable zone planets} {Abiotic oxygen-dominated atmospheres on terrestrial
  habitable zone planets}.{\BBCQ}
\newblock
\APACjournalVolNumPages{The Astrophysical Journal Letters}{785}{2}{L20}.
\PrintBackRefs{\CurrentBib}

\bibitem [\protect \citeauthoryear {%
Worsham%
\ \BBA {} Kleine%
}{%
Worsham%
\ \BBA {} Kleine%
}{%
{\protect \APACyear {2021}}%
}]{%
Worsham+Kleine2021}
\APACinsertmetastar {%
Worsham+Kleine2021}%
\begin{APACrefauthors}%
Worsham, E\BPBI A.%
\BCBT {}\ \BBA {} Kleine, T.%
\end{APACrefauthors}%
\unskip\
\newblock
\APACrefYearMonthDay{2021}{}{}.
\newblock
{\BBOQ}\APACrefatitle {Late accretionary history of {E}arth and {M}oon
  preserved in lunar impactites} {Late accretionary history of {E}arth and
  {M}oon preserved in lunar impactites}.{\BBCQ}
\newblock
\APACjournalVolNumPages{Science Advances}{7}{44}{eabh2837}.
\PrintBackRefs{\CurrentBib}

\bibitem [\protect \citeauthoryear {%
Yoshioka%
, Wiedenbeck%
, Shcheka%
\BCBL {}\ \BBA {} Keppler%
}{%
Yoshioka%
\ \protect \BOthers {.}}{%
{\protect \APACyear {2018}}%
}]{%
Yoshioka+2018}
\APACinsertmetastar {%
Yoshioka+2018}%
\begin{APACrefauthors}%
Yoshioka, T.%
, Wiedenbeck, M.%
, Shcheka, S.%
\BCBL {}\ \BBA {} Keppler, H.%
\end{APACrefauthors}%
\unskip\
\newblock
\APACrefYearMonthDay{2018}{}{}.
\newblock
{\BBOQ}\APACrefatitle {Nitrogen solubility in the deep mantle and the origin of
  {E}arth's primordial nitrogen budget} {Nitrogen solubility in the deep mantle
  and the origin of {E}arth's primordial nitrogen budget}.{\BBCQ}
\newblock
\APACjournalVolNumPages{Earth and Planetary Science Letters}{488}{}{134--143}.
\PrintBackRefs{\CurrentBib}

\bibitem [\protect \citeauthoryear {%
Zahnle%
, Gacesa%
\BCBL {}\ \BBA {} Catling%
}{%
Zahnle%
\ \protect \BOthers {.}}{%
{\protect \APACyear {2019}}%
}]{%
Zahnle2019}
\APACinsertmetastar {%
Zahnle2019}%
\begin{APACrefauthors}%
Zahnle, K\BPBI J.%
, Gacesa, M.%
\BCBL {}\ \BBA {} Catling, D\BPBI C.%
\end{APACrefauthors}%
\unskip\
\newblock
\APACrefYearMonthDay{2019}{}{}.
\newblock
{\BBOQ}\APACrefatitle {Strange messenger: A new history of hydrogen on {E}arth,
  as told by {X}enon} {Strange messenger: A new history of hydrogen on {E}arth,
  as told by {X}enon}.{\BBCQ}
\newblock
\APACjournalVolNumPages{Geochimica et Cosmochimica Acta}{244}{}{56--85}.
\PrintBackRefs{\CurrentBib}

\bibitem [\protect \citeauthoryear {%
Zahnle%
\ \BBA {} Kasting%
}{%
Zahnle%
\ \BBA {} Kasting%
}{%
{\protect \APACyear {1986}}%
}]{%
Zahnle+1986}
\APACinsertmetastar {%
Zahnle+1986}%
\begin{APACrefauthors}%
Zahnle, K\BPBI J.%
\BCBT {}\ \BBA {} Kasting, J\BPBI F.%
\end{APACrefauthors}%
\unskip\
\newblock
\APACrefYearMonthDay{1986}{}{}.
\newblock
{\BBOQ}\APACrefatitle {Mass fractionation during transonic escape and
  implications for loss of water from {M}ars and {V}enus} {Mass fractionation
  during transonic escape and implications for loss of water from {M}ars and
  {V}enus}.{\BBCQ}
\newblock
\APACjournalVolNumPages{Icarus}{68}{3}{462--480}.
\PrintBackRefs{\CurrentBib}

\bibitem [\protect \citeauthoryear {%
Zerkle%
}{%
Zerkle%
}{%
{\protect \APACyear {2018}}%
}]{%
Zerkle2018}
\APACinsertmetastar {%
Zerkle2018}%
\begin{APACrefauthors}%
Zerkle, A\BPBI L.%
\end{APACrefauthors}%
\unskip\
\newblock
\APACrefYearMonthDay{2018}{}{}.
\newblock
{\BBOQ}\APACrefatitle {Biogeodynamics: bridging the gap between surface and
  deep {E}arth processes} {Biogeodynamics: bridging the gap between surface and
  deep {E}arth processes}.{\BBCQ}
\newblock
\APACjournalVolNumPages{Philosophical Transactions of the Royal Society A:
  Mathematical, Physical and Engineering Sciences}{376}{2132}{20170401}.
\PrintBackRefs{\CurrentBib}

\bibitem [\protect \citeauthoryear {%
Zerkle%
\ \BBA {} Mikhail%
}{%
Zerkle%
\ \BBA {} Mikhail%
}{%
{\protect \APACyear {2017}}%
}]{%
Zerkle+Mikhail2017}
\APACinsertmetastar {%
Zerkle+Mikhail2017}%
\begin{APACrefauthors}%
Zerkle, A\BPBI L.%
\BCBT {}\ \BBA {} Mikhail, S.%
\end{APACrefauthors}%
\unskip\
\newblock
\APACrefYearMonthDay{2017}{}{}.
\newblock
{\BBOQ}\APACrefatitle {The geobiological nitrogen cycle: {F}rom microbes to the
  mantle} {The geobiological nitrogen cycle: {F}rom microbes to the
  mantle}.{\BBCQ}
\newblock
\APACjournalVolNumPages{Geobiology}{15}{3}{343--352}.
\PrintBackRefs{\CurrentBib}

\end{thebibliography}

%
%
%
%
%

\end{document}